\documentclass[aip,jcp,amsmath,reprint]{revtex4-1}
\usepackage{mathptmx,jpcrd}
\usepackage{textcomp,graphicx,dcolumn,xspace}
\newcolumntype{e}[1]{D{.}{.}{#1}}

\usepackage{hyperref}
\hypersetup{bookmarksnumbered=true,bookmarksopen,bookmarksopenlevel=1,
            colorlinks,allcolors=blue,urlcolor=black,
			bookmarksdepth=2,pdfpagemode=UseOutlines,pdfprintscaling=None,
            pdfauthor={Vincent Holten, Jan V. Sengers, and Mikhail A. Anisimov},
            pdftitle={Equation of state for supercooled water at pressures up to 400 MPa}}

\usepackage[T1]{fontenc}
\usepackage[ansinew]{inputenc}

\newcommand{\ea}{\textit{et al.}\xspace}
\newcommand{\figref}[1]{Fig.~\ref{#1}}
\newcommand{\ffigref}[1]{Figure~\ref{#1}}

\renewcommand*{\eqref}[1]{Eq.~(\ref{#1})}
\newcommand*{\eeqref}[1]{Equation~(\ref{#1})}
\newcommand*{\eqsref}[2]{Eqs.~(\ref{#1}) and (\ref{#2})}

\newcommand*{\secref}[1]{Sec.~\ref{#1}}
\newcommand{\tabref}[1]{Table~\ref{#1}}
\newcommand{\pypxl}[2]{{\partial{#1}/\partial{#2}}}
\newcommand{\pypxd}[3]{\left(\frac{\partial{#1}}{\partial{#2}}\right)_{#3}}
\newcommand{\di}{\mathrm{d}}

\newcommand*{\ts}[1]{_\text{#1}}
\newcommand*{\tu}[1]{^\text{#1}}

\newcommand*{\Pv}{P_{\sigma}}
\newcommand*{\THN}{T\ts{H}}

\begin{document}
\title{Equation of state for supercooled water at pressures up to 400 MPa}
\affiliation{Institute for Physical Science and Technology and
  Department of Chemical and Biomolecular Engineering,\\
  University of Maryland, College Park, Maryland 20742, USA}
\author{Vincent Holten}
\author{Jan V. Sengers}
\author{Mikhail A. Anisimov}
\email[Author to whom correspondence should be addressed; electronic mail: ]
{anisimov@umd.edu}
\date{\today}

\begin{abstract}
An equation of state is presented for the thermodynamic properties of cold and
supercooled water. It is valid for temperatures from the homogeneous ice nucleation
temperature up to 300~K and for pressures up to 400~MPa, and can be extrapolated up to
1000~MPa. The equation of state is compared with experimental data for the density,
expansion coefficient, isothermal compressibility, speed of sound, and heat capacity.
Estimates for the accuracy of the equation are given. The melting curve of ice I is
calculated from the phase-equilibrium condition between the proposed equation and an
existing equation of state for ice I.
\end{abstract}

\maketitle

\tableofcontents

\section{Introduction}
Supercooled water has been of interest to science since it was first described by
Fahrenheit in 1724.\cite{fahrenheit1724} At atmospheric pressure, water can exist as a
metastable liquid down to 235~K, and supercooled water has been observed in clouds down
to this temperature.\cite{rosenfeld2000,heymsfield1993} Properties of supercooled water
are important for meteorological and climate
models\cite{schmelzerhellmuthbook,skogseth2009} and for
cryobiology.\cite{nashbook1966,song2010} Furthermore, the thermodynamic properties of
cold and supercooled water at high pressure are needed for the design of food
processing.\cite{otero2002}

It is well known that several properties of supercooled water -- such as the isobaric
heat capacity, the expansion coefficient, and the isothermal compressibility -- show
anomalous behavior; they increase or decrease rapidly with cooling. A liquid--liquid
phase transition, terminated by a critical point, hidden below the homogeneous ice
nucleation temperature has been proposed to explain this anomalous thermodynamic
behavior.\cite{poole1992,mishima1998review}

Several equations of state for supercooled water have been published. Sato\cite{sato1990}
proposed an equation of state for water in the liquid phase including the metastable
state, valid up to 100~MPa. Jeffery and Austin\cite{jeffery1997,jeffery1999} developed an
analytic equation of state of H$_2$O that also covers the supercooled region. Kiselev and
Ely\cite{kiselev2002} made an early attempt to describe supercooled water in terms of an
equation of state incorporating critical behavior. Anisimov and
coworkers\cite{fuentevilla2006,bertrand2011,holtenSCW,holtentwostate} also based their
equations of state on an assumed liquid--liquid critical point. Since the publication of
these equations, new experimental data have become available that enable development of
an equation of state with a significantly improved accuracy.

The equation of state of this work was developed with the following aims:

1. It should represent the experimental data of liquid water in the metastable region as
well as possible. This work only considers supercooled water above the homogeneous
nucleation temperature. The equation does not cover the glassy state of water (below
136~K at atmospheric pressure\cite{loerting2011PCCP}).

2. The current reference for the thermodynamic properties of water is the IAPWS-95
formulation.\cite{wag02nonote,iapws95} IAPWS-95 is, strictly speaking, valid only at
temperatures above the melting curve. When extrapolated into the supercooled region,
IAPWS-95 also yields a good description of the data in the supercooled region that were
available at the time the formulation was developed. For practical use, a new formulation
for the thermodynamic properties of supercooled water should smoothly connect with the
IAPWS-95 formulation at higher temperatures without significant discontinuities at the
point of switching.

3. The correlation should allow extrapolation up to 1000~MPa. There are only a few data
in the supercooled region above 400~MPa, but smooth extrapolation up to 1000~MPa would be
desirable.

\section{Experimental Data}
Most of the experimental data that were considered in this work have been reviewed
before.\cite{angell1982book,angell83,sato1991,wag02nonote,deben03,holtenSCW} In this
section, we mainly discuss new data and data that were treated differently than in our
previous work.\cite{holtenSCW,holtenMF2012,holtentwostate}

\subsection{Density}\label{sec:densitydata}

\begin{figure*}
\includegraphics{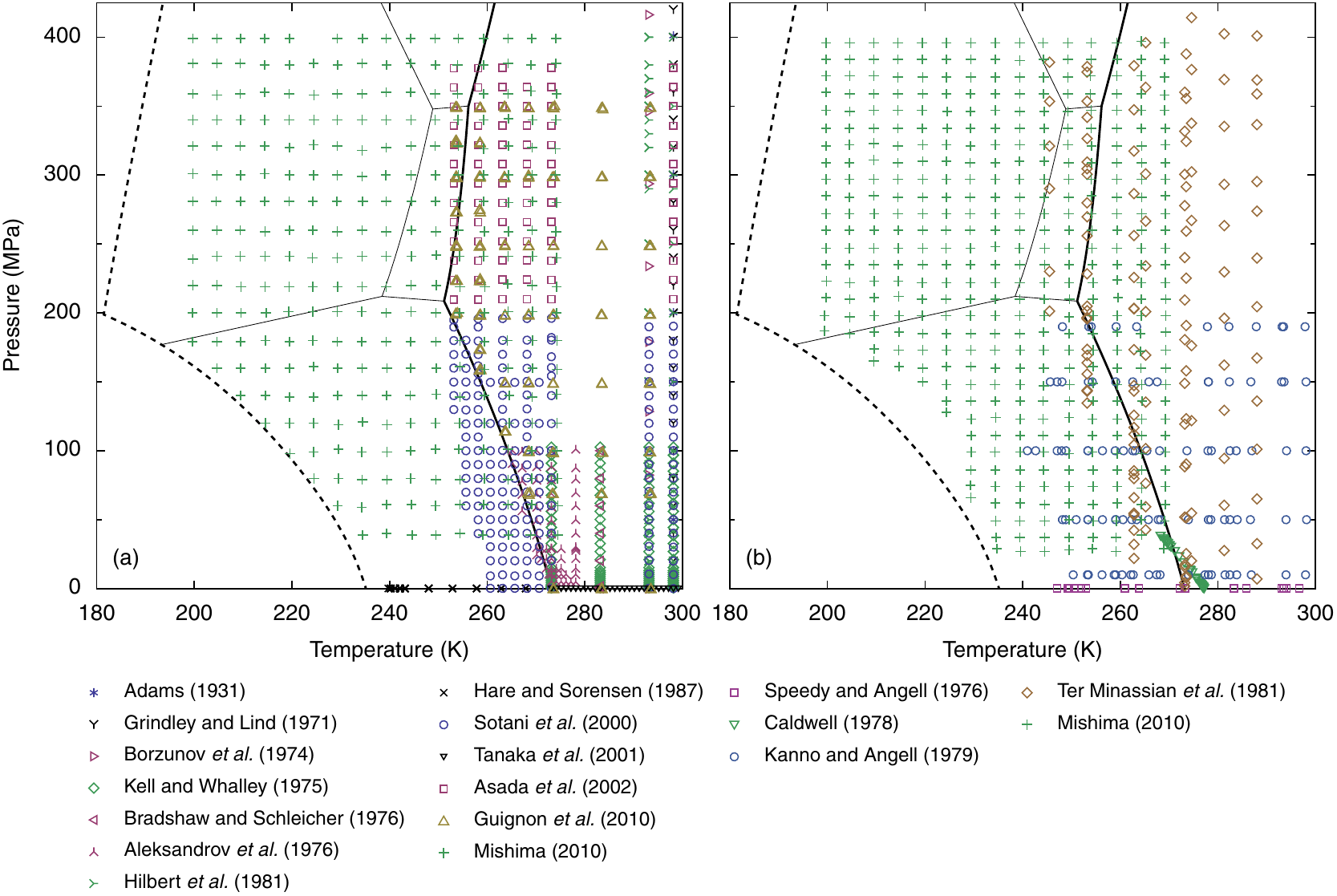}
\caption{\label{fig:ExpPT}(a) Location of experimental H$_2$O density data considered in
this work.
\cite{adams1931,grindley1971,borzunov1974,kellwhalley1975,bradshaw1976,%
aleksandrov1976rhorussian,*aleksandrov1976rho,%
hilbert1981,hare87,sotani2000,tanaka2001,asada2002,mishima2010,guignon2010} The thick
solid curve is the melting curve,\cite{iapwsmeltsub2011,wagner2011} the dashed curve is
the homogeneous ice nucleation limit (see Appendix~\ref{app:TH}), and the thin solid
curves are the ice phase boundaries.\cite{bridgman1912,kell1968} (b) Location of the
experimental H$_2$O density-derivative data. Ter Minassian \ea\cite{terminassian1981} and
Caldwell\cite{caldwell1978} have measured the expansivity; other authors
\cite{speedy1976,kanno1979,mishima2010} have measured the isothermal compressibility.}
\end{figure*}

\begin{table*}\caption{\label{tab:density}Experimental density data}
\begin{ruledtabular}
\begin{tabular}{lllllll}
                                 &              & Temperature & Pressure    & Density\\
Reference                        & Year         & range (K) & range (MPa)   & uncertainty (\%)  & Source%
\footnote{T = table from article, S = table from supplement, G = extracted from graph}  & Included in fit\\
\hline
Adams\cite{adams1931}               & 1931  & 298       & 0.1--900      & 0.1           & T     & --\\
Grindley \& Lind\cite{grindley1971} & 1971  & 298--423  & 20--800       & 0.02          & T     & Yes\\
Borzunov \ea\cite{borzunov1974}     & 1974  & 293--338  & 0--923        & 0.1           & T     & --\\
Kell \& Whalley\cite{kellwhalley1975}& 1975 & 273--423  & 0.5--103      & 0.001--0.003  & T     & Yes\\
Bradshaw \& Schleicher\cite{bradshaw1976}& 1976 & 283   & 0.1--100      & 0.007         & T     & --\\
Aleksandrov \ea\cite{aleksandrov1976rhorussian}
                                    & 1976  & 264--278  & 5--101        & 0.1\footnotemark[3]  & T     & --\\
Hare \& Sorensen\cite{hare87}       & 1987  & 240--268  & 0.101325      & 0.02          & T     & Yes\\
Sotani \ea\cite{sotani1998}         & 1998  & 253--293  & 0--200        & 0.05          & --\footnote{Superseded by Sotani \ea\cite{sotani2000}} & --\\
Sotani \ea\cite{sotani2000}         & 2000  & 253--298  & 0--196        & 0.03\footnotemark[3]  & G & Yes\\
Tanaka \ea\cite{tanaka2001}         & 2001  & 273--313  & 0.101325      & 0.0001        & T     & --\\
Asada \ea\cite{asada2002}           & 2002  & 253--298  & 210--378      & 0.1           & G     & Yes\\
Guignon \ea\cite{guignon2010}       & 2010  & 254--323  & 0.1--350      & 0.2           & T     & --\\
Mishima\cite{mishima2010}           & 2010  & 200--275  & 39--399       & 0.5\footnotemark[4] & S & Yes\\
\end{tabular}
\end{ruledtabular}
\footnotetext[3]{Estimated by Wagner and Thol\cite{wagnerthol2013}}%
\footnotetext[4]{Uncertainty is unknown below 253~K, see the text}
\end{table*}

The experimental density data that were considered in this work are listed in
\tabref{tab:density} and shown in \figref{fig:ExpPT}(a). Additional references to older
data can be found in the articles of Teká\v{c} \ea\cite{tekac1985} and Wagner and
Pruß.\cite{wag02nonote} In a large part of the supercooled region, the only available
density data are those of Mishima.\cite{mishima2010} As a result, it is difficult to
estimate the systematic error of these data at low temperatures. In a graph in his
article,\cite{mishima2010} Mishima showed the random (type A) uncertainty for each data
point, which is 0.2\% on average and at most 0.5\%. The systematic (type B) uncertainty
can only be estimated above 253 K, in the region of overlap with density data of Kell and
Whalley,\cite{kellwhalley1975} Sotani \ea,\cite{sotani2000} and Asada \ea\cite{asada2002}
In this region, the densities of Mishima deviate systematically by up to 0.4\% from these
other data. Below 253~K, the systematic uncertainty is unknown. As in earlier
work,\cite{holtenSCW} we adjusted the density values of Mishima, under the assumption
that the systematic deviation at low temperatures, where it is not known, is the same as
at higher temperatures, where it can be calculated. It was found that the adjusted data
of Mishima do not completely agree with the expansivity measurements of Ter Minassian
\ea,\cite{terminassian1981} which we consider to be more accurate. Therefore, the
adjusted data of Mishima were included in the fit of the equation of state with a
relatively low weight.

The only experimental density data at atmospheric pressure that were included in the fit
are those of Hare and Sorensen,\cite{hare87} which are considered to be the best
available. For pressures higher than atmospheric, we included data from Sotani
\ea,\cite{sotani2000} Asada \ea,\cite{asada2002} and Kell and
Whalley.\cite{kellwhalley1975} To enable extrapolation of the equation above 400~MPa,
density data from Grindley and Lind\cite{grindley1971} up to 800~MPa were included in the
fit.

\subsection{Density derivatives}
Several data sets exist for temperature and pressure derivatives of the density $\rho$.
The cubic expansion coefficient $\alpha_P$, also known as expansivity, is defined as
\begin{equation}
    \alpha_P = -\frac{1}{\rho}\pypxd{\rho}{T}{P},
\end{equation}
where $T$ is the temperature and $P$ is the pressure. The isothermal compressibility
$\kappa_T$ is defined as
\begin{equation}
    \kappa_T = \frac{1}{\rho}\pypxd{\rho}{P}{T}.
\end{equation}
The data sets listed in \tabref{tab:compexp} were all included in the fit, with the
exception of the compressibility data of Mishima.\cite{mishima2010} Mishima's data were
not included because they may be affected by systematic errors of unknown size at low
temperatures. In previous work,\cite{holtenSCW,holtentwostate} expansivities reported by
Hare and Sorensen\cite{hare87} were included in the fit. However, Hare and Sorensen did
not measure the expansivity directly, but derived it from a fit to their density data.
Because we already included Hare and Sorensen's density data in our fit, their
expansivity data were not used in the fit. Expansivity values from Ter Minassian
\ea\cite{terminassian1981} were calculated from their empirical correlation. The accuracy
of their correlation is not given; the relative difference with expansivities calculated
from IAPWS-95 is at most 3.2\% in the range of 300~K to 380~K and 0~MPa to 400~MPa.

At points in the phase diagram where the expansivity is zero, the density has a maximum
with respect to temperature. The temperature at which this occurs is usually referred to
as the temperature of maximum density (TMD). Caldwell\cite{caldwell1978} measured the TMD
for pressures up to 38~MPa, and these measurements were included in the expansivity data
set of the fit as $\alpha_P = 0$ points. The recent TMD measurements of Hiro
\ea\cite{hiro2013} were not used, because they deviate systematically by about 1.5~K from
more accurate data.

\begin{table} \caption{\label{tab:compexp}Experimental data on compressibility and expansivity}
\begin{ruledtabular}
\begin{tabular}{lllll}
              &                                 & Temperature & Pressure\\
Reference     & Year                            & range (K) & range (MPa)   & Source%
\footnote{T = table from article, S = table from supplement, G = extracted from graph}\\
\hline
\multicolumn{5}{c}{\emph{Compressibility data}}\\
Speedy \& Angell\cite{speedy1976}           & 1976  & 247--297  & 0.101325      & G\\
Kanno \& Angell\cite{kanno1979}             & 1979  & 241--298  & 10--190       & G\\
Mishima\cite{mishima2010}                   & 2010  & 199--269  & 27--397       & S\\
\multicolumn{5}{c}{\emph{Expansivity data}}\\
Caldwell\cite{caldwell1978}                 & 1978  & 268--277  & 0.1--38       & T\footnotemark[2]\\
Ter Minassian \ea\cite{terminassian1981}    & 1981  & 246--410  & 2--636        & G\footnotemark[2]\\
\end{tabular}
\end{ruledtabular}
\footnotetext[2]{An empirical correlation is also provided}
\end{table}

\subsection{Speed of sound}
\begin{figure}
\includegraphics{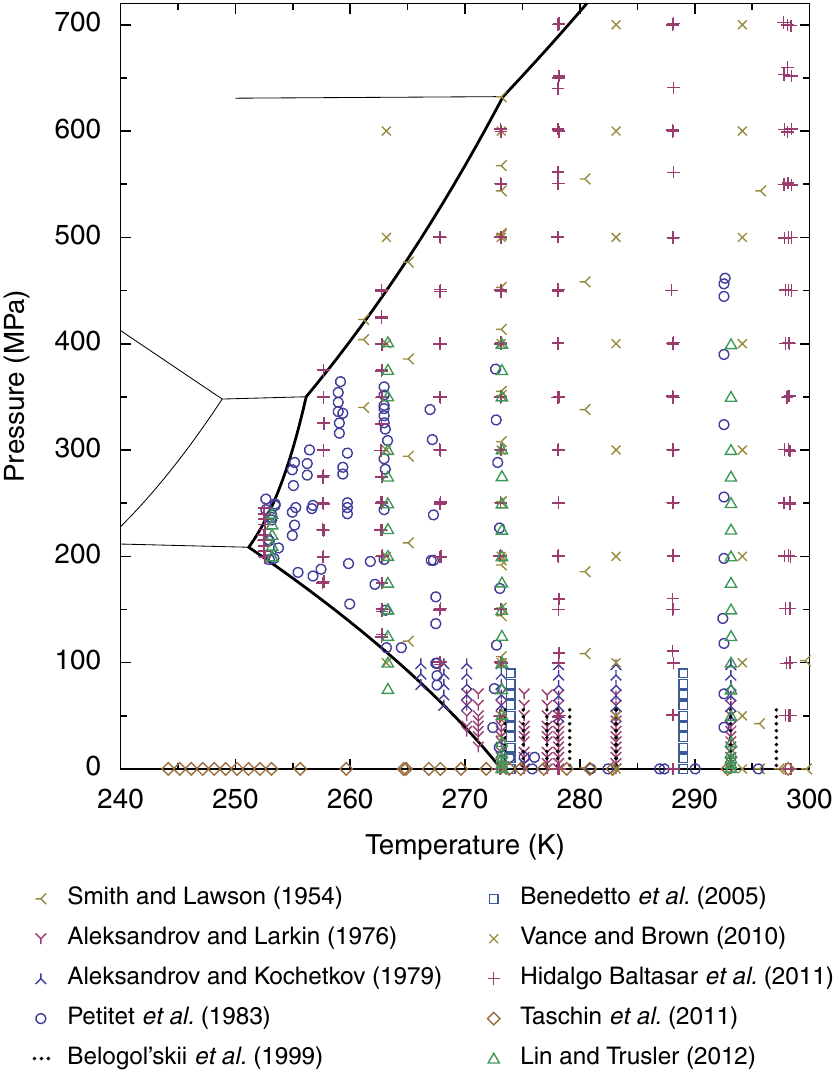}
\caption{\label{fig:wPT}Location of experimental data on the speed of sound considered
in this work.
\cite{smithlawson1954,aleksandrov1976,aleksandrov1979,petitet1983,belogolskii1999,%
benedetto2005,vance2010,hidalgobaltasar2011,taschin2011,lintrusler2012}
The thick curve is the melting curve,\cite{iapwsmeltsub2011,wagner2011} and the thin curves are
the ice phase boundaries.\cite{bridgman1912,kell1968}
Belogol'skii \ea\cite{belogolskii1999} did not publish their individual data points;
their reported isotherms are shown as dotted lines.}
\end{figure}

The experimental data on the speed of sound considered in this article are given in
\tabref{tab:w} and shown in \figref{fig:wPT}. Recent data that were not considered in
previous work are the accurate measurements of the speed of sound by Lin and
Trusler\cite{lintrusler2012} down to of 253~K and from 1~MPa to 400~MPa. Although there
are few data points in the supercooled region, the accuracy of 0.03\%--0.04\% makes this
an important data set. Lin and Trusler also derived densities and isobaric heat
capacities by integrating their speed-of-sound data. We have not considered these derived
properties in the development of the equation of state in this work for the following
reason. To enable integration of the speed of sound, Lin and Trusler represented their
experimental data on the speed of sound by an empirical correlation. The experimental
data of Lin and Trusler are closer to the prediction of our equation of state than to
their correlation, in the temperature range considered here (\secref{sec:wcompare})
Therefore, densities and heat capacities calculated from our equation of state are more
accurate than the values derived by Lin and Trusler.

The work of Smith and Lawson\cite{smithlawson1954} deserves mention because they were
likely the first to measure the speed of sound below 273~K at elevated pressures.
However, their pressure calibration has an uncertainty of about 1\%, as discussed by
Holton \ea,\cite{holton1968} and their data were not further considered for this work.

The most accurate measurements of the speed of sound in the range from 273~K to 300~K and
up to 60~MPa are those of Belogol'skii \ea\cite{belogolskii1999,*[See also: ]leroy2008}
They presented a correlation that represents their data with a standard deviation of
0.003\% in the speed of sound. We estimated the accuracy of this correlation by comparing
it to the experimental data of Lin and Trusler. For this comparison, Lin and Trusler's
speeds of sound on each of their isotherms were corrected to compensate for their small
deviation at atmospheric pressure. For each isotherm, this correction involved fitting a
third-degree polynomial to the isothermal data in the range of 1~MPa to 100~MPa and
extrapolating this fit to 0.101325~MPa, where the ratio with the speed of sound computed
from IAPWS-95 was calculated, after which all speed-of-sound values on the isotherm were
divided by that ratio. After this correction, the difference between the data of Lin and
Trusler and the correlation of Belogol'skii \ea is at most 0.01\%, which suggests that
the correlation of Belogol'skii \ea has an accuracy of 0.01\% or better in the speed of
sound. Measurements of Aleksandrov and
Larkin\cite{aleksandrov1976,*aleksandrov1976russian} in this temperature and pressure
range have a slightly higher uncertainty of 0.02\%. The data presented by
Mamedov\cite{mamedov1979,*mamedov1979russian} are not considered here, because Mamedov
published rounded data of Aleksandrov and Larkin.\cite{aleksandrov1976} Aleksandrov and
Kochetkov\cite{aleksandrov1979,*aleksandrov1979russian} used the setup described by
Aleksandrov and Larkin\cite{aleksandrov1976} to measure the speed of sound down to 266~K
and up to 100~MPa. A comparison with the data of Lin and Trusler\cite{lintrusler2012}
suggests that the accuracy of the data of Aleksandrov and Kochetkov\cite{aleksandrov1979}
is about~0.1\%.

To improve the extrapolation behavior of the equation above 400~MPa, data from Vance and
Brown\cite{vance2010} up to 700~MPa were included in the fit. The data from Hidalgo
Baltasar \ea,\cite{hidalgobaltasar2011} which also extend up to 700~MPa, were not
included because they systematically deviate from other data (\secref{sec:wcompare})

At atmospheric pressure in the supercooled region, the data of Taschin
\ea\cite{taschin2011} seem to be the best available; they are consistent with other
thermodynamic properties.\cite{taschin2011} Above 273.15~K, the data deviate at most
0.15\% from the IAPWS-95 formulation, and the uncertainty below 260~K is 0.7\%.

\begin{table*}\caption{\label{tab:w}Experimental data on the speed of sound}
\begin{ruledtabular}
\begin{tabular}{llllllll}
                                                &       & Temperature & Pressure    & Frequency & Speed-of-sound    &         & Included\\
Reference                                       & Year  & range (K)   & range (MPa) & (MHz)     & uncertainty (\%)  & Source%
\footnote{T = table from article, A = data provided by authors, G = extracted from graph, C = calculated from correlation} & in fit\\
\hline
Smith \& Lawson\cite{smithlawson1954}           & 1954  & 261--402  & 0.1--923      & 12        & --                & T & --\\
Wilson\cite{wilson1959}                         & 1959  & 274--364  & 0.1--97       & 5         & 0.1%
                    \footnote{Wilson\cite{wilson1959} estimated the uncertainty at 0.01\%. The estimate of 0.1\% is from Sato \ea\cite{sato1991}}                                                          & T & --\\
Del Grosso \& Mader\cite{delgrosso1972}         & 1972  & 273--368  & 0.101325      & 5         & 0.001             & T & --\\
Aleksandrov \& Larkin\cite{aleksandrov1976}     & 1976  & 270--647  & 0.1--71       & 3         & 0.02%
                    \footnote{Uncertainty below 303~K, estimated from comparison with values from Belogol'skii \ea\cite{belogolskii1999}}
                                                                                                                    & T & Yes\\
Trinh \& Apfel
\cite{trinhapfel1978JASA,trinhapfel1978JCP}     & 1978  & 256--283  & 0.101325      & 2--3      & 0.2               & G & --\\
Aleksandrov \& Kochetkov\cite{aleksandrov1979}   & 1979  & 266--423  & 6--99         & 2.5, 5.6  & 0.1               & T & --\\
Bacri \& Rajaonarison\cite{bacri1979}           & 1979  & 247--280  & 0.101325      & 925       & --                & G & --\\
Trinh \& Apfel\cite{trinhapfel1980}             & 1980  & 240--256  & 0.101325      & 0.054     & 1.3               & G & --\\
Petitet \ea\cite{petitet1983}                   & 1983  & 253--296  & 0.1--462      & 10        & 0.1               & T & --\\
Fujii \& Masui\cite{fujii1993}                  & 1993  & 293--348  & 0.101325      & 16        & 0.001             & T & --\\
Belogol'skii \ea\cite{belogolskii1999}          & 1999  & 273--313  & 0.1--60       & 5--10     & 0.01%
                    \footnote{Estimated from comparison with data from Lin \& Trusler\cite{lintrusler2012}
                    after correcting for systematic deviations at atmospheric pressure}                             & C & Yes\\
Benedetto \ea\cite{benedetto2005}               & 2005  & 274--394  & 0.1--90       & 5         & 0.05              & T & --\\
Vance \& Brown\cite{vance2010}                  & 2010  & 263--371  & 0.1--700      & 400--700  & 0.2--0.3          & T & Yes\\
Taschin \ea\cite{taschin2006,taschin2011}       & 2011  & 244--363  & 0.101325      & 140       & 0.7               & A & Yes\\
Hidalgo Baltasar \ea\cite{hidalgobaltasar2011}  & 2011  & 252--350  & 0.1--705      & 2         & 0.2--0.3          & T & --\\
Lin \& Trusler\cite{lintrusler2012}             & 2012  & 253--473  & 1--401        & 5         & 0.03--0.04        & T & Yes\\
\end{tabular}
\end{ruledtabular}
\end{table*}

\subsection{Heat capacity}\label{sec:cpdata}
The isobaric heat capacity $c_P$ of cold and supercooled water at atmospheric pressure
has been measured by several investigators; a list is given in \tabref{tab:cp}. There are
two data sets that extend down to 236~K, those of Angell \ea\cite{angell1982} and Archer
and Carter.\cite{archer1993} The difference between the data sets increases with
decreasing temperature, and at 236~K, the heat capacity reported by Angell
\ea\cite{angell1982} is 5\% higher than that found by Archer and Carter.\cite{archer1993}
Because it is not known which data set is best, the equation of the current work was
initially not fitted to any heat-capacity data in the supercooled region. It was found
that most of the preliminary equations predicted heat capacities in agreement with the
data of Angell \ea,\cite{angell1982} and were close to values calculated from the
extrapolated IAPWS-95 formulation. However, in some cases, the predicted heat capacities
were slightly higher than those Angell \ea\cite{angell1982} Therefore, to reduce the
difference with the experimental data, values calculated from IAPWS-95 were added as
input for the fit.

There are only few measurements of $c_P$ at elevated pressures. The data of Sirota
\ea\cite{sirota1970} at pressures up to 98~MPa were included in the fit. The data of
Czarnota\cite{czarnota1984} were not considered accurate enough to be included in the
fit.

Recently, Manyà \ea \cite{manya2011} have measured $c_P$ at 4~MPa from 298~K to 465~K.
The results of Manyà \ea\ imply that the derivative $(\pypxl{c_P}{P})_T$ is positive for
pressures lower than 4~MPa, which contradicts the thermodynamic relation
$(\pypxl{c_P}{P})_T = -T (\partial^2 v / \partial T^2)_P$, where $v$ is the specific
volume. The sign of the second derivative in this relation is well known from isobaric
volumetric data. Hence, the data of Manyà \ea\ were not considered in this work.

\begin{table} \caption{\label{tab:cp}Experimental heat-capacity data}
\begin{ruledtabular}
\begin{tabular}{lllll}
                &                             & Temperature  & Pressure  \footnote{Data are at 0.101325~MPa unless otherwise specified}\\
Reference         & Year                      & range (K)    & range (MPa)  & Source%
\footnote{T = table from article, A = data provided by authors, G = extracted from graph}\\
\hline
Osborne \ea\cite{osborne1939}     & 1939      & 274--368  && T\\
Sirota \ea\cite{sirota1970}       & 1970      & 272--306  & 20--98  & T \\
Anisimov \ea\cite{anisimov1972}   & 1972      & 266--304  && G\\
Angell \ea\cite{angell1973}       & 1973      & 235--273  && T\footnote{Superseded by Angell \ea\cite{angell1982}}\\
Angell \ea\cite{angell1982}       & 1982      & 236--290  && T\\
Czarnota\cite{czarnota1984}       & 1984      & 299--300  & 224--1032 & T\\
Bertolini \ea\cite{bertolini1985} & 1985      & 247--254  && G\\
Tombari \ea\cite{tombari1999}     & 1999      & 245--283  && A\\
Archer \& Carter\cite{arc00}      & 2000      & 236--285  && T\\
\end{tabular}
\end{ruledtabular}
\end{table}

\subsection{Values from IAPWS-95}
\begin{figure}
\includegraphics{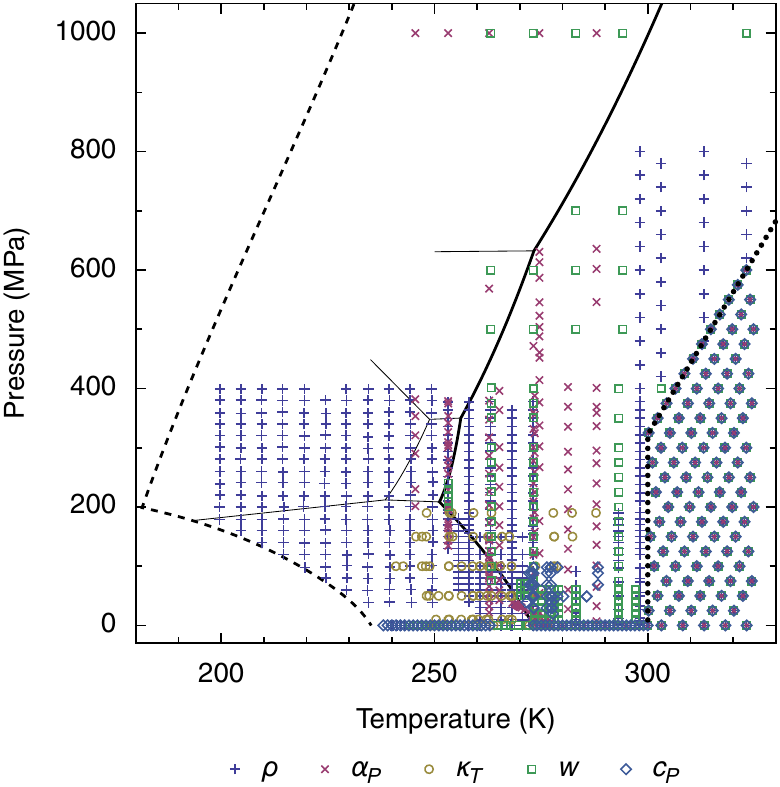}
\caption{\label{fig:ExpPTfit}Location of the experimental data on density $\rho$,
thermal expansivity $\alpha_P$, isothermal compressibility $\kappa_T$, speed of sound
$w$, and isobaric heat capacity $c_P$ that were selected as input for the fit.
On the dotted line and to the right of it, the source of the data is the IAPWS-95 formulation.
The thick solid curve is the
melting curve,\cite{iapwsmeltsub2011,wagner2011} the dashed curve is the homogeneous ice
nucleation limit (see Appendix~\ref{app:TH}), and the thin solid curves are the ice phase
boundaries.\cite{bridgman1912,kell1968}}
\end{figure}

To ensure a smooth connection to the IAPWS-95 formulation, the equation of state from
this work was fitted to property values calculated from IAPWS-95 in the temperature and
pressure range defined by
\begin{equation}\label{eq:iapwsrange}
\begin{gathered}
    T/\text{K} \geq 273.15 + (P/\text{MPa} - 0.1)/12 ,\\
    300 \leq T/\text{K} \leq 325.
\end{gathered}
\end{equation}
This range, shown in \figref{fig:ExpPTfit}, was determined from the differences between
values calculated from IAPWS-95 and from preliminary fits, as well as the deviations from
experimental data. Within the range defined by \eqref{eq:iapwsrange}, only IAPWS-95
values were included in the final fit. In addition, the equation of state was also fitted
to values from IAPWS-95 at atmospheric pressure from 273.15~K to 300~K. The locations of
all data that were included in the fit are shown in \figref{fig:ExpPTfit}.

\subsection{Adjustment of data}
\label{sec:dataadjustment}Temperatures in this work are expressed on the international
temperature scale of 1990 (ITS-90).\cite{prestonthomas1990} Temperatures on the IPTS-68
scale were converted to ITS-90 according to the equation of Rusby.\cite{rusby91}
Temperatures on the IPTS-48 scale were first converted to IPTS-68 and then to ITS-90. The
conversion from IPTS-48 to IPTS-68 was performed according to the equations given by
Bedford and Kirby;\cite{bedford1969} these conversion equations were found to agree with
those of Douglas.\cite{douglas1969} In the temperature range considered in this work, the
ITS-27 and IPTS-48 scales can be considered as identical,\cite{goldberg1992} so
temperatures on the ITS-27 scale were treated as IPTS-48 temperatures. In principle, the
values of quantities that depend on temperature intervals, such as the expansion
coefficient and the heat capacity, should also be
converted.\cite{rusby91,archer1993,weir1996} In this work, such an adjustment was only
found to be necessary for the accurate heat-capacity measurements at atmospheric pressure
close to the melting temperature;\cite{osborne1939,anisimov1972} the changes in heat
capacity due to the conversion were less than 0.1\%.

The absolute volumes measured by Bradshaw and Schleicher\cite{bradshaw1976} were
converted to densities, where the mass of the samples was calculated from the IAPWS-95
value for the density at atmospheric pressure. The density data of Grindley and
Lind\cite{grindley1971} show a systematic deviation from the more accurate data of Kell
and Whalley,\cite{kellwhalley1975} as was noted by Wagner and Pruß.\cite{wag02nonote} For
the isotherms in the range of 298~K to 323~K, which are considered in this work, the
relative density difference between the data of Grindley and Lind\cite{grindley1971} and
the data of Kell and Whalley\cite{kellwhalley1975} is roughly proportional to the
pressure. To prevent this deviation from affecting the fit, the data from Grindley and
Lind were corrected. The corrected density $\rho\ts{corr}$ was computed from the original
density $\rho$ as
\begin{equation}
    \rho\ts{corr} = \rho \times [1 - 1.1\times 10^{-6} (P/\text{MPa})].
\end{equation}
This adjustment is largest at 800~MPa, the highest pressure in the data of Grindley and
Lind, where the densities were reduced by 0.088\%. In the figures in this article, the
data of Grindley and Lind are shown without this adjustment. The speed of sound
measurements of Lin and Trusler\cite{lintrusler2012} at 273.21~K show a small systematic
deviation from more accurate data,\cite{delgrosso1972,belogolskii1999} and speed-of-sound
values on this isotherm were increased by 0.025\% to compensate for this deviation. In
the figures, the original values of Lin and Trusler\cite{lintrusler2012} are shown.
Differences between vapor pressures over liquid water and over ice measured by
Bottomley\cite{bottomley1978} were converted to absolute vapor pressures over liquid
water by adding the ice sublimation pressure calculated from the IAPWS
expression.\cite{iapwsmeltsub2011,wagner2011}

Many data for supercooled water are available only in graphical form and have not been
published as numerical values. In the case of recent publications, we requested the
authors to provide us with data in tabular form. In the case of older data or when the
authors could not be reached, the data were extracted from graphs. For all references,
the data source that we used is indicated in the tables in this section. All data are
provided in tabular form in the supplemental material.\cite{SCWsupplement2013}

\subsection{Values for extrapolation}
To enable extrapolation of the equation of state to 1000~MPa, it was found necessary to
guide the fit at high pressures by including estimated values for the expansivity and
speed of sound at 1000~MPa (\figref{fig:ExpPTfit}).

\clearpage
\section{Equation of State}
\subsection{Structure of the equation}
\newcommand*{\Th}{\hat{T}}
\newcommand*{\Gh}{\hat{g}}
\newcommand*{\GA}{\Gh\tu{A}}
\newcommand*{\GB}{\Gh\tu{B}}
\newcommand*{\Pc}{P_\text{c}}
\newcommand*{\Tc}{T_\text{c}}
\newcommand*{\Vref}{v_0}
\newcommand*{\rhoref}{\rho_0}
\newcommand*{\Ph}{\hat{P}}
\newcommand*{\Dt}{t}
\newcommand*{\Dp}{p}
\newcommand*{\DG}{\Delta \hat{g}}
\newcommand*{\Vh}{\hat{v}}
\newcommand*{\Sh}{\hat{s}}

\newcommand*{\xe}{x\ts{e}}
\newcommand*{\Lt}{L_{\Th}}
\newcommand*{\Lp}{L_{\Ph}}
\newcommand*{\Ltt}{L_{\Th\Th}}
\newcommand*{\Ltp}{L_{\Th\Ph}}
\newcommand*{\Lpp}{L_{\Ph\Ph}}
\newcommand{\Bp}{\GA_{\Ph}}
\newcommand{\Bt}{\GA_{\Th}}
\newcommand{\Bpp}{\GA_{\Ph\Ph}}
\newcommand{\Btp}{\GA_{\Th\Ph}}
\newcommand{\Btt}{\GA_{\Th\Th}}
\newcommand*{\kap}{\hat{\kappa}_T}
\newcommand*{\alp}{\hat{\alpha}_P}
\newcommand*{\Cph}{\hat{c}_P}
\newcommand*{\Cvh}{\hat{c}_v}

\newcommand*{\tb}{\tau}
\newcommand*{\pb}{\pi}

The thermodynamic formulation presented here is a mean-field version of an equation of
state developed in Ref.~\onlinecite{holtentwostate}. It is based on the so-called
two-state model, in which it is assumed that liquid water is a mixture of a high-density
structure A and a low-density structure B. There is experimental evidence for the
existence of two distinct local structures in water.\cite{nilsson2012,taschin2013}

Competition between these structures naturally explains the density anomaly and other
thermodynamic anomalies in cold water. In particular, if the excess Gibbs energy of
mixing of these two structures is positive, the nonideality of the ``mixture'' can be
sufficient to cause liquid--liquid separation, or, at least, to significantly reduce the
stability of the homogeneous liquid phase and consequently generate the anomalies in the
thermodynamic response functions. However, since experimental data are not yet available
beyond the homogeneous ice nucleation limit, the possibility of a liquid--liquid
transition in water must be postulated and is to be examined by indirect means. The
location of the hypothesized liquid--liquid critical point, characterized by the critical
temperature $\Tc$ and critical pressure $\Pc$, is obtained from the extrapolation of the
properties far away from the transition, thus making it very
uncertain.\cite{holtenSCW,holtentwostate}

We introduce the dimensionless quantities
\begin{gather}
\Th=\frac{T}{\Tc},\qquad \Ph=\frac{P\Vref}{R\Tc}, \qquad \Gh=\frac{g}{R\Tc},\qquad\Vh = \frac{v}{\Vref},\\
\Sh=\frac{s}{R},\qquad\Dt=\frac{T-\Tc}{\Tc},\qquad \Dp=\frac{(P-\Pc)\Vref}{R\Tc},
\end{gather}
where $T$ is the temperature, $P$ is the pressure, $g$ is the specific Gibbs energy, $R$
is the specific gas constant, $v$ is the specific volume, $\Vref$ is a reference volume,
and $s$ is the specific entropy. We adopt the equation of state for the Gibbs energy in
the form of ``athermal mixing'', suggested in
Ref.~\onlinecite{bertrand2011,holtentwostate},
\begin{equation}\label{eq:eos}
    \Gh = \GA + \Th\bigl[x L + x\ln x + (1-x)\ln(1-x) + \omega x(1-x)\bigr],
\end{equation}
where $\GA$ is the Gibbs energy of the hypothetical pure high-density structure, $x$ is
the fraction of the low-density structure, $\omega$ is an interaction parameter, and
\begin{equation}
    L = \frac{\GB-\GA}{\Th},
\end{equation}
with $\GB$ the Gibbs energy of the hypothetical pure low-density structure. The
difference in Gibbs energy between the pure components $\GB - \GA$ is related to the
equilibrium constant $K$ of the ``reaction'' $A \rightleftharpoons B$,
\begin{equation}
    \ln K \equiv L.
\end{equation}
For the interaction parameter $\omega$ in \eqref{eq:eos}, a linear pressure dependence is
taken,
\begin{equation}
    \omega = 2 + \omega_0 \Dp.
\end{equation}
A low-density/high-density phase-transition curve is located at $L=0$, for $\omega>2$,
and the phase separation occurs upon increase of pressure. This phase transition lies
below the homogeneous ice nucleation temperature and cannot be observed in experiments.
Experiments by Mishima\cite{mishima2000,mishima2010} suggest that the phase transition
curve, if it exists, lies close to the homogeneous nucleation curve, and has the same
shape. In this work, we use a hyperbola for the $L=0$ phase-transition curve, as in
previous work,\cite{holtenSCW,holtentwostate}
\begin{equation}\label{eq:LLT}
    t + k_0 p + k_1 t p = 0,
\end{equation}
\begin{figure}
\includegraphics{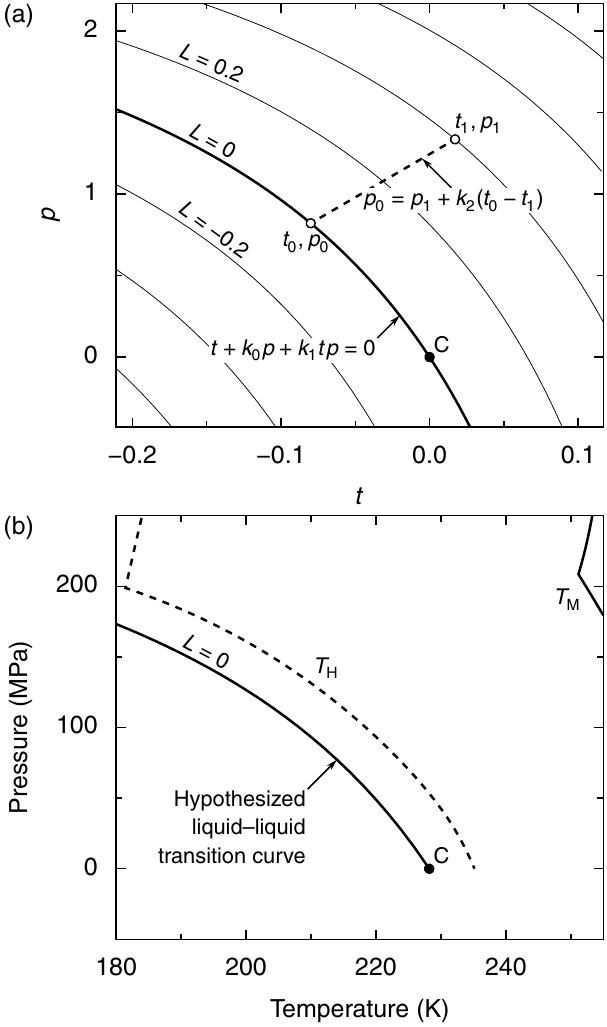}
\caption{\label{fig:field}(a) Construction of the field $L$. Solid curves are contour lines of
constant $L$, with the $L=0$ contour, defined by \eqref{eq:LLT}, drawn thicker.
The point $(t_1,p_1)$ is projected on the $L=0$ curve along the dashed line with slope
$\di p/\di t = k_2$, given by \eqref{eq:fieldprojection}, yielding the point $(t_0,p_0)$.
The field $L$ is taken proportional to the distance between the two points.
The critical point is indicated by C. All contour lines have the same shape,
but are shifted in the $p$--$t$ plane in the direction of the dashed line.
(b) Predicted location of the low-density/high-density phase transition in the phase diagram.
The curves labeled $T\ts{M}$ and $T\ts{H}$ represent the experimental melting temperature
and temperature of homogeneous ice nucleation, respectively.
}
\end{figure}%
where $k_0$ is the slope $\di t/\di p$ of the $L=0$ curve at the critical point, and
$k_1$ determines the curvature. In this work, the expression for $L(t,p)$ is constructed
as proportional to the distance to the $L=0$ curve in the $\Dp$--$\Dt$ diagram
[\figref{fig:field}(a)]. Consider a point in the phase diagram with dimensionless
coordinates $(t,p)$. The projection $(t_0,p_0)$ of this point on the $L=0$ curve, along a
line with slope $\di p/\di t = k_2$, is found by the solution of the equations
\begin{gather}
    t_0 + k_0 p_0 + k_1 t_0 p_0 = 0,\\
    p_0 = p + k_2(t_0-t),\label{eq:fieldprojection}
\end{gather}
which are illustrated in \figref{fig:field}(a). The field $L$ is taken proportional to
the distance between the points $(t,p)$ and $(t_0,p_0)$ with proportionality factor
$L_0$, which results in
\begin{equation}\label{eq:field}
    L = L_0 \frac{K_2}{2k_1 k_2}
        \bigl[ 1 + k_0 k_2 + k_1(p + k_2 t) - K_1 \bigr].
\end{equation}
with
\begin{align}
    K_1 &= \bigl\{[1 + k_0 k_2 + k_1 (p - k_2 t)]^2
        - 4 k_0 k_1 k_2 (p - k_2 t) \bigr\}^{1/2},\notag\\
    K_2 &= (1+k_2^2)^{1/2}.\label{eq:fieldK}
\end{align}
The expression for $L$ in \eqref{eq:field} yields $L=0$ if \eqref{eq:LLT} is satisfied,
as can be verified by solving for $\Dp$ in \eqref{eq:LLT} and substituting the result in
\eqsref{eq:field}{eq:fieldK}. In previous work,\cite{holtenSCW,holtentwostate} the
simpler expression
\begin{equation}\label{eq:previousL}
    L = L_0(t + k_0 p + k_1 t p)
\end{equation}
was used, which yields the same location [\eqref{eq:LLT}] for the $L=0$ curve.
\eeqref{eq:previousL} is not valid for large $L$, corresponding to large pressures (about
1000~MPa at 310~K), where this equation generates an additional, unphysical root.

At any pressure and temperature, the equilibrium value $\xe$ of the fraction $x$ is found
from the condition
\begin{equation}
    \left( \frac{\partial g}{\partial {x}}\right) _{T,P}=0 \quad \text{at}\quad x=\xe,
\end{equation}
which yields
\begin{equation}\label{eq:xe}
    L+\ln \frac{\xe}{1-\xe}+\omega (1-2\xe)=0.
\end{equation}
This equation must be solved numerically for the fraction $\xe$. The location of the
critical point is defined by
\begin{equation}\label{eq:critpoint}
    \biggl(\frac{\partial^2 g}{\partial x^2}\biggr)_{T,P} = 0, \qquad
    \biggl(\frac{\partial^3 g}{\partial x^3}\biggr)_{T,P} = 0.
\end{equation}
In the theory of critical thermodynamic behavior, the thermodynamic properties are
expressed in terms of the order parameter and the ordering field. In our equation of
state, $L$ is the ordering field, and the order parameter $\phi$ is given
by\cite{holtentwostatesupplement}
\begin{equation}
    \phi = 2\xe-1.
\end{equation}
The susceptibility $\chi$ defines the liquid--liquid stability limit (spinodal) as
\begin{equation}
    \chi^{-1} = \frac{1}{2\Th} \biggl(\frac{\partial^2 \Gh}{\partial x^2}\biggr)_{T,P} = 0,
\end{equation}
and is given by
\begin{equation}
    \chi = \biggl(\frac{2}{1-\phi^2} - \omega\biggr)^{-1}.
\end{equation}
The dimensionless volume and entropy can then be written as
\begin{align}
    \Vh &= \frac{\Th}{2}\left[\frac{\omega_0}{2}(1-\phi^2) + \Lp(\phi+1)\right] + \Bp,\\
    \Sh &= -\frac{\Th\Lt}{2}(\phi+1) - \frac{\Gh-\GA}{\Th} - \Bt,
\end{align}
with subscripts $\Th$ and $\Ph$ indicating partial derivatives with respect to the
subscripted quantity. Expressions for the derivatives of the field $L$ and the Gibbs
energy $\GA$ are given in Appendix~\ref{sec:derivatives}.

The dimensionless response functions, namely isothermal compressibility $\kap$, expansion
coefficient $\alp$, and isobaric heat capacity $\Cph$, are given by
\begin{align}
    \kap &= \frac{1}{\Vh}\biggl\{\frac{\Th}{2}\left[\chi(\Lp - \omega_0\phi)^2
                -(\phi+1)\Lpp\right]  - \Bpp\biggr\},\\
    \alp &= \frac{1}{\Vh}\biggl\{\frac{\Ltp}{2}\Th(\phi + 1)
                + \frac{1}{2}\left[\frac{\omega_0}{2}(1-\phi^2) + \Lp(\phi + 1)\right]\notag\\
          &\qquad      - \frac{\Th\Lt}{2}\chi(\Lp - \omega_0\phi) + \Btp\biggr\},\\
    \Cph &= -\Lt\Th(\phi + 1) +\frac{\Th^2}{2}\bigl[\Lt^2\chi -\Ltt(\phi+1)\bigr] - \Th \Btt.
\end{align}
These dimensionless quantities are defined as
\begin{equation}
    \rho     = \frac{\rhoref}{\Vh},\quad
    \kappa_T = \frac{\kap}{\rhoref R\Tc},\quad
    \alpha_P = \frac{\alp}{\Tc},\quad
    c_P      = R \Cph,
\end{equation}
with the density $\rho=1/v$, $\rhoref = 1/\Vref$, and $c_P$ the isobaric specific heat
capacity. The isochoric specific heat capacity $c_v$ is found from the thermodynamic
relation
\begin{equation}
    c_v = c_P - \frac{T \alpha_P^2}{\rho\kappa_T},
\end{equation}
and the speed of sound $w$ is found from
\begin{equation}
    w = \left(\rho\kappa_T\frac{c_v}{c_P}\right)^{-1/2}
      = \left(\rho\kappa_T - \frac{T\alpha_P^2}{c_P}\right)^{-1/2}.
\end{equation}
\eeqref{eq:eos} is a mean-field equation of state which neglects effects of
fluctuations.\cite{holtentwostatesupplement} In particular, not taking these effects into
account results in a lower critical pressure,\cite{holtentwostate} see
\tabref{tab:physparam} in Appendix~\ref{app:tables}. As shown
earlier,\cite{holtenSCW,holtenMF2012} a mean-field equation describes the experimental
data for supercooled water equally well as a nonanalytic equation based on critical
scaling theory. The reason for the good description by a mean-field approximation is that
the region asymptotically close to the hidden critical point, where scaling theory would
be necessary, is not experimentally accessible [\figref{fig:field}(b)]. Moreover, in
practice, a mean-field equation of state is more convenient for computational use than
one incorporating scaling theory.

The Gibbs energy of the high-density structure $\GA$ is a function of dimensionless
temperature and pressure $\tb$ and $\pb$, and serves as a background function in the
two-state model. We selected the empirical form
\begin{equation}\label{eq:backgr}
    \GA(\tb,\pb) = \sum_{i=1}^{n}c_i \tb^{a_i} \pb^{b_i}\mathrm{e}^{-d_i \pb},
\end{equation}
where $n$ is the number of terms and $a_i$, $b_i$, $c_i$, and $d_i$ are adjustable
parameters. In previous work,\cite{holtentwostate} the exponents $a_i$ and $b_i$ were
integers, and the dimensionless temperature and pressure were defined as $\tb=\Dt$ and
$\pb=\Dp$. In this work, this definition could not be used, because $\Dt$ and $\Dp$
become negative below the critical point and powers of negative numbers with real
exponents are generally complex numbers. To avoid negative $\tb$ and $\pb$, they were
redefined as
\begin{equation}
    \tb = \Th = \frac{T}{\Tc}, \qquad \pb = \frac{(P-P_0)\Vref}{R\Tc},
\end{equation}
where the offset $P_0 = -300$~MPa was chosen to enable extrapolation to negative
pressures.

The equation of state describes liquid water as a ``mixture'' of two structures, but
these structures do not exist in isolation. The hypothetical ``pure'' low-density
structure in particular may be unstable in a part of the phase diagram. Such behavior is
not unexpected, because the fraction of the low-density structure increases as the
homogeneous ice nucleation limit is approached, where liquid water becomes kinetically
unstable with respect to ice. Therefore, the low-density structure could have a negative
compressibility in a certain range of temperatures and pressures to destabilize the
liquid state. Moreover, beyond the homogeneous ice nucleation limit, supercooled water
cannot exist as a bulk metastable state and the proposed equation of state may not be
valid. Nevertheless, in the entire region of validity, our equation of state describes
liquid water (stable and metastable) with positive isothermal compressibility and
positive heat capacity.

\subsection{Optimization method}
The aim of the least-squares optimization used in this work was to obtain a fit that
minimizes $\chi^2$, the sum of squared deviations of the fit from experimental data. To
make these deviations dimensionless, the differences of experimental and calculated
values were divided by the experimental uncertainty. The fit of \eqref{eq:eos} requires
optimization of the parameters $L_0$, $\rhoref$, $\omega_0$, and $k_2$ as well as the
parameters $a_i$, $b_i$, $c_i$, and $d_i$ for the Gibbs energy $\GA$ in
\eqref{eq:backgr}. The optimization was carried out in two steps, and is roughly based on
the procedure of Lemmon.\cite{lemmon2009} In the first step, a bank of 135 terms of the
form
\begin{equation}
    \tb^{a_i} \pb^{b_i}\mathrm{e}^{-d_i \pb}
\end{equation}
was created, where the exponents $a_i$ and $b_i$ were restricted to integers in the range
of 0 to 8, with $a_i + b_i \leq 8$. The coefficient $d_i$ was restricted to the values 0,
0.6, and 1. The aim of the first optimization step was to find a good selection of 20 to
25 terms out of the 135 terms for the Gibbs energy $\GA$. Initially, $\GA$ contained a
few manually selected terms with low values of the exponents $a_i$ and $b_i$. The
algorithm then determined the best term to add to $\GA$, in the following way. The first
term in the bank of terms that was not in the selection was added, and all parameters
except $a_i$, $b_i$, and $d_i$ were optimized. The newly added term was then removed,
another term was added, and the parameters were optimized again. This procedure was
repeated for all terms and a value of $\chi^2$ was computed for the addition of each
term. The term that resulted in the lowest $\chi^2$ was then permanently added to $\GA$.
By repeating the addition procedure, the number of terms in $\GA$ was increased to about
20. The quality was then further improved by deleting terms that could be deleted without
significantly increasing $\chi^2$, and adding new terms to replace the deleted terms.

In the second optimization step, the parameters $a_i$, $b_i$, and $d_i$ were also taken
adjustable and optimized simultaneously with the other parameters. The additional degrees
of freedom made it possible to delete terms while still improving the fit. Terms with
similar values of the exponents were combined if that was possible without deteriorating
the fit.

The shape of the liquid--liquid phase transition curve, the $L=0$ curve of
\eqref{eq:LLT}, was not taken adjustable. The parameters that determine this curve, $k_0$
and $k_1$ in \eqref{eq:LLT}, were derived from the shape of the experimental homogeneous
nucleation curve, described in Appendix~\ref{app:TH}. The initial location of the
liquid--liquid critical point was taken from the mean-field equation of state by Holten
\ea\cite{holtentwostate} and was later adjusted to a slightly lower temperature to
improve the description of experimental data. The numerical values of all parameters are
listed in Appendix~\ref{app:tables}, and computer code for the equation of state is
included in the supplemental material.\cite{SCWsupplement2013}

\clearpage
\section{Comparison with Experimental Data}
\subsection{Density}\label{sec:densitycomparison}
\begin{figure}[b]
\includegraphics{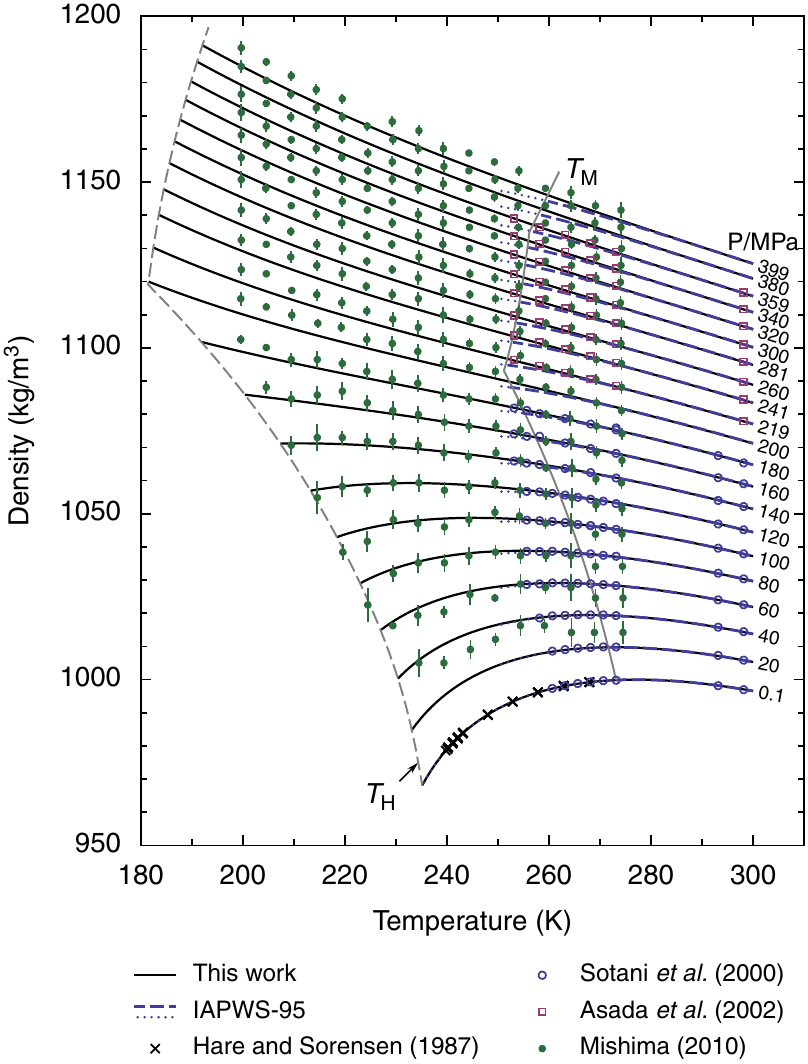}
\caption{\label{fig:densityT}Density of liquid water as a function of temperature and pressure.
Solid black curves are calculated from \eqref{eq:eos}, and symbols represent experimental
data.\cite{mishima2010,sotani2000,asada2002,hare87}
Vertical lines through the points of Mishima represent reported random uncertainties,
and do not take systematic errors into account.
$T\ts{M}$ indicates the melting temperature and
$T\ts{H}$ the homogeneous nucleation temperature.
In this figure, the densities of Asada \ea\cite{asada2002} are interpolated values
that match the isobar pressures of Mishima.\cite{mishima2010}
Values calculated from IAPWS-95 are shown for comparison; dashed in the stable region
and dotted in the metastable region.}
\end{figure}
In \figref{fig:densityT}, the density calculated from \eqref{eq:eos} is plotted as a
function of temperature for several isobars, and compared with experimental data. Below
250~K, Mishima's data are the only data that are available. As described in
\secref{sec:densitydata}, \eqref{eq:eos} was fitted to these data with a low weight. The
equation reproduces the trend of Mishima's data, and most of these densities are
reproduced by \eqref{eq:eos} to within 0.5\%. A comparison of Mishima's data with the
more accurate data of Sotani \ea\cite{sotani2000} and Asada \ea\cite{asada2002} shows
that Mishima's data are systematically too low at low pressures and too high at high
pressures, with deviations of up to 0.4\%. The calculated density isobars in
\figref{fig:densityT} below 150~MPa curve down at low temperature, while the isobars at
higher pressures slightly curve upwards, and an inflection point is present for isobars
above a certain pressure. This inflection is related to a minimum in the expansivity, as
will be discussed in \secref{sec:expansivity}. For comparison purposes, figures in this
section also include values calculated from the extrapolated IAPWS-95 formulation, down
to 235~K at atmospheric pressure and down to 250~K at higher pressures.

\begin{figure}
\includegraphics{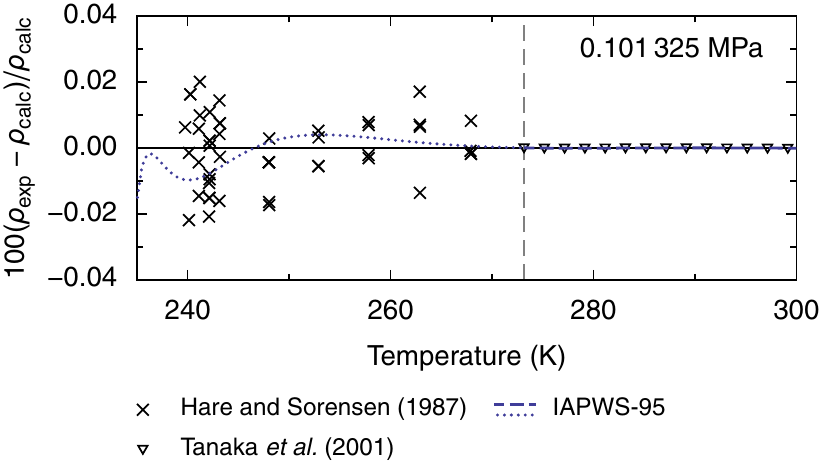}
\caption{\label{fig:densdev1atm}Percentage deviations of experimental density
data\cite{hare87,tanaka2001} at
atmospheric pressure from values calculated from \eqref{eq:eos}.
Values calculated from IAPWS-95 are shown for comparison, dashed in the stable-liquid region
and dotted in the metastable region. The vertical dashed line indicates the melting temperature.}
\end{figure}

In \figref{fig:densdev1atm}, experimental density values at atmospheric pressure are
compared with values calculated from \eqref{eq:eos}. Both \eqref{eq:eos} and the
extrapolated IAPWS-95 formulation represent the data of Hare and Sorensen within the
experimental scatter of about 0.02\%. In the stable region, \eqref{eq:eos} differs less
than 0.0001\% (1 part per million) from the densities recommended by Tanaka
\ea\cite{tanaka2001}

\begin{figure*}
\includegraphics{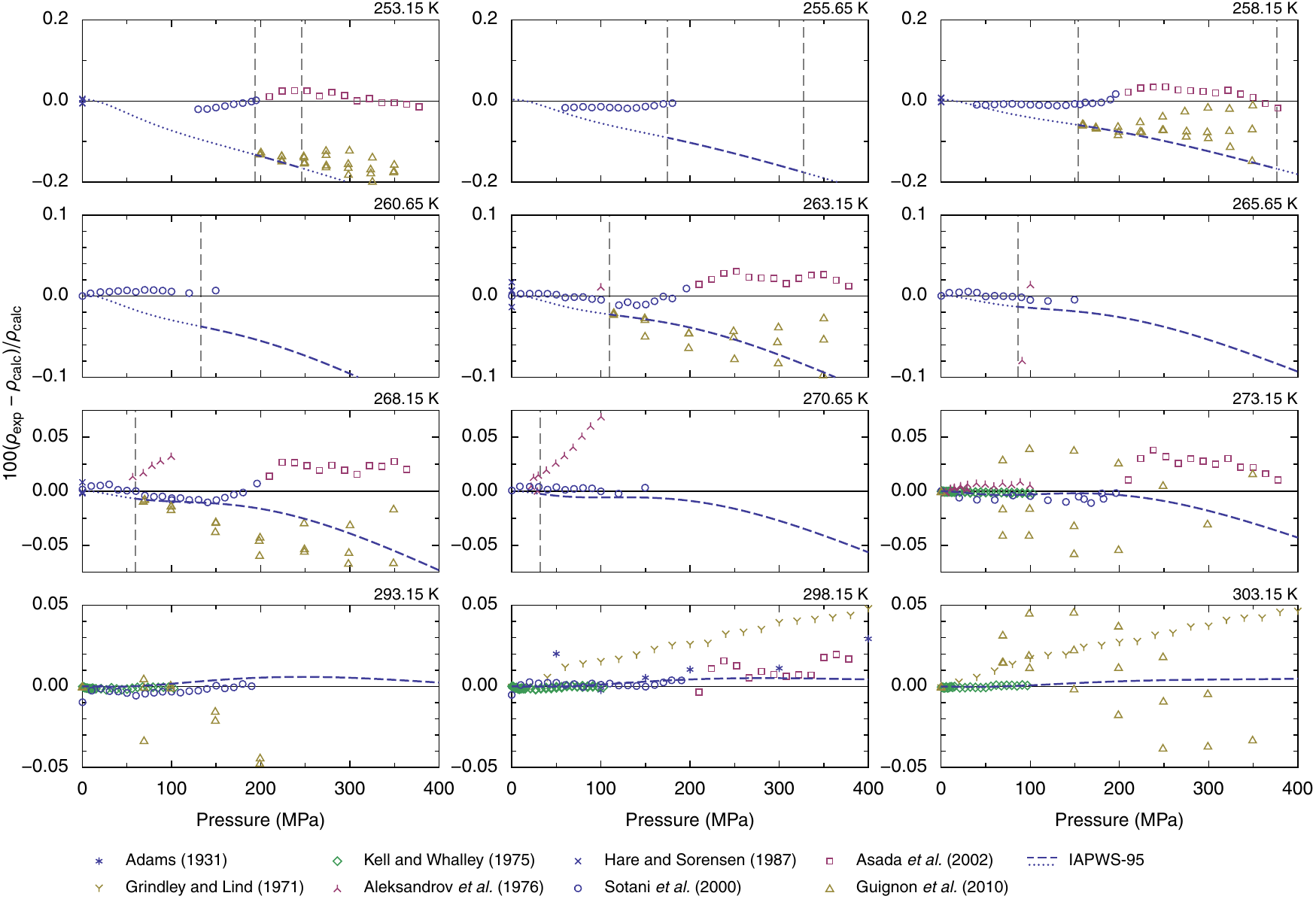}
\caption{\label{fig:densdev}Percentage deviations of experimental density
data\cite{adams1931,grindley1971,kellwhalley1975,hare87,sotani2000,asada2002,guignon2010}
from values calculated from \eqref{eq:eos}. Values calculated from IAPWS-95 are plotted
for comparison; dashed in the stable-liquid region and dotted in the metastable region.
The vertical dashed lines indicate the melting pressure.}
\end{figure*}

\ffigref{fig:densdev} shows differences between experimental densities up to 400~MPa and
values calculated from \eqref{eq:eos}. The proposed equation represents the data of
Sotani \ea\cite{sotani2000} to within 0.02\%, which is within the uncertainty of those
data of 0.03\% as estimated by Wagner and Thol.\cite{wagnerthol2013} The data of Asada
\ea,\cite{asada2002} with an uncertainty of 0.1\%, are reproduced to within 0.04\%. The
data of Guignon \ea\cite{guignon2010} are represented to within the uncertainty of 0.2\%.
The density data of Grindley and Lind\cite{grindley1971} deviate systematically from
other data, as was described in Sec.~\ref{sec:dataadjustment}. At 298~K, IAPWS-95 agrees
well with \eqref{eq:eos} and with the densities of Sotani \ea\cite{sotani2000} and Asada
\ea,\cite{asada2002} which were not available when IAPWS-95 was developed. The density
data of Aleksandrov \ea\cite{aleksandrov1976rhorussian} at temperatures above 271~K are
in satisfactory agreement with the proposed equation of state. For example, at 273.15~K,
the densities of Aleksandrov \ea differ less than 0.01\% from the values calculated from
\eqref{eq:eos}. However, at lower temperatures, such as at 267~K and 270~K, their data
show systematic deviations of up to 0.07\% from the data of Sotani \ea\cite{sotani2000}
and, hence, from the proposed equation.

\begin{figure}
\includegraphics{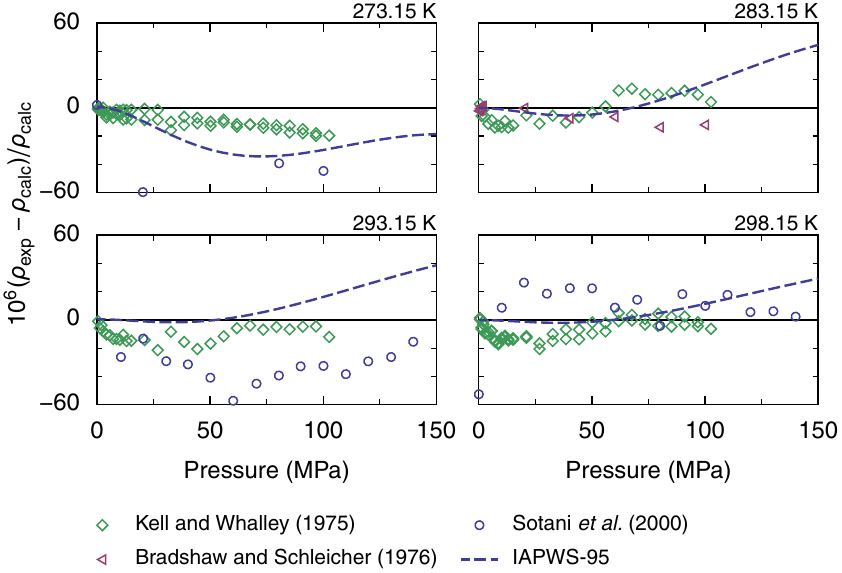}
\caption{\label{fig:densdevKW}Parts-per-million differences between experimental density
data\cite{kellwhalley1975,bradshaw1976,sotani2000} and
values calculated from \eqref{eq:eos}. Values calculated from IAPWS-95 are
plotted for comparison.}
\end{figure}

In the stable region up to 100~MPa, there are accurate density measurements of Kell and
Whalley,\cite{kellwhalley1975} with an uncertainty of 10~parts per million (ppm) at low
pressures and 30~ppm at 100~MPa. \ffigref{fig:densdevKW} shows that \eqref{eq:eos}
represents these density data to within this uncertainty. Kell and Whalley adjusted their
density data to bring them in agreement with the speed-of-sound data of
Wilson,\cite{wilson1959} which were the best available at the time. In
\secref{sec:wcompare}, we show that the speeds of sound of Wilson deviate up to 0.08\%
from more accurate data at 273~K. This deviation may be the reason for the small
systematic deviation of the density data of Kell and Whalley from \eqref{eq:eos} at
273~K, seen in \figref{fig:densdevKW}.

\subsection{Expansivity}\label{sec:expansivity}
\begin{figure}
\includegraphics{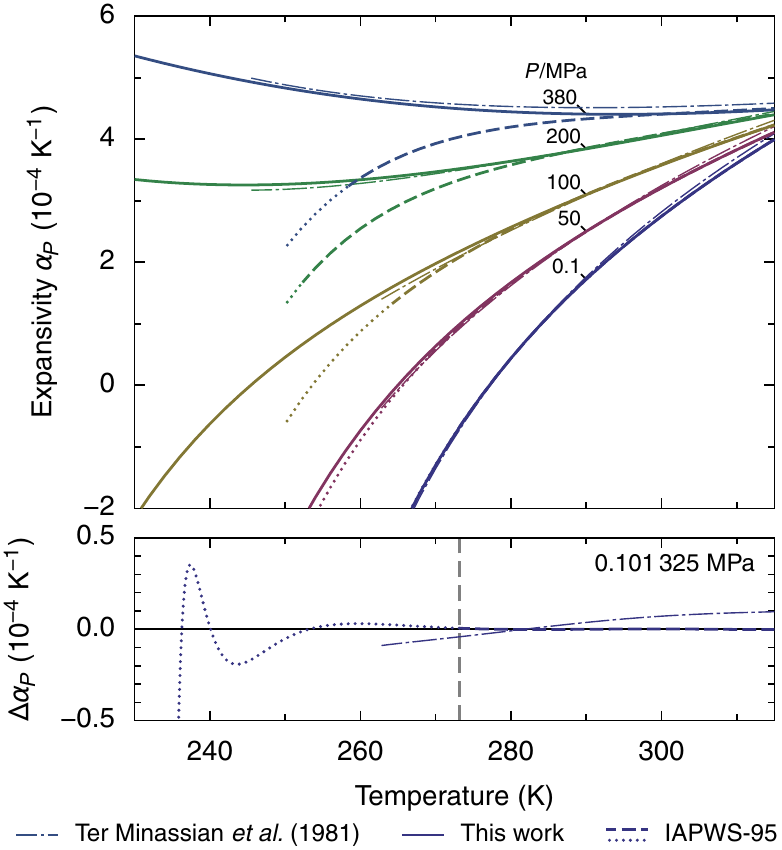}
\caption{\label{fig:expansivity}
Expansivity calculated from \eqref{eq:eos}, the correlation of
Ter Minassian \ea,\cite{terminassian1981} and IAPWS-95 (dashed in the stable-liquid region
and dotted in the metastable region).
The bottom panel shows the expansivity difference at atmospheric pressure,
where values from \eqref{eq:eos}
have been subtracted from values from IAPWS-95 and the correlation of
Ter Minassian \ea\cite{terminassian1981}
The vertical dashed line indicates the melting temperature.}
\end{figure}

\begin{figure}
\includegraphics{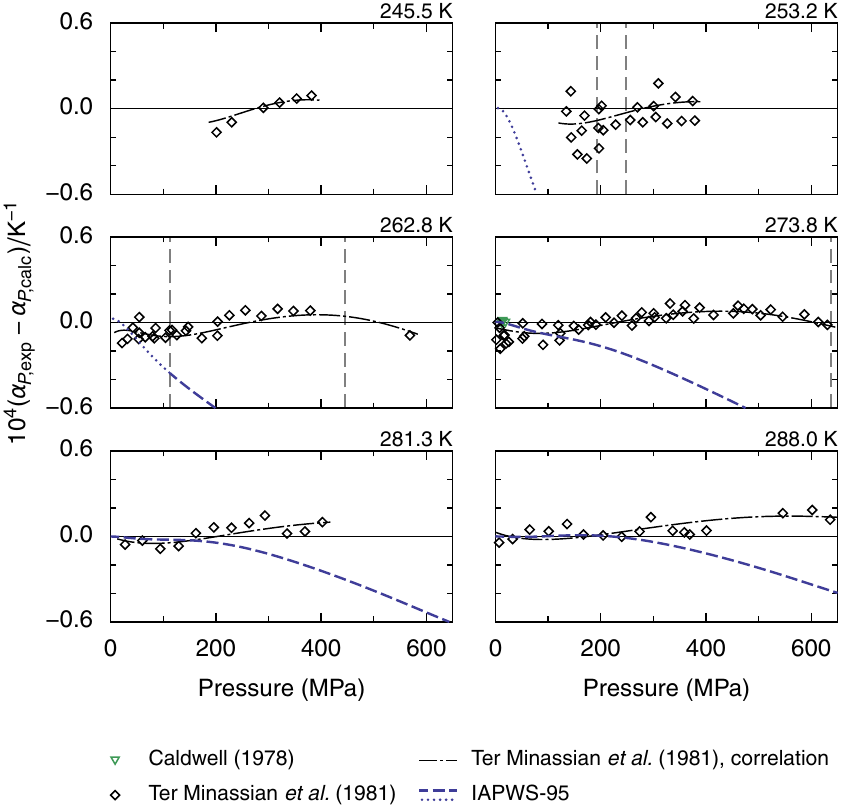}
\caption{\label{fig:alphadev}
Deviations of experimental expansivity data\cite{terminassian1981,caldwell1978} from
values calculated from \eqref{eq:eos}. Vertical dashed lines indicate the melting pressure.
Values calculated from IAPWS-95 are shown for comparison, dashed in the stable-liquid region
and dotted in the metastable region. The correlation that Ter Minassian \ea\cite{terminassian1981}
fitted to their data is also shown.}
\end{figure}

The expansivity calculated from \eqref{eq:eos} is compared with values calculated from
the correlation of Ter Minassian \ea\cite{terminassian1981} and IAPWS-95 in
\figref{fig:expansivity}. At atmospheric pressure, there is little difference in the
expansivity values of \eqref{eq:eos} and the extrapolated IAPWS-95 formulation down to
250~K. At higher pressures, \eqref{eq:eos} follows the correlation of Ter Minassian \ea,
to which it was fitted. More detailed deviations of the data of Ter Minassian \ea from
\eqref{eq:eos} are shown in \figref{fig:alphadev}. These deviation plots use absolute
instead of relative differences, because the expansivity passes through zero in the
temperature and pressure range considered. \eeqref{eq:eos} represents the correlation of
Ter Minassian \ea and most of their data points to within $2\times 10^{-5}$~K$^{-1}$.

\begin{figure}
\includegraphics{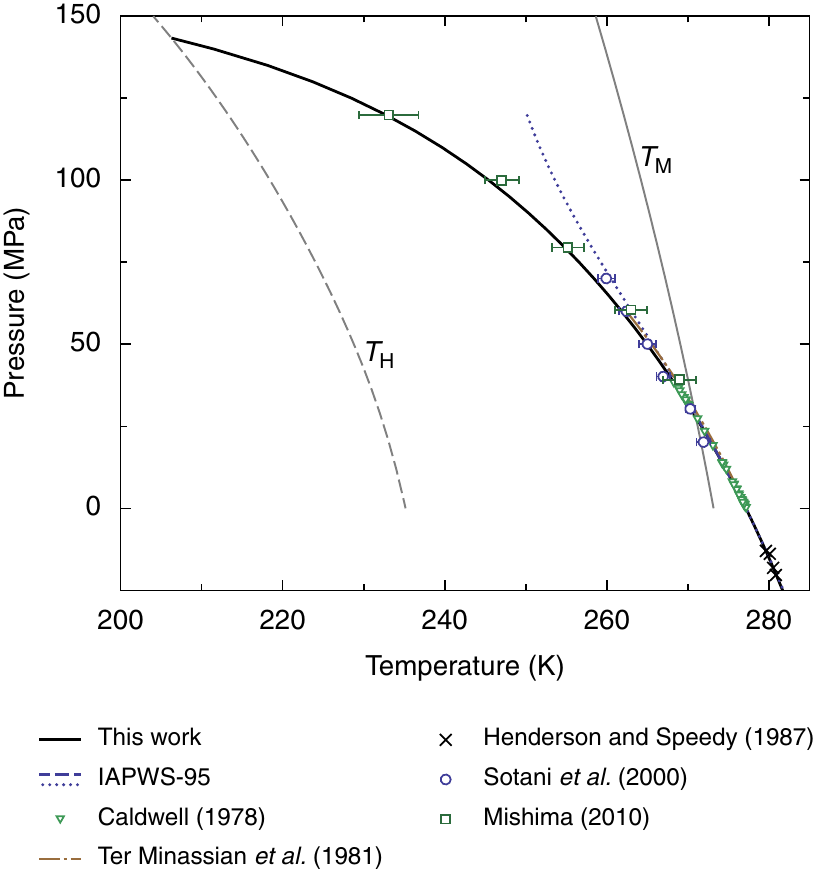}
\caption{\label{fig:TMD}
Temperature of maximum density (TMD) and locus of zero expansivity,
calculated from \eqref{eq:eos}, the correlation of
Ter Minassian \ea,\cite{terminassian1981} and IAPWS-95 (dashed in the stable-liquid region
and dotted in the metastable region). Caldwell\cite{caldwell1978} and
Henderson \& Speedy\cite{henderson1987a} reported the TMD itself;
the TMD of the data of Mishima\cite{mishima2010} and Sotani \ea\cite{sotani2000}
was calculated from polynomial fits to their density data.
The melting temperature and temperature of homogeneous nucleation are indicated
by $T\ts{M}$ and $T\ts{H}$, respectively.}
\end{figure}

\begin{figure}
\includegraphics{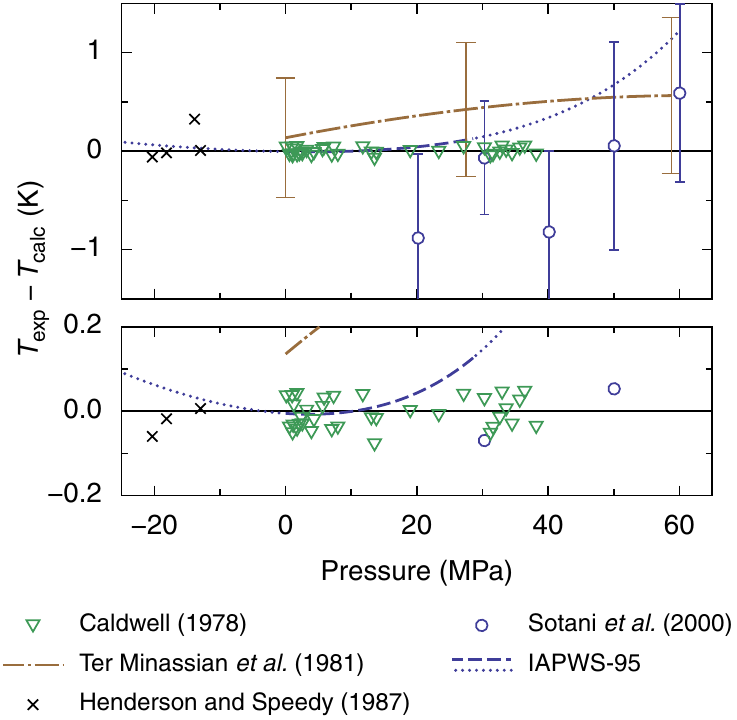}
\caption{\label{fig:TMDdev}
Differences between experimental temperatures of maximum density\cite{caldwell1978,henderson1987a}
and values calculated from \eqref{eq:eos}. Values calculated from density data
of Sotani \ea,\cite{sotani2000}
the correlation of Ter Minassian \ea,\cite{terminassian1981}
and IAPWS-95 (dashed in the stable-liquid region and dotted in the metastable region)
are shown for comparison. The error bars on the curve of Ter Minassian \ea\cite{terminassian1981}
represent the temperature uncertainty, derived from the uncertainty
of their expansivity.
}
\end{figure}

Experimental and calculated values for the temperature of maximum density (TMD) are shown
in \figref{fig:TMD}. The TMD moves to lower temperatures with increasing pressure, and
the rate at which it does so also increases with pressure. The TMD corresponding to
Mishima's data is relatively uncertain because of the scatter in his density data.
Deviations of the experimental TMD values from \eqref{eq:eos} are plotted in
\figref{fig:TMDdev}. The data of Caldwell\cite{caldwell1978} are represented to within
0.08~K. The uncertainty $\delta T$ in the TMD values calculated from the expansivity
correlation of Ter Minassian \ea\cite{terminassian1981} can be estimated as
\begin{equation}\label{eq:TMDerrorTM}
    \delta T \approx \biggl| \pypxd{\alpha_P}{T}{P} \biggr|^{-1} \delta\alpha_P,
\end{equation}
where the temperature derivative of the expansivity is calculated from the correlation of
Ter Minassian \ea\cite{terminassian1981} With an expansivity uncertainty $\delta\alpha_P$
of at least $10^{-5}$~K$^{-1}$ (estimated from \figref{fig:alphadev}),
\eqref{eq:TMDerrorTM} gives an uncertainty $\delta T$ of 0.6~K at 0~MPa, increasing to
0.8~K at 60~MPa. When one takes these uncertainties into account, the correlation of Ter
Minassian \ea\cite{terminassian1981} is consistent with the data of
Caldwell\cite{caldwell1978} and with \eqref{eq:eos}. The data of Henderson and
Speedy\cite{henderson1987a} are represented fairly well by both \eqref{eq:eos} and
IAPWS-95, when these equations are extrapolated to negative pressure.

\begin{figure}
\includegraphics{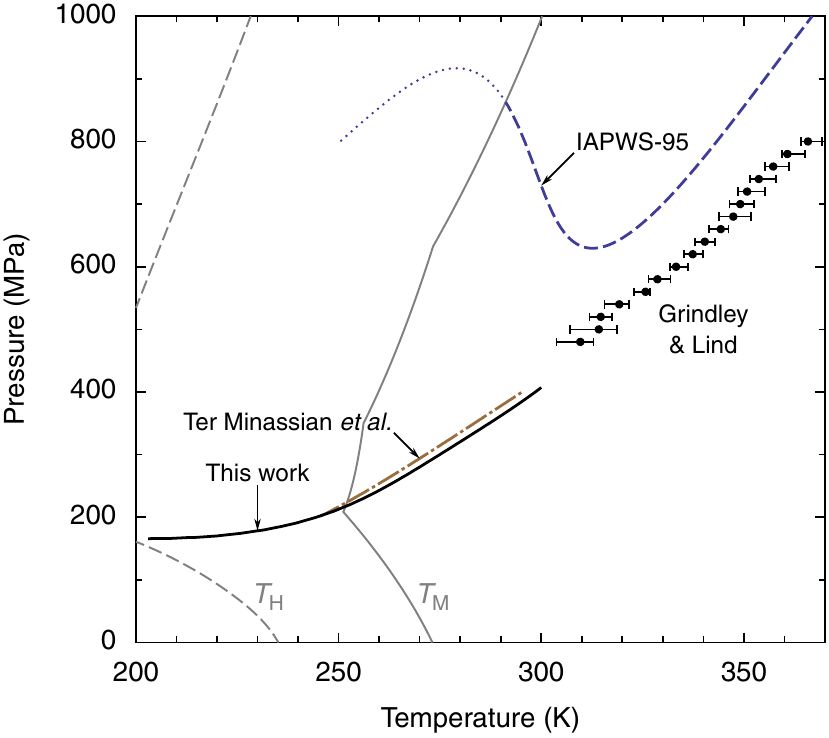}
\caption{\label{fig:alphamin}Location at which the temperature derivative of the
expansivity $(\partial\alpha_P/\partial T)_P$ is zero, for the correlation from this work,
the data of Grindley and Lind,\cite{grindley1971} the correlation of
Ter Minassian \ea,\cite{terminassian1981} and the IAPWS-95 formulation.
These locations represent minima of the expansivity, except for IAPWS-95 between 280~K and 310~K,
where the extremum is a maximum.
Values from IAPWS-95 are shown dotted in the region where IAPWS-95 was extrapolated.
$T\ts{H}$ and $T\ts{M}$ indicate the homogenous nucleation temperature and the melting temperature,
respectively.
}
\end{figure}

The existence of a minimum in the expansivity,
\begin{equation}
    \pypxd{\alpha_P}{T}{P} = 0,
\end{equation}
was noticed by Ter Minassian \ea\cite{terminassian1981} and by Mishima\cite{mishima2010}
for temperatures lower than 300~K. The minimum in the expansivity is related to the
inflection points in curves of the density versus temperature, as is visible in
\figref{fig:densityT} above 200~MPa. The minimum is seen in the expansivity correlation
of Ter Minassian \ea\cite{terminassian1981} in \figref{fig:expansivity}. An expansivity
minimum is also present in the expansivity derived from the volume data of Grindley and
Lind,\cite{grindley1971} as shown in \figref{fig:alphamin}. We obtained the location of
the expansivity minimum in the data of Grindley and Lind by fitting several polynomials
of different order to all their volume data, and also to data on each isobar separately.
For every isobar, this results in several estimated temperatures of the expansivity
minimum. The differences in these temperatures were used to estimate the uncertainty of
the minimum. The locus of expansivity minima of Grindley and Lind thus obtained can be
smoothly connected to that of Ter Minassian \ea\cite{terminassian1981} Our equation of
state closely follows the expansivity minimum obtained by Ter Minassian
\ea,\cite{terminassian1981} as seen in \figref{fig:alphamin}.

\subsection{Isothermal compressibility}
\begin{figure}
\includegraphics{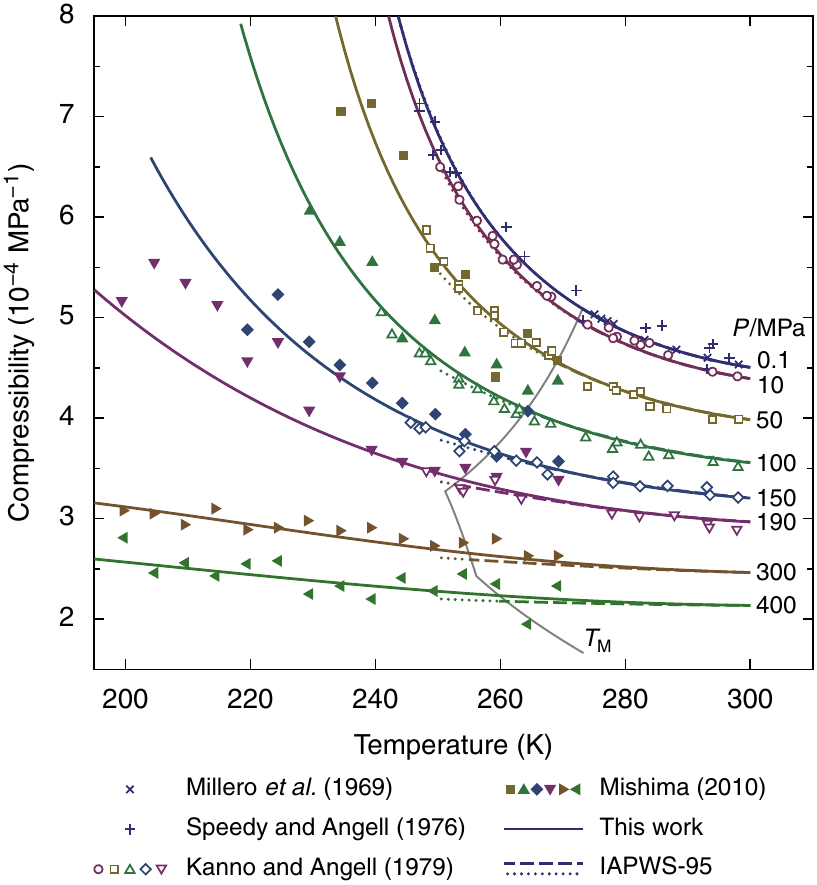}
\caption{\label{fig:compressibility}
Isothermal compressibility calculated from \eqref{eq:eos} (solid curves) and
IAPWS-95 (dashed in the stable-liquid region and dotted in the metastable region).
Symbols represent experimental data.\cite{millero1969,speedy1976,kanno1979,mishima2010}
Solid and open symbols with the same shape belong to the same isobar.
The curve marked $T\ts{M}$ represents the melting temperature.}
\end{figure}

\begin{figure}
\includegraphics{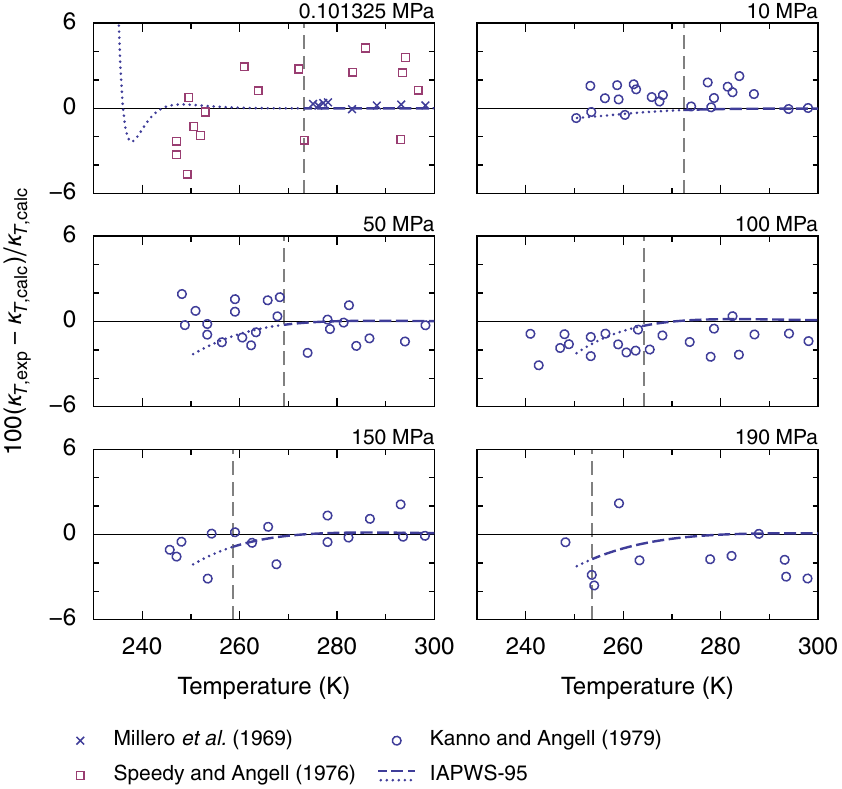}
\caption{\label{fig:kappadev}
Percentage deviations of experimental isothermal-compressibility data of
Millero \ea,\cite{millero1969}
Speedy and Angell,\cite{speedy1976} and Kanno and Angell\cite{kanno1979} from values
calculated from \eqref{eq:eos}. Values from IAPWS-95 are shown for comparison
(dashed in the stable-liquid region and dotted in the metastable region).
Dashed vertical lines represent melting temperatures.}
\end{figure}

\begin{figure}
\includegraphics{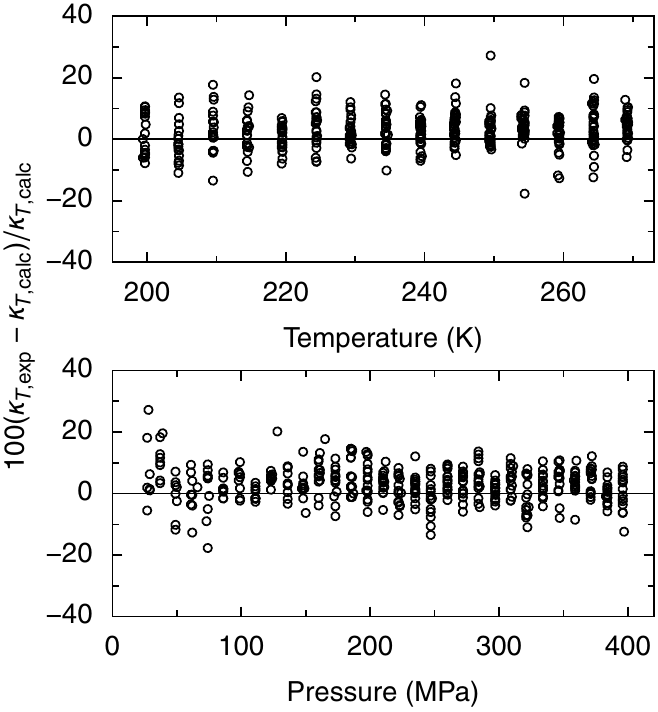}
\caption{\label{fig:kappadevmishima}
Percentage deviations of experimental isothermal-compressibility data of
Mishima\cite{mishima2010} from values calculated from \eqref{eq:eos},
as a function of temperature and pressure.}
\end{figure}

Experimental data on the isothermal compressibility are shown in
\figref{fig:compressibility} together with values calculated from \eqref{eq:eos} and
IAPWS-95. The data of Mishima\cite{mishima2010} exhibit more scatter than those of Angell
and coworkers,\cite{speedy1976,kanno1979} but they show a consistent trend of a decrease
in the anomalous behavior of the compressibility with increasing pressure. Deviations of
the experimental data from \eqref{eq:eos} are plotted in \figref{fig:kappadev}. It can be
seen in \figref{fig:kappadev} that \eqref{eq:eos} represents the compressibility data of
Speedy and Angell\cite{speedy1976} and Kanno and Angell\cite{kanno1979} to within their
scatter. The difference between the extrapolated IAPWS-95 formulation and \eqref{eq:eos}
increases with decreasing temperature; this difference is related to the density
difference between the two equations shown in \figref{fig:densityT}. The deviation of the
compressibilities measured by Mishima\cite{mishima2010} from \eqref{eq:eos} is shown in
\figref{fig:kappadevmishima}. Although \eqref{eq:eos} was not fitted to Mishima's
compressibility data, it represents them fairly well when one takes into account the
experimental scatter of about \textpm 10\%.

\subsection{Speed of sound}\label{sec:wcompare}
\begin{figure}
\includegraphics{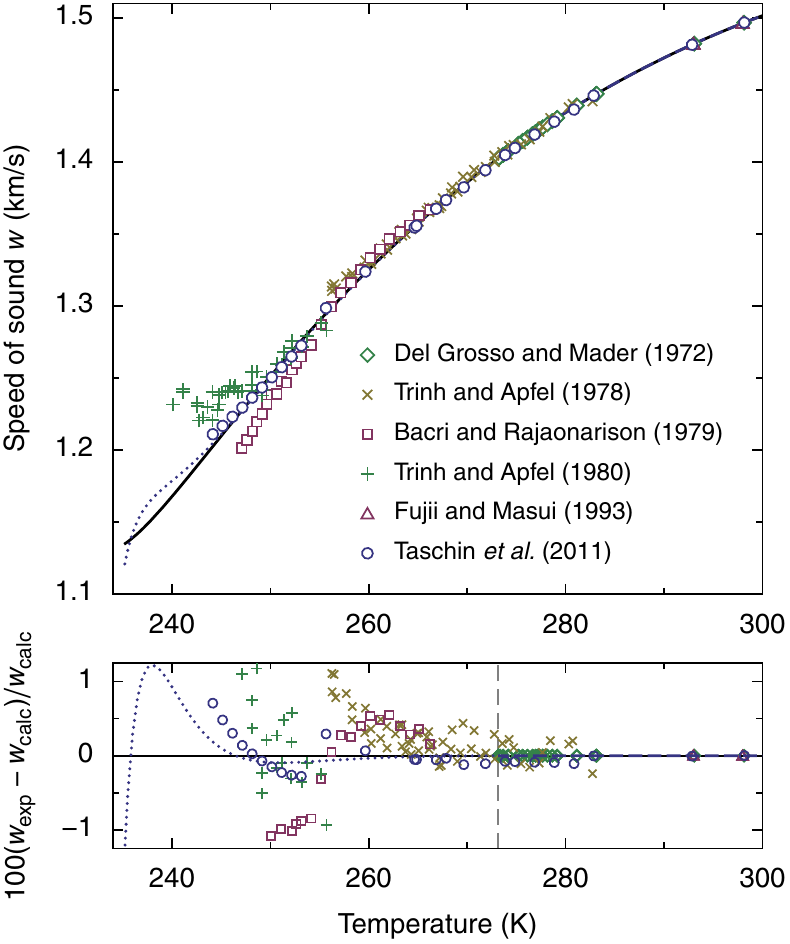}
\caption{\label{fig:w1atm}Experimental data on the speed of sound
\cite{delgrosso1972,trinhapfel1978JASA,trinhapfel1978JCP,bacri1979,trinhapfel1980,fujii1993,taschin2011}
at 0.101325~MPa, together with values calculated from \eqref{eq:eos} (solid curve)
and IAPWS-95 (dashed in the stable-liquid region and dotted in the metastable region).
The bottom graph shows percentage deviations of experimental data from \eqref{eq:eos}.
The vertical dashed line indicates the melting temperature.}
\end{figure}

\begin{figure*}
\includegraphics{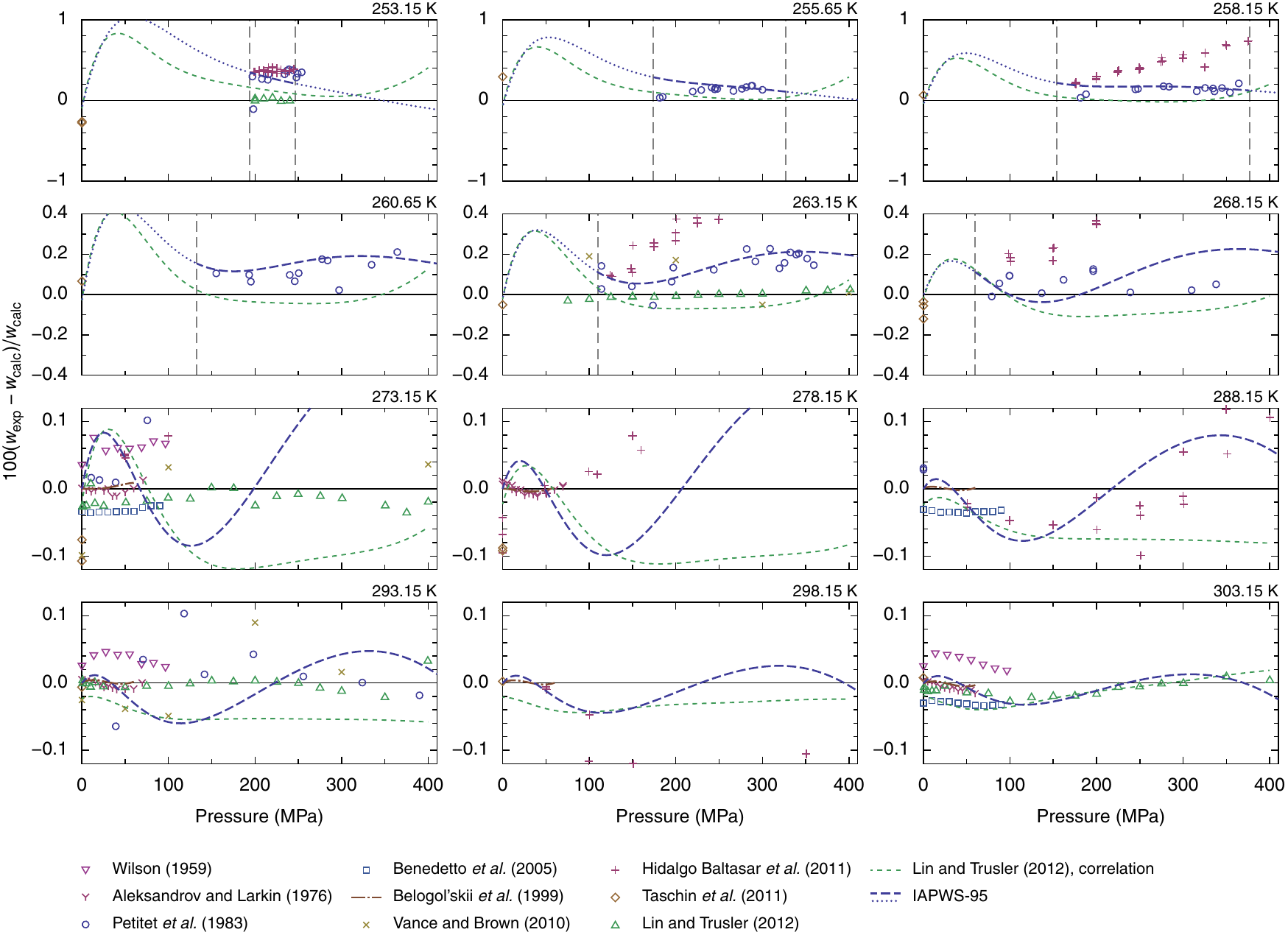}
\caption{\label{fig:wdev}Percentage deviations of experimental data on the speed of
sound\cite{wilson1959,aleksandrov1976,petitet1983,benedetto2005,%
vance2010,hidalgobaltasar2011,taschin2011,lintrusler2012} from values calculated from
\eqref{eq:eos}. Values calculated from IAPWS-95 are plotted for comparison; dashed in the
stable-liquid region and dotted in the metastable region. The correlations that
Belogol'skii\cite{belogolskii1999} and Lin and Trusler\cite{lintrusler2012} fitted to
their data are also shown. The vertical dashed lines indicate the melting pressure.}
\end{figure*}

Experimental data on the speed of sound at atmospheric pressure are shown in
\figref{fig:w1atm} together with values calculated from \eqref{eq:eos} and IAPWS-95. In
the supercooled region, \eqref{eq:eos} represents the data of Taschin
\ea\cite{taschin2011} to within their uncertainty of 0.7\%. In the stable region,
\eqref{eq:eos} represents the speed-of-sound data of Del Grosso and
Mader\cite{delgrosso1972} to within their uncertainty of 0.001\%, and the data of Fujii
and Masui\cite{fujii1993} deviate less than 0.004\% from \eqref{eq:eos}.

Speed-of-sound data up to 400~MPa are compared with values calculated from \eqref{eq:eos}
in \figref{fig:wdev}. The proposed equation represents the data of Lin and
Trusler\cite{lintrusler2012} to within 0.04\%. For comparison, the correlation of Lin and
Trusler has deviations of up to 0.2\% from their data in the temperature range considered
here. The IAPWS-95 formulation was fitted to the speed-of-sound data of Petitet
\ea,\cite{petitet1983} which systematically deviate from the data of Lin and Trusler by
up to 0.2\%. This deviation is the reason for the difference between IAPWS-95 and
\eqref{eq:eos} in the stable region in the range of 253~K to 263~K. Speed of sound values
from the correlation of Belogol'skii \ea\cite{belogolskii1999} differ less than 0.01\%
from values from \eqref{eq:eos}, and the data of Aleksandrov and
Larkin\cite{aleksandrov1976} are represented to within 0.02\%.

In the metastable region from 253~K to 265~K and for pressures around 50~MPa, there is a
rather large difference between the extrapolated IAPWS-95 formulation and \eqref{eq:eos}
(\figref{fig:wdev}). At 253~K, the difference is more than 1\%. To investigate the nature
of this difference, a test equation was forced to follow extrapolated IAPWS-95 values in
the region of the difference. The density calculated from this test equation showed
systematic deviations from both the data of Sotani \ea\cite{sotani2000} and Asada
\ea,\cite{asada2002} outside the experimental uncertainty. The difference in speed of
sound between extrapolated IAPWS-95 and \eqref{eq:eos} is therefore related to the
difference in density between IAPWS-95 values and the data of Sotani \ea\cite{sotani2000}
and Asada \ea\cite{asada2002} As can be seen in \figref{fig:wdev}, the two points of Lin
and Trusler\cite{lintrusler2012} in the metastable region at 263~K support the behavior
of \eqref{eq:eos} in this region. The correlation of Lin and Trusler behaves similarly to
IAPWS-95 in the metastable region; as a result, the densities derived by Lin and Trusler
from their correlation are close to the IAPWS-95 values.

\subsection{Heat capacity}\label{sec:cpcompare}
\begin{figure}
\includegraphics{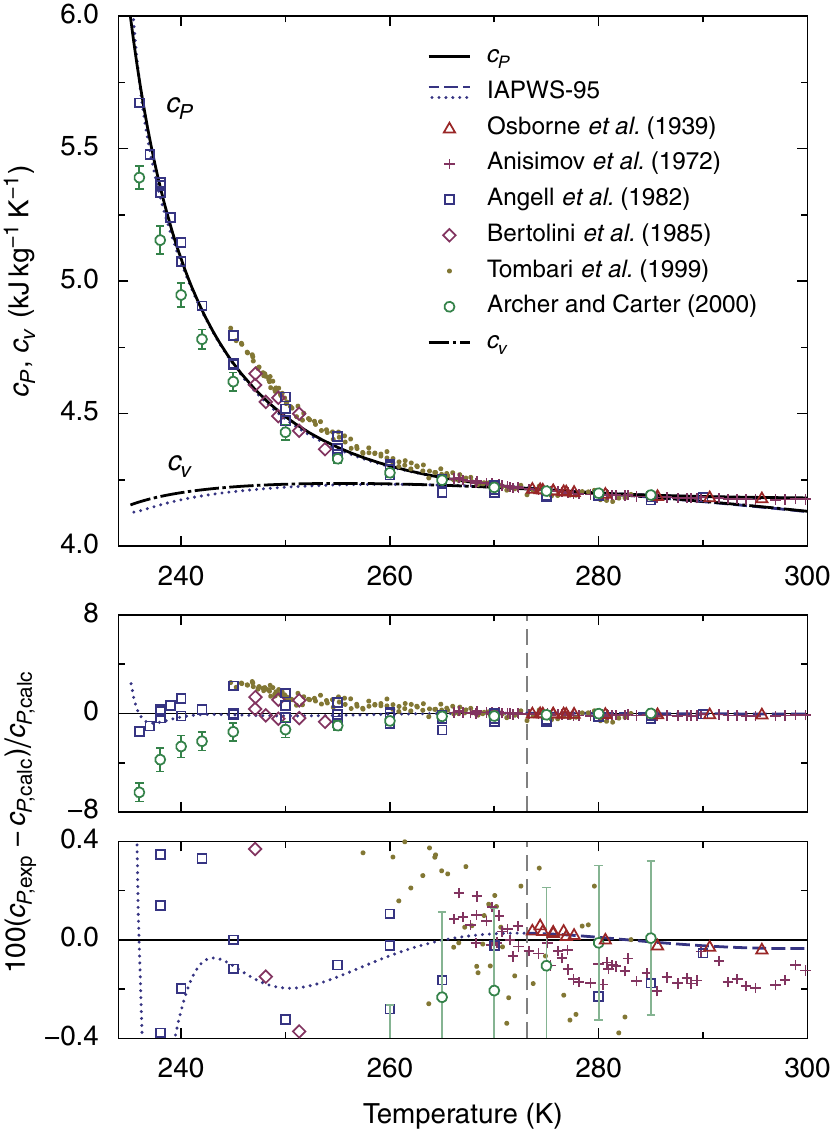}
\caption{\label{fig:Cp1atm}Heat capacity at 0.101\,325~MPa calculated from \eqref{eq:eos}
(solid curve: $c_P$, dash-dotted curve: $c_v$). Symbols represent
experimental data.\cite{osborne1939,anisimov1972,angell1982,bertolini1985,tombari1999,arc00} Values
from IAPWS-95 are plotted for comparison; dashed in the stable-liquid region
and dotted in the metastable region.
The bottom two graphs show deviations of experimental $c_P$ data from \eqref{eq:eos}.
The vertical dashed line indicates the melting temperature.}
\end{figure}

In \figref{fig:Cp1atm}, values for the isobaric heat capacity calculated from
\eqref{eq:eos} are compared with experimental data at atmospheric pressure. There are two
sets of experimental data that extend down to 236~K, that of Angell \ea\cite{angell1982}
and that of Archer and Carter.\cite{arc00} Both \eqref{eq:eos} and IAPWS-95 agree better
with the data of Angell \ea than with the data of Archer and Carter. In the case of
IAPWS-95, this is expected, as it was fitted to the data of Angell \ea\cite{angell1982}
As described in \secref{sec:cpdata}, \eqref{eq:eos} was fitted to values computed from
IAPWS-95.

The data of Bertolini \ea\cite{bertolini1985} agree with those of Angell \ea after a
correction that is described in Ref.~\onlinecite{holtenSCW}. The data of Tombari
\ea\cite{tombari1999} suggest even larger $c_P$ values in the supercooled region than the
data of Angell \ea In the stable region, \eqref{eq:eos} represents the accurate data of
Osborne \ea\cite{osborne1939} to within 0.1\%. For the isochoric heat capacity $c_v$,
both \eqref{eq:eos} and the extrapolated IAPWS-95 formulation predict a weak temperature
dependence in the supercooled region at atmospheric pressure.

\begin{figure}
\includegraphics{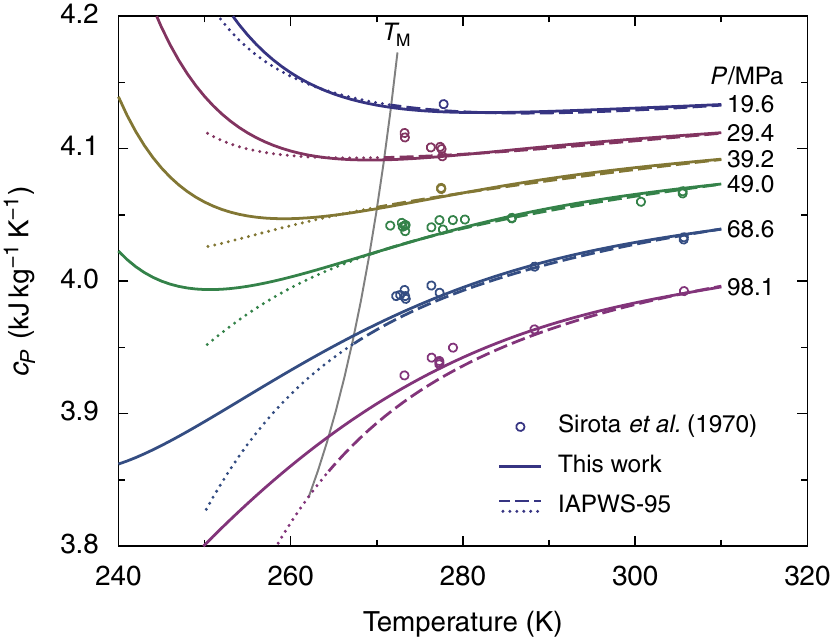}
\caption{\label{fig:cpsirota}Isobaric heat capacity calculated from \eqref{eq:eos}
(solid curves) together with experimental data from Sirota \ea\cite{sirota1970}
Values from IAPWS-95 are plotted for comparison; dashed in the stable-liquid region
and dotted in the metastable region. $T\ts{M}$ denotes the melting temperature.}
\end{figure}

Sirota \ea\cite{sirota1970} measured the isobaric heat capacity at pressures up to
100~MPa in the stable region. These data are compared with \eqref{eq:eos} and with
IAPWS-95 in \figref{fig:cpsirota}. The data of Sirota \ea show systematic deviations from
both IAPWS-95 and \eqref{eq:eos}.

\subsection{Extrapolation to 1000 MPa}
There are no experimental data in the supercooled region above 400~MPa, except for one
expansivity data point of Ter Minassian \ea\cite{terminassian1981} at 263~K and 569~MPa
(\figref{fig:alphadev}), and two speed-of-sound measurements of Vance and
Brown\cite{vance2010} at 263~K (up to 600~MPa, see \figref{fig:wPT}). The speed-of-sound
measurements of Hidalgo Baltasar \ea\cite{hidalgobaltasar2011} in the supercooled region
at 278~K and 700~MPa seem to have been affected by ice formation, because they deviate
from the trend of their data in the stable-liquid region.

\begin{figure}
\includegraphics{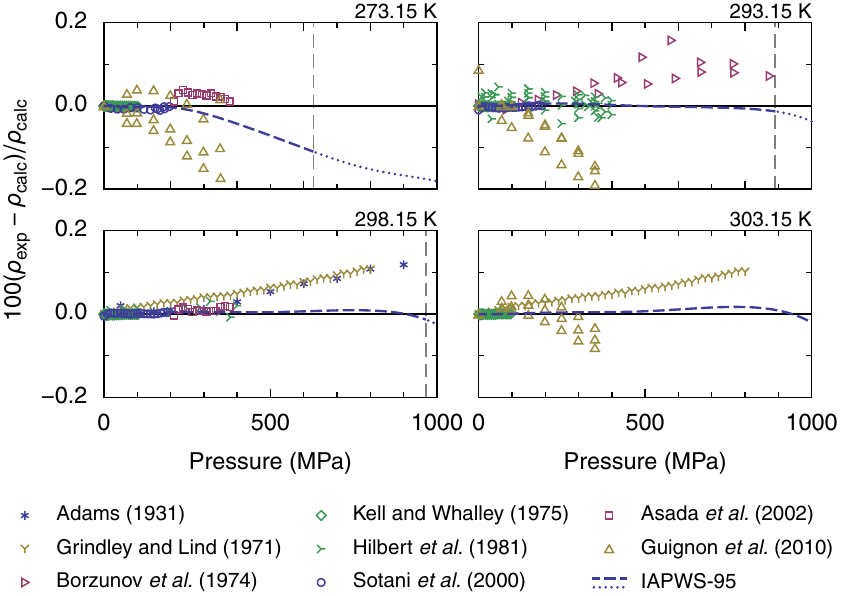}
\caption{\label{fig:densdevhighp}Percentage deviations of experimental density
data\cite{adams1931,grindley1971,borzunov1974,kellwhalley1975,%
hilbert1981,sotani2000,asada2002,guignon2010}
from values calculated from \eqref{eq:eos}. Values calculated from IAPWS-95 are
plotted for comparison; dashed in the stable-liquid region and dotted in the metastable
region. The vertical dashed lines indicate the melting pressure.}
\end{figure}

In the stable-liquid region below 300~K, there do exist data above 400~MPa. The
expansivity data of Ter Minassian \ea\cite{terminassian1981} extend up to 635~MPa
(\figref{fig:alphadev}). Grindley and Lind\cite{grindley1971} measured densities up to
800~MPa. \ffigref{fig:densdevhighp} shows the deviations of experimental densities and
IAPWS-95 values from \eqref{eq:eos} up to 1000~MPa. Above 293~K, \eqref{eq:eos} follows
IAPWS-95 closely, while the data of Grindley and Lind\cite{grindley1971} and
Adams\cite{adams1931} show systematic deviations from both equations of state that
increase with increasing pressure.

\begin{figure}
\includegraphics{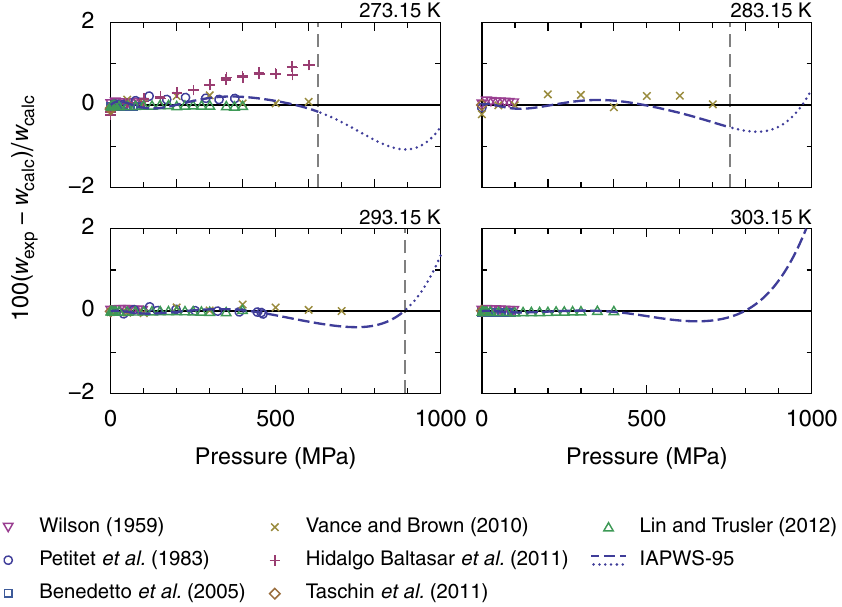}
\caption{\label{fig:wdevhighp}Percentage deviations of experimental data on the speed of
sound\cite{wilson1959,petitet1983,benedetto2005,vance2010,hidalgobaltasar2011,taschin2011,lintrusler2012}
from values calculated from \eqref{eq:eos}. Values calculated from IAPWS-95 are
plotted for comparison; dashed in the stable-liquid region and dotted in the metastable
region. The vertical dashed lines indicate the melting pressure.}
\end{figure}

Both Vance and Brown\cite{vance2010} and Hidalgo Baltasar \ea\cite{hidalgobaltasar2011}
have measured the speed of sound up to 700~MPa. Wang \ea\cite{wang2013} determined the
speed of sound at 293~K up to the melting pressure of about 900~MPa. Their data
systematically deviate by about 3\% from the data of Vance and Brown,\cite{vance2010} and
will not be considered here. The differences between speed-of-sound data and
\eqref{eq:eos} up to 1000~MPa are shown in \figref{fig:wdevhighp}.

\begin{figure}
\includegraphics{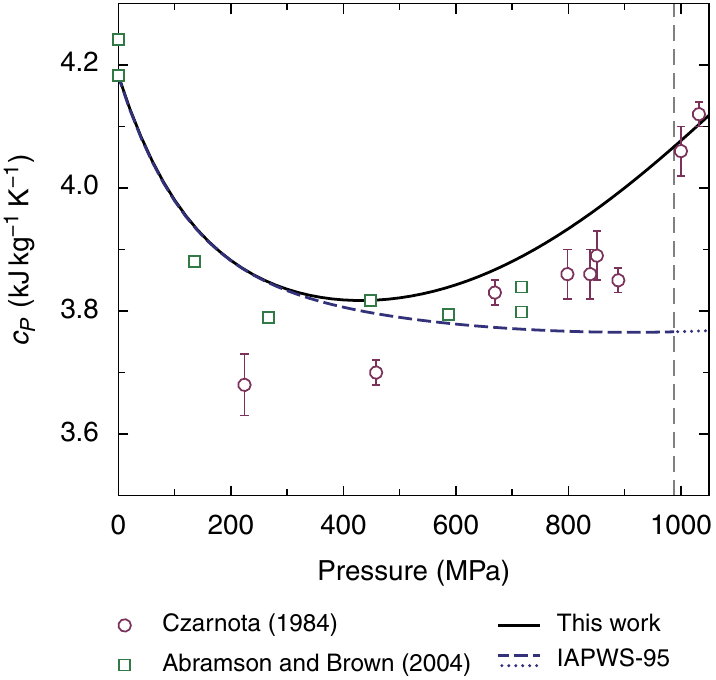}
\caption{\label{fig:cpczarnota}Isobaric heat capacities $c_P$ at 299.5~K calculated from
\eqref{eq:eos} and IAPWS-95
(dashed in the stable-liquid region and dotted in the metastable region).
Also shown are experimental data of Czarnota\cite{czarnota1984} in the range of
298.8~K to 300.1~K and data of Abramson and Brown\cite{abramson2004} at 298.15~K,
derived from thermal diffusivity and thermal conductivity measurements.
The vertical dashed line indicates the melting pressure.}
\end{figure}

The isobaric heat capacity $c_P$ was measured by Czarnota\cite{czarnota1984} at 300~K up
to 1000~MPa. Abramson and Brown\cite{abramson2004} derived $c_P$ values at 298~K up to
700~MPa from thermal diffusivity and thermal conductivity data. These data are compared
with values calculated from \eqref{eq:eos} and IAPWS-95 in \figref{fig:cpczarnota}. Two
data points of Czarnota are above the melting pressure, but Czarnota reported that the
water was still liquid for those measurements.

\subsection{Connection to IAPWS-95}
Because \eqref{eq:eos} was fitted to values calculated from IAPWS-95 in a part of the
temperature and pressure range (see \eqref{eq:iapwsrange} and \figref{fig:ExpPTfit}), the
differences between the two equations of state are small in that region. Therefore, there
are no large discontinuities when one switches from \eqref{eq:eos} to IAPWS-95 there. For
example, one can switch from one equation to the other at the isotherm
\begin{equation}\label{eq:mergeT}
    T = 320~\text{K},
\end{equation}
The differences between \eqref{eq:eos} and IAPWS-95 along this isotherm are given in
\tabref{tab:merge}.

\begin{table}
\caption{\label{tab:merge}Differences between \eqref{eq:eos} and IAPWS-95 along
\eqref{eq:mergeT} in the $P$--$T$ plane, for pressures from 0~MPa to 400~MPa}
\begin{ruledtabular}
\begin{tabular}{lll}
Quantity                & Mean\footnotemark[1]   & Maximum\footnotemark[2]\\
\hline
Density                 & 0.0006\%	& 0.0017\%        \\
Expansivity             & 0.010 K$^{-1}$&	0.021 K$^{-1}$  \\
Compressibility         & 0.02\%    & 0.05\%          \\
Heat capacity $c_P$     & 0.02\%	& 0.05\%          \\
Speed of sound          & 0.005\%	& 0.012\%         \\
\end{tabular}
\end{ruledtabular}
\footnotetext[1]{Average absolute difference}%
\footnotetext[2]{Maximum absolute difference}
\end{table}

\subsection{Uncertainty estimates}
\begin{figure}
\includegraphics{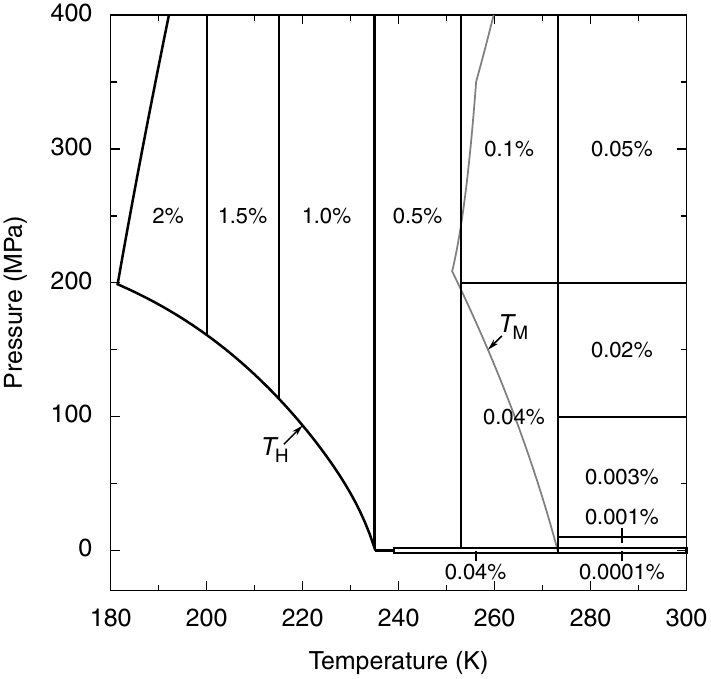}
\caption{\label{fig:densityuncert}Uncertainties in density estimated for \eqref{eq:eos}.
The thin rectangles around zero pressure refer to atmospheric pressure (0.101325~MPa).
$T\ts{M}$ indicates the melting temperature and
$T\ts{H}$ the homogeneous nucleation temperature (Appendix~\ref{app:TH}).
The melting curve does not separate uncertainty regions.}
\end{figure}

\begin{figure}
\includegraphics{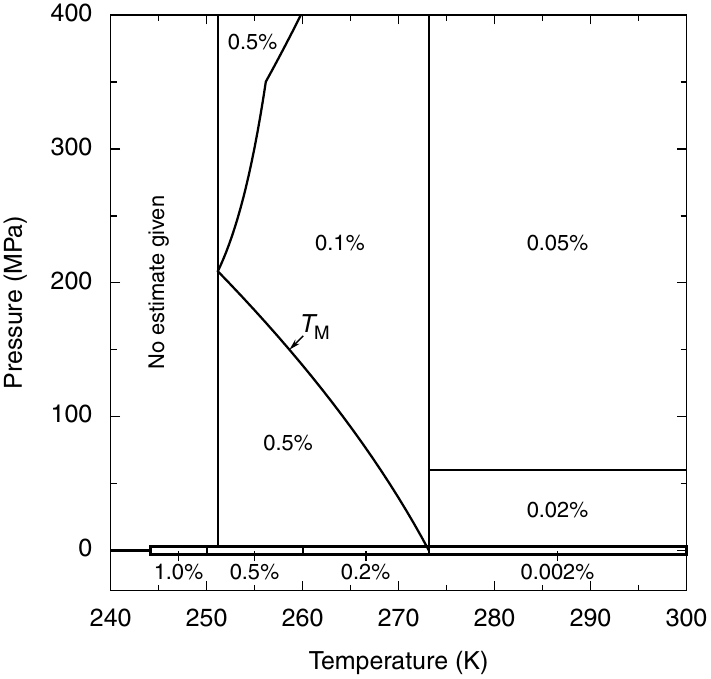}
\caption{\label{fig:wuncert}Uncertainties in speed of sound estimated for \eqref{eq:eos}.
The thin rectangles around zero pressure refer to atmospheric pressure (0.101325~MPa).
$T\ts{M}$ indicates the melting temperature. In the region labeled ``No estimate given'',
there are no experimental data for the speed of sound; this region extends down to the
homogeneous nucleation temperature.}
\end{figure}

Uncertainty estimates for the density calculated from \eqref{eq:eos} are shown in
\figref{fig:densityuncert}. These estimates are based on the differences between
\eqref{eq:eos} and experimental data, as well as on the uncertainty of the data. In a
large region of the phase diagram below 253~K, only Mishima's data are available. The
estimates in that region are conservative to account for the unknown systematic error of
Mishima's data. Uncertainty estimates for the speed of sound calculated from
\eqref{eq:eos} are shown in \figref{fig:wuncert}. In the region above atmospheric
pressure and below 253~K, no estimates are given because of the absence of experimental
data.

\subsection{Ice I melting curve}
As an additional test of the accuracy of the equation of
state, the melting curve of ice I was calculated from the phase-equilibrium condition
\begin{equation}\label{eq:I-L}
    g(T,P) = g\ts{I}(T,P).
\end{equation}
Here, $g\ts{I}$ is the Gibbs energy of ice I, which was calculated from the equation of
state of Feistel and Wagner.\cite{feistel2006} The Gibbs energy of liquid water $g$ was
calculated from \eqref{eq:eos}, with zero points of entropy $s$ and Gibbs energy chosen
such that
\begin{align}
    s(T\ts{t},P\ts{t}) &= 0,\label{eq:zeroentropy}\\
    g(T\ts{t},P\ts{t}) &= g\ts{I}(T\ts{t},P\ts{t}),\label{eq:gibbstriple}
\end{align}
where $T\ts{t}$ and $P\ts{t}$ are the temperature and pressure at the ice
I--liquid--vapor triple point, with\cite{feistel2008}
\begin{align}
    T\ts{t} &= 273.16~\text{K},\\
    P\ts{t} &= 611.654\,771\,007\,894~\text{Pa}.
\end{align}
The value for the triple-point pressure given here is not the experimental value, but was
calculated\cite{feistel2008} from the IAPWS-95 formulation and the equation of state of
ice I.\cite{feistel2006} The calculated value agrees with the experimental value of
$(611.657 \pm 0.010)$~Pa.\cite{guildner1976} \eeqref{eq:zeroentropy} represents the
convention that the specific entropy $s$ of liquid water is zero at the triple
point.\cite{wag02nonote} \eeqref{eq:gibbstriple} ensures that the melting curve
calculated from \eqref{eq:I-L} crosses the triple point.

\begin{figure}
\includegraphics{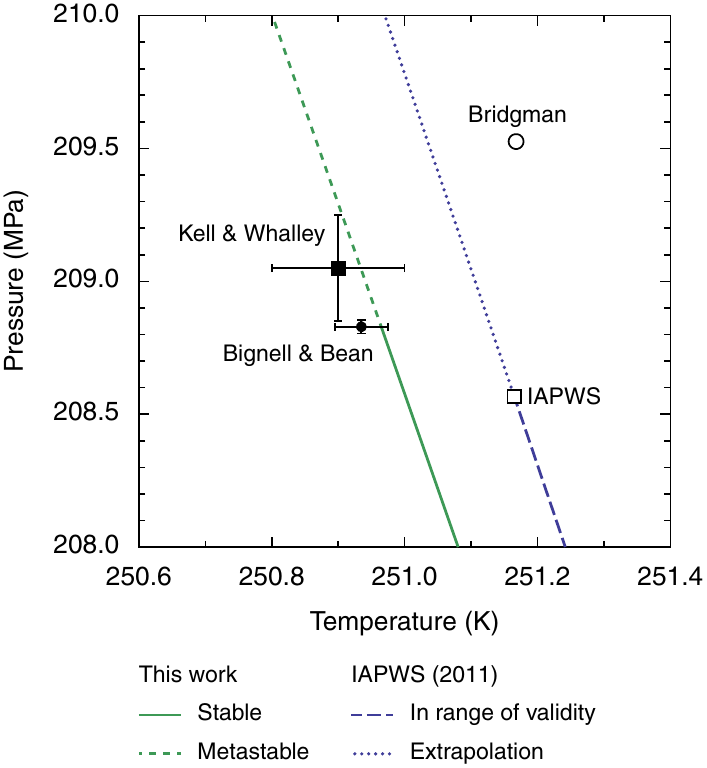}
\caption{\label{fig:I-III-L}Locations of the ice I--ice III--liquid triple point reported by
Bridgman,\cite{bridgman1912} Kell and Whalley,\cite{kell1968} and Bignell and Bean.\cite{bignell1988}
The green curve is the ice I melting curve calculated from \eqref{eq:I-L}; solid in the stable
region and dashed in the metastable region.
The triple point and melting curve from IAPWS\cite{wagner2011,iapwsmeltsub2011} are shown
for comparison.
}
\end{figure}

The melting curve of ice I crosses the triple point of ice I, ice III and liquid water at
about 209~MPa. Experimental locations of the I-III-L triple point are shown in
\figref{fig:I-III-L}. Bridgman\cite{bridgman1912} located the triple point in 1912. As
described by Babb\cite{babb1963} and La Mori,\cite{lamori1965} Bridgman's pressure values
are about 1\% low. For this work, Bridgman's pressures were multiplied by the correction
factor 1.0102, which follows from the current value of the mercury melting
pressure.\cite{molinar1980} Kell and Whalley\cite{kell1968} reported the location of the
triple point as part of their investigation of the ice I--ice III phase transition line.
Bignell and Bean\cite{bignell1988} determined the triple-point pressure with metrological
accuracy (0.01\%). Their measurements of the triple-point pressure and temperature are
currently the best available. As seen in \figref{fig:I-III-L}, the melting curve
calculated from \eqref{eq:I-L} agrees with the measurement of Bignell and
Bean.\cite{bignell1988}

\begin{figure}
\includegraphics{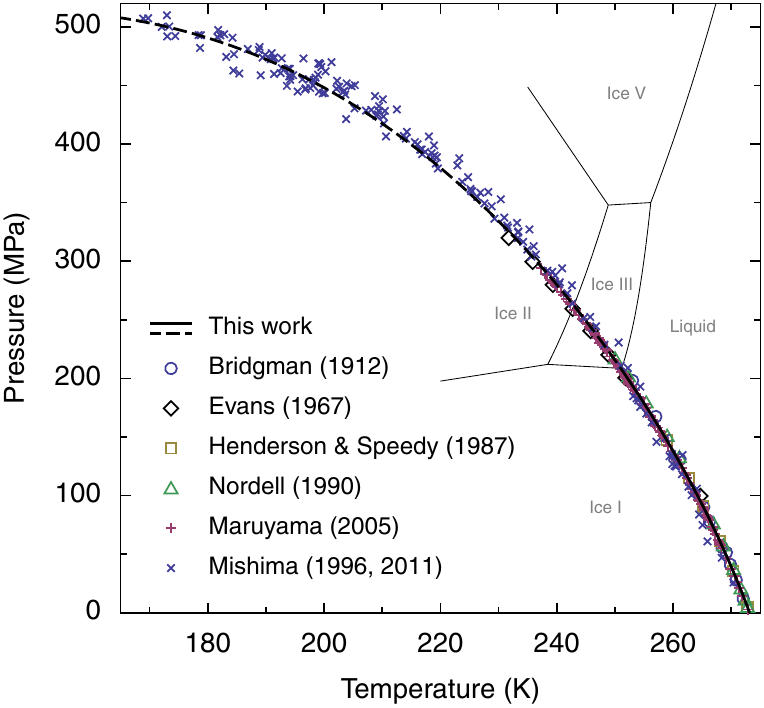}
\caption{\label{fig:I-L}
Measurements of the ice I melting curve.%
\cite{bridgman1912,evans1967,henderson1987b,nordell1990,maruyama2005,mishima1996,mishima2011}
The thick curve is the predicted melting curve, calculated from \eqref{eq:I-L} (solid in the
stable region and dashed in the metastable region).
Thin curves are the boundaries
of other ices.\cite{bridgman1912,kell1968,wagner2011}}
\end{figure}

\begin{figure}
\includegraphics{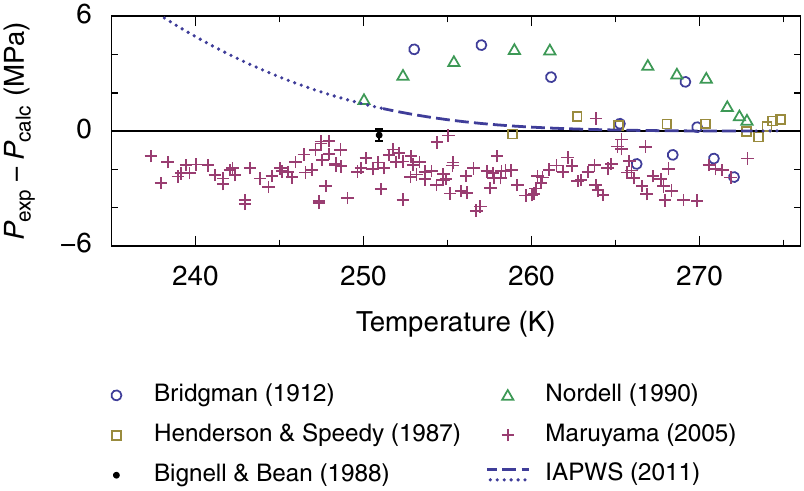}
\caption{\label{fig:I-Ldev}
Difference between experimental pressures on the ice I melting
curve\cite{bridgman1912,henderson1987b,bignell1988,nordell1990,maruyama2005}
and the pressure calculated from \eqref{eq:I-L}.
The melting curve from IAPWS\cite{wagner2011,iapwsmeltsub2011}
is shown for comparison (dashed in the stable region and dotted in the metastable region).
Note that the expression from IAPWS is valid only in the stable region;
the values in the metastable region were obtained by extrapolation.
}
\end{figure}

Bridgman located 14 points on the ice I melting curve.\cite{bridgman1912} Of these, four
points are unreliable according to Bridgman, so they will not be considered here. About
60 points on the melting curve were determined by Kishimoto and
Maruyama,\cite{kishimoto1998} who found a discontinuity in the melting curve at 160~MPa.
In a follow-up study, Maruyama\cite{maruyama2005} did not observe the discontinuity, and
suggested that it could have been an artifact of the previous experimental setup.
Mishima\cite{mishima1996,mishima2011} determined the course of the melting curve in the
range where ice I is metastable. \eeqref{eq:I-L} was used to calculate the melting curve
in this range by extrapolating both the equation of state of ice and of that of
supercooled water. \ffigref{fig:I-L} shows that the calculated melting curve agrees
fairly well with Mishima's data. \ffigref{fig:I-Ldev} shows deviations of the
experimental data from values computed from \eqref{eq:I-L}. The data from Henderson and
Speedy\cite{henderson1987b} are the most accurate and differ less than 1~MPa from
\eqref{eq:I-L}. The data of Maruyama\cite{maruyama2005} systematically lie 2~MPa below
\eqref{eq:I-L}. The maximum pressure difference in the stable region between
\eqref{eq:I-L} and values from the IAPWS correlation\cite{wagner2011} is 0.6\%, which is
well within the uncertainty of 2\% of the IAPWS correlation.

\subsection{Vapor pressure}
\begin{figure}
\includegraphics{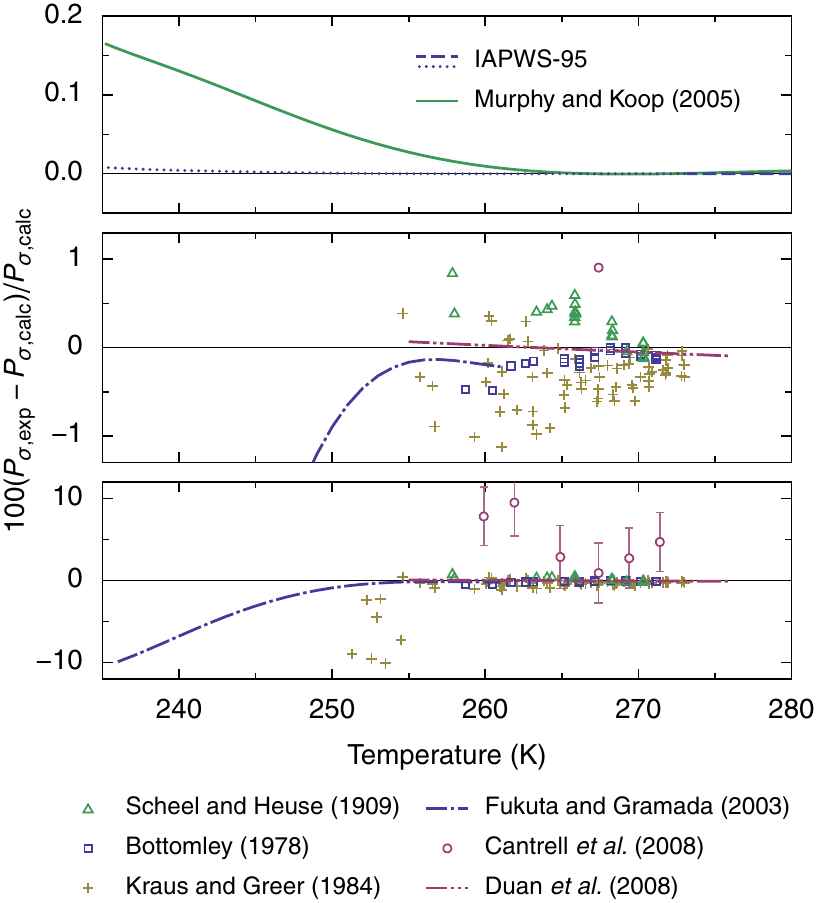}
\caption{\label{fig:vaporpressure}
Percentage deviations of correlations\cite{mur05} and experimental vapor pressure
data\cite{scheel1909,bottomley1978,kraus1984,fukuta2003,cantrell2008,*cantrell2009,duan2008}
from vapor pressure values calculated from \eqref{eq:liq-vap}. The vapor pressure calculated
from IAPWS-95 is shown dashed in the stable region and dotted in the metastable region.}
\end{figure}

The vapor pressure $\Pv(T)$ of stable and supercooled water was computed from the
equation of state, by using the phase-equilibrium condition
\begin{equation}\label{eq:liq-vap}
    g(T,\Pv) = g\ts{vap}(T,\Pv),
\end{equation}
where $g\ts{vap}$ is the specific Gibbs energy of water vapor, which was calculated from
the IAPWS-95 formulation. In \figref{fig:vaporpressure}, the calculated vapor pressure is
compared with other correlations and with experimental data. The vapor pressure computed
from IAPWS-95 is essentially the same as that computed from \eqref{eq:liq-vap}; the
maximum difference is 0.008\%. Murphy and Koop\cite{mur05} derived their vapor pressure
correlation from the heat capacities measured by Archer and Carter,\cite{arc00} which are
lower than heat capacities predicted by \eqref{eq:eos} and IAPWS-95
(\figref{fig:Cp1atm}). As a result, the vapor pressures in the supercooled region
predicted by Murphy and Koop\cite{mur05} are higher than those calculated from
\eqref{eq:liq-vap}.

Above 255~K, the experimental data agree to within about 1\% with \eqref{eq:liq-vap},
except the data of of Cantrell \ea,\cite{cantrell2008,*cantrell2009} which have larger
uncertainties. Below 255~K, the vapor pressures measured by Kraus and
Greer\cite{kraus1984} are anomalously low, which they suspect to be caused by freezing of
some of the droplets in their experiment. The correlation that Fukuta and
Gramada\cite{fukuta2003} obtained from a fit to their data deviates by up to 10\% from
\eqref{eq:liq-vap} at low temperature. As Murphy and Koop\cite{mur05} remarked, such low
values for the vapor pressure can only be explained if the isobaric heat capacity $c_P$
of supercooled water were about a factor of three higher than has been measured, which is
unlikely.

\section{Conclusion}
We have developed an equation of state for cold and supercooled water, explicit in the
Gibbs energy, valid from the homogeneous nucleation temperature to 300~K and for
pressures up to 400~MPa. The equation is based on a two-state model of water, combined
with empirical background terms. It is the first equation of state that represents the
density data of Sotani \ea\cite{sotani2000} and Asada \ea\cite{asada2002} as well as the
speed-of-sound data of Lin and Trusler\cite{lintrusler2012} in the considered temperature
range. In part of the stable region of liquid water, the equation can be connected to the
IAPWS-95 formulation with minimal discontinuities in the property values.

To improve the accuracy of future equations of state, density measurements with an
accuracy of 0.02\% or better below 250~K up to 400~MPa are desirable. Currently, this
area is only covered by Mishima's data. For the speed of sound, there are only a few
measurements in the metastable region for pressures higher than atmospheric. Experimental
data are needed especially down to 250~K and up to 200~MPa. Also, there are no data for
the heat capacity of supercooled water above atmospheric pressure, while such data are
highly desirable.

\section*{Acknowledgments}
The research has been supported by the Division of Chemistry of the US National Science
Foundation under Grant No. CHE-1012052. The research of V.H. was also supported by the
International Association for the Properties of Water and Steam. We have benefited from
discussions in the IAPWS Task Group on metastable water. A report of W. Wagner and M.
Thol\cite{wagnerthol2013} on the behavior of IAPWS-95 has also been helpful. We thank O.
Mishima for pointing out to us the existence of a minimum in the expansivity derived from
the data of Grindley and Lind, and for sending us his data on the melting curves of ice.
We are grateful to M. Maruyama for making available to us his data on the ice I melting
curve, and thank W. Cantrell for sending us data on the vapor pressure of supercooled
water.

\appendix

\section{Homogeneous Nucleation Curve}\label{app:TH}
\begin{figure}[b]
\includegraphics{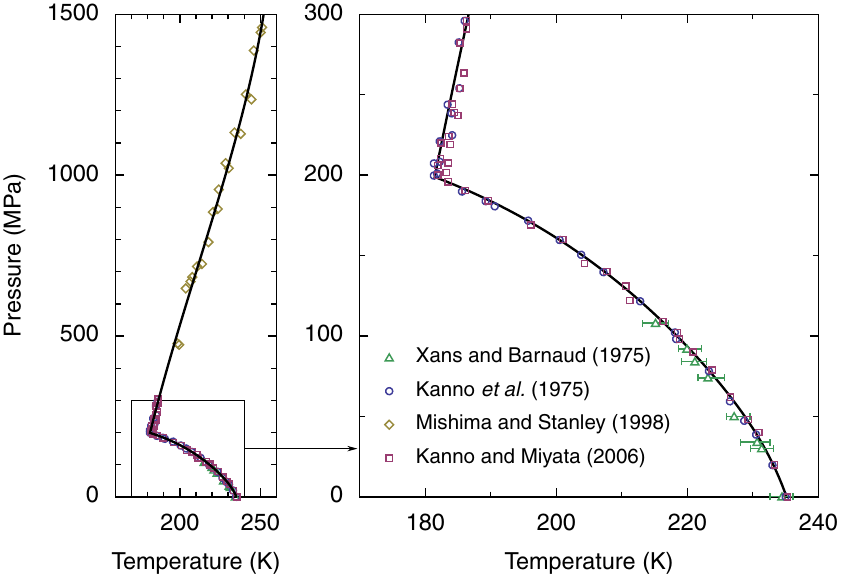}
\caption{\label{fig:TH}Temperature of homogenous ice nucleation for micrometer-size samples.
Symbols represent experimental data,\cite{xans1975,kanno1975,kanno2006,mishima1998}
and the curve is given by \eqsref{eq:THH2OlowP}{eq:THH2OhighP}.}
\end{figure}

Liquid water can be supercooled down to the homogenous ice nucleation temperature $\THN$,
which is about 235~K at atmospheric pressure. At higher pressures, $\THN$ is lower, with
a minimum of 181~K at 200~MPa. The pressure dependence of $\THN$ has been measured by
Xans and Barnaud\cite{xans1975}, Kanno \ea\cite{kanno1975} and Kanno and
Miyata\cite{kanno2006} at pressures below 300~MPa; see \figref{fig:TH}. Mishima and
Stanley\cite{mishima1998} have measured $\THN$ at pressures from 500~MPa to 1500~MPa. At
about 200~MPa, there is a break in the $\THN$ curve as a result of nucleation of a
different kind of ice above this pressure (ice III according to Kanno \ea
\cite{kanno1975,kanno1977} and ice II according to Kanno and Miyata\cite{kanno2006}). For
pressures below the break point, the shape of the $\THN$ curve can be described by an
equation of the Simon type, which was also used by Wagner \ea\cite{wagner1994} to
describe the melting curves of water. A fit of such an equation to the data of Kanno
\ea\cite{kanno1975} and Kanno and Miyata\cite{kanno2006} yields the pressure $P\ts{H}$ on
the homogeneous nucleation curve as a function of the temperature~$T$,
\begin{equation}\label{eq:THH2OlowP}
    P\ts{H}/P_0 = 1 + 2282.7 (1 - \theta^{6.243}) + 157.24 (1 - \theta^{79.81}),
\end{equation}
where $\theta = T/T_0$, $T_0 = 235.15$~K, and $P_0 = 0.1$~MPa. Above the break point, a
third-order polynomial was fitted to the data, including the data of Mishima and Stanley
up to 1500~MPa, which resulted in
\begin{equation}
\begin{aligned}\label{eq:THH2OhighP}
    \THN/\text{K} = &172.82 + 0.03718 p_1 + 3.403 \times 10^{-5} p_1^2\\
                    & - 1.573 \times 10^{-8} p_1^3,
\end{aligned}
\end{equation}
with $p_1 = P/\text{MPa}$. This polynomial is based on the assumption
that the $\THN$ curve is smooth at pressures above the break point. However, because
there are only few data in the range of 300~MPa to 600~MPa, the existence of other break
points in the curve cannot be excluded. It must also be noted that unlike the melting
curve, the homogenous nucleation curve is a kinetic limit and depends on the size and
time scale of the experiment. The experimental homogeneous nucleation temperatures
described in this section were obtained with samples having a diameter of several
micrometers.

\section{Derivatives}\label{sec:derivatives}
The derivatives of the field $L$, given by \eqref{eq:field}, are
\begin{gather}
    \Lt = \frac{L_0 K_2}{2}\left(1+\frac{1 - k_0 k_2 + k_1 (p - k_2 t)}{K_1}\right),\\
    \Lp = \frac{L_0 K_2(K_1 + k_0 k_2 - k_1 p + k_1 k_2 t - 1)}{2k_2 K_1},\\
    \Ltt = -\frac{2 L_0 K_2 k_0 k_1 k_2^2}{K_1^3},\\
    \Ltp = \frac{2 L_0 K_2 k_0 k_1 k_2}{K_1^3},\\
    \Lpp = -\frac{2 L_0 K_2 k_0 k_1}{K_1^3}.
\end{gather}
The derivatives of the Gibbs energy $\GA$ of the hypothetical pure high-density
structure, given by \eqref{eq:backgr}, are
\begin{gather}
    \Bt(\tb,\pb) =  \sum_{i=1}^{n} c_i a_i \tb^{a_i-1} \pb^{b_i}\mathrm{e}^{-d_i \pb},\\
    \Bp(\tb,\pb) =  \sum_{i=1}^{n} c_i \tb^{a_i} \pb^{b_i-1}(b_i-d_i \pb)\mathrm{e}^{-d_i \pb},\\
    \Btt(\tb,\pb) = \sum_{i=1}^{n} c_i a_i(a_i-1) \tb^{a_i-2} \pb^{b_i}\mathrm{e}^{-d_i \pb},\\
    \Btp(\tb,\pb) = \sum_{i=1}^{n} c_i a_i \tb^{a_i-1} \pb^{b_i-1} (b_i-d_i \pb) \mathrm{e}^{-d_i \pb},\\
    \Bpp(\tb,\pb) = \sum_{i=1}^{n} c_i \tb^{a_i} \pb^{b_i-2}[(d_i\pb-b_i)^2-b_i]\mathrm{e}^{-d_i \pb}.
\end{gather}

\section{Tables}\label{app:tables}
The values of the parameters that are necessary to evaluate \eqref{eq:eos} are listed in
\tabref{tab:physparam} and \tabref{tab:backgrparam}. For the verification of computer
programs, \tabref{tab:verification} lists calculated values for several properties. The
values are given with more digits than their accuracy justifies, to enable a more
detailed verification. Computer code for the equation of state is included in the
supplemental material.\cite{SCWsupplement2013}

\begin{table} \caption{\label{tab:physparam}Parameter values for the equation of state}
\begin{ruledtabular}
\begin{tabular}{le{7}lle{9}}
Parameter & \multicolumn{1}{c}{Value} & Unit & Parameter & \multicolumn{1}{c}{Value}\\
\hline
$\Tc$	&	228.2	&	K	&	$\omega_0$	&	0.521\,226\,9	\\
$\Pc$	&	0\footnotemark[1]	&	MPa	&	$L_0$	&	0.763\,179\,54	\\
$\rhoref$	&	1\,081.648\,2	&	kg m$^{-3}$	&	$k_0$	&	0.072\,158\,686	\\
$R$	&	461.523\,087\footnotemark[2]	&	J kg$^{-1}$ K$^{-1}$	&	$k_1$	&	-0.315\,692\,32	\\
	&	 	&		&	$k_2$	&	5.299\,260\,8	\\
\end{tabular}
\end{ruledtabular}
\footnotetext[1]{Mean-field value of the critical pressure. The actual location of the
hypothesized critical point is uncertain. Accounting for critical fluctuations may
increase this value by about 15~MPa. Correspondingly, the value for the critical
temperature will decrease by about 2~K.\cite{holtentwostate}}\footnotetext[2]{Equal to
the ratio of the molar gas constant\cite{mohr2012} $R\ts{m} =
8.314\;462\;1$~$\mathrm{J\,mol^{-1}\,K^{-1}}$ and the molar mass\cite{wag02nonote} $M =
18.015\;268$~$\mathrm{g\,mol^{-1}}$.}
\end{table}

\begin{table} \caption{\label{tab:backgrparam}Parameter values for \eqref{eq:backgr}}
\begin{ruledtabular}
\begin{tabular}{le{12}e{4}e{4}e{4}}
$i$ & \multicolumn{1}{c}{$c_i$} & a_i & b_i & d_i\\
\hline
1	&	-8.157\;068\;138\;165\;5	&	0	&	0	&	0	\\
2	&	1.287\;503\;2	&	0	&	1	&	0	\\
3	&	7.090\;167\;359\;801\;2	&	1	&	0	&	0	\\
4	&	-3.277\;916\;1\times 10^{-2}	&	-0.2555	&	2.1051	&	-0.0016	\\
5	&	7.370\;394\;9\times 10^{-1}	&	1.5762	&	1.1422	&	0.6894	\\
6	&	-2.162\;862\;2\times 10^{-1}	&	1.6400	&	0.9510	&	0.0130	\\
7	&	-5.178\;247\;9	&	3.6385	&	0	&	0.0002	\\
8	&	4.229\;351\;7\times 10^{-4}	&	-0.3828	&	3.6402	&	0.0435	\\
9	&	2.359\;210\;9\times 10^{-2}	&	1.6219	&	2.0760	&	0.0500	\\
10	&	4.377\;375\;4	&	4.3287	&	-0.0016	&	0.0004	\\
11	&	-2.996\;777\;0\times 10^{-3}	&	3.4763	&	2.2769	&	0.0528	\\
12	&	-9.655\;801\;8\times 10^{-1}	&	5.1556	&	0.0008	&	0.0147	\\
13	&	3.759\;528\;6	&	-0.3593	&	0.3706	&	0.8584	\\
14	&	1.263\;244\;1	&	5.0361	&	-0.3975	&	0.9924	\\
15	&	2.854\;269\;7\times 10^{-1}	&	2.9786	&	2.9730	&	1.0041	\\
16	&	-8.599\;494\;7\times 10^{-1}	&	6.2373	&	-0.3180	&	1.0961	\\
17	&	-3.291\;615\;3\times 10^{-1}	&	4.0460	&	2.9805	&	1.0228	\\
18	&	9.001\;961\;6\times 10^{-2}	&	5.3558	&	2.9265	&	1.0303	\\
19	&	8.114\;972\;6\times 10^{-2}	&	9.0157	&	0.4456	&	1.6180	\\
20	&	-3.278\;821\;3	&	1.2194	&	0.1298	&	0.5213	\\
\end{tabular}
\end{ruledtabular}
\footnotesize\raggedright Note: the values of $c_1$ and $c_3$ do not affect measurable
properties, but only the zero points of energy and entropy. The values shown here were
calculated to satisfy \eqsref{eq:zeroentropy}{eq:gibbstriple}.
\end{table}

\begin{table*} \caption{\label{tab:verification}Calculated property values for computer program verification}
\begin{ruledtabular}
\begin{tabular}{le{6}e{5}e{6}e{6}e{5}e{5}e{10}e{9}}
$\displaystyle\frac{T}{\mathrm{K}}$ &
\multicolumn{1}{c}{$\dfrac{P}{\mathrm{MPa}}$} &
\multicolumn{1}{c}{$\dfrac{\rho}{\mathrm{kg\,m^{-3}}}$} &
\multicolumn{1}{c}{$\dfrac{\alpha_P}{10^{-4}\,\mathrm{K^{-1}}}$} &
\multicolumn{1}{c}{$\dfrac{\kappa_T}{10^{-4}\,\mathrm{MPa^{-1}}}$} &
\multicolumn{1}{c}{$\dfrac{c_P}{\mathrm{J\,kg^{-1}\,K^{-1}}}$} &
\multicolumn{1}{c}{$\dfrac{w}{\mathrm{m\,s^{-1}}}$} &
\multicolumn{1}{c}{$\xe$} &
\multicolumn{1}{c}{$L$}\\
\hline
273.15	&	0.101325	&	999.842\;29	&	-0.683\;042	&	5.088\;499	&	4\;218.300\;2	&	1\;402.388\;6	&	0.096\;654\;72	&	0.621\;204\;74	\\
235.15	&	0.101325	&	968.099\;99	&	-29.633\;816	&	11.580\;785	&	5\;997.563\;2	&	1\;134.585\;5	&	0.255\;102\;86	&	0.091\;763\;676	\\
250	&	200	&	1\;090.456\;77	&	3.267\;768	&	3.361\;311	&	3\;708.390\;2	&	1\;668.202\;0	&	0.030\;429\;27	&	0.723\;770\;81	\\
200	&	400	&	1\;185.028\;00	&	6.716\;009	&	2.567\;237	&	3\;338.525	&	1\;899.329\;4	&	0.007\;170\;08	&	1.155\;396\;5	\\
250	&	400	&	1\;151.715\;17	&	4.929\;927	&	2.277\;029	&	3\;757.214\;4	&	2\;015.878\;2	&	0.005\;358\;84	&	1.434\;514\;5	\\
\end{tabular}
\end{ruledtabular}
\end{table*}

\bibliography{vincentnew,supercooled}

\begin{thebibliography}{132}%
\makeatletter
\providecommand \@ifxundefined [1]{%
 \@ifx{#1\undefined}
}%
\providecommand \@ifnum [1]{%
 \ifnum #1\expandafter \@firstoftwo
 \else \expandafter \@secondoftwo
 \fi
}%
\providecommand \@ifx [1]{%
 \ifx #1\expandafter \@firstoftwo
 \else \expandafter \@secondoftwo
 \fi
}%
\providecommand \natexlab [1]{#1}%
\providecommand \enquote  [1]{``#1''}%
\providecommand \bibnamefont  [1]{#1}%
\providecommand \bibfnamefont [1]{#1}%
\providecommand \citenamefont [1]{#1}%
\providecommand \href@noop [0]{\@secondoftwo}%
\providecommand \href [0]{\begingroup \@sanitize@url \@href}%
\providecommand \@href[1]{\@@startlink{#1}\@@href}%
\providecommand \@@href[1]{\endgroup#1\@@endlink}%
\providecommand \@sanitize@url [0]{\catcode `\\12\catcode `\$12\catcode
  `\&12\catcode `\#12\catcode `\^12\catcode `\_12\catcode `\%12\relax}%
\providecommand \@@startlink[1]{}%
\providecommand \@@endlink[0]{}%
\providecommand \url  [0]{\begingroup\@sanitize@url \@url }%
\providecommand \@url [1]{\endgroup\@href {#1}{\urlprefix }}%
\providecommand \urlprefix  [0]{URL }%
\providecommand \Eprint [0]{\href }%
\providecommand \doibase [0]{http://dx.doi.org/\ignorespaces}%
\providecommand \selectlanguage [0]{\@gobble}%
\providecommand \bibinfo  [0]{\@secondoftwo}%
\providecommand \bibfield  [0]{\@secondoftwo}%
\providecommand \translation [1]{[#1]}%
\providecommand \BibitemOpen [0]{}%
\providecommand \bibitemStop [0]{}%
\providecommand \bibitemNoStop [0]{.\EOS\space}%
\providecommand \EOS [0]{\spacefactor3000\relax}%
\providecommand \BibitemShut  [1]{\csname bibitem#1\endcsname}%
\let\auto@bib@innerbib\@empty
\bibitem [{\citenamefont {Fahrenheit}(1724)}]{fahrenheit1724}%
  \BibitemOpen
  \bibfield  {author} {\bibinfo {author} {\bibfnamefont {D.~G.}\ \bibnamefont
  {Fahrenheit}},\ }\href {\doibase10.1098/rstl.1724.0016} {\bibfield  {journal}
  {\bibinfo  {journal} {Phil. Trans.}\ }\textbf {\bibinfo {volume} {33}},\
  \bibinfo {pages} {78} (\bibinfo {year} {1724})}\BibitemShut {NoStop}%
\bibitem [{\citenamefont {Rosenfeld}\ and\ \citenamefont
  {Woodley}(2000)}]{rosenfeld2000}%
  \BibitemOpen
  \bibfield  {author} {\bibinfo {author} {\bibfnamefont {D.}~\bibnamefont
  {Rosenfeld}}\ and\ \bibinfo {author} {\bibfnamefont {W.~L.}\ \bibnamefont
  {Woodley}},\ }\href@noop {} {\bibfield  {journal} {\bibinfo  {journal}
  {Nature}\ }\textbf {\bibinfo {volume} {405}},\ \bibinfo {pages} {440}
  (\bibinfo {year} {2000})}\BibitemShut {NoStop}%
\bibitem [{\citenamefont {Heymsfield}\ and\ \citenamefont
  {Miloshevich}(1993)}]{heymsfield1993}%
  \BibitemOpen
  \bibfield  {author} {\bibinfo {author} {\bibfnamefont {A.~J.}\ \bibnamefont
  {Heymsfield}}\ and\ \bibinfo {author} {\bibfnamefont {L.~M.}\ \bibnamefont
  {Miloshevich}},\ }\href@noop {} {\bibfield  {journal} {\bibinfo  {journal}
  {J. Atmos. Sci.}\ }\textbf {\bibinfo {volume} {50}},\ \bibinfo {pages} {2335}
  (\bibinfo {year} {1993})}\BibitemShut {NoStop}%
\bibitem [{\citenamefont {Schmelzer}\ and\ \citenamefont
  {Hellmuth}(2013)}]{schmelzerhellmuthbook}%
  \BibitemOpen
  \bibinfo {editor} {\bibfnamefont {J.~W.~P.}\ \bibnamefont {Schmelzer}}\ and\
  \bibinfo {editor} {\bibfnamefont {O.}~\bibnamefont {Hellmuth}},\ eds.,\
  \href@noop {} {\emph {\bibinfo {title} {Nucleation Theory and Applications:
  Special Issues}}},\ Vol.~\bibinfo {volume} {1}\ (\bibinfo  {publisher} {Joint
  Institute for Nuclear Research},\ \bibinfo {address} {Dubna},\ \bibinfo
  {year} {2013})\BibitemShut {NoStop}%
\bibitem [{\citenamefont {Skogseth}, \citenamefont {Nilsen},\ and\
  \citenamefont {Smedsrud}(2009)}]{skogseth2009}%
  \BibitemOpen
  \bibfield  {author} {\bibinfo {author} {\bibfnamefont {R.}~\bibnamefont
  {Skogseth}}, \bibinfo {author} {\bibfnamefont {F.}~\bibnamefont {Nilsen}}, \
  and\ \bibinfo {author} {\bibfnamefont {L.~H.}\ \bibnamefont {Smedsrud}},\
  }\href {\doibase10.3189/002214309788608840} {\bibfield  {journal} {\bibinfo
  {journal} {J. Glaciology}\ }\textbf {\bibinfo {volume} {55}},\ \bibinfo
  {pages} {43} (\bibinfo {year} {2009})}\BibitemShut {NoStop}%
\bibitem [{\citenamefont {Nash}(1966)}]{nashbook1966}%
  \BibitemOpen
  \bibfield  {author} {\bibinfo {author} {\bibfnamefont {T.}~\bibnamefont
  {Nash}},\ }\href@noop {} {\emph {\bibinfo {title} {Cryobiology}}}\ (\bibinfo
  {publisher} {Academic},\ \bibinfo {address} {New York},\ \bibinfo {year}
  {1966})\BibitemShut {NoStop}%
\bibitem [{\citenamefont {Song}\ \emph {et~al.}(2010)\citenamefont {Song},
  \citenamefont {Sharp}, \citenamefont {Lu},\ and\ \citenamefont
  {Hassan}}]{song2010}%
  \BibitemOpen
  \bibfield  {author} {\bibinfo {author} {\bibfnamefont {Y.}~\bibnamefont
  {Song}}, \bibinfo {author} {\bibfnamefont {R.}~\bibnamefont {Sharp}},
  \bibinfo {author} {\bibfnamefont {F.}~\bibnamefont {Lu}}, \ and\ \bibinfo
  {author} {\bibfnamefont {M.}~\bibnamefont {Hassan}},\ }\href@noop {}
  {\bibfield  {journal} {\bibinfo  {journal} {Cryobiology}\ }\textbf {\bibinfo
  {volume} {60}},\ \bibinfo {pages} {S60} (\bibinfo {year} {2010})}\BibitemShut
  {NoStop}%
\bibitem [{\citenamefont {Otero}, \citenamefont {Molina-García},\ and\
  \citenamefont {Sanz}(2002)}]{otero2002}%
  \BibitemOpen
  \bibfield  {author} {\bibinfo {author} {\bibfnamefont {L.}~\bibnamefont
  {Otero}}, \bibinfo {author} {\bibfnamefont {A.~D.}\ \bibnamefont
  {Molina-García}}, \ and\ \bibinfo {author} {\bibfnamefont {P.~D.}\
  \bibnamefont {Sanz}},\ }\href@noop {} {\bibfield  {journal} {\bibinfo
  {journal} {Crit. Rev. Food Sci. Nutrit.,}\ }\textbf {\bibinfo {volume}
  {42}},\ \bibinfo {pages} {339} (\bibinfo {year} {2002})}\BibitemShut
  {NoStop}%
\bibitem [{\citenamefont {Poole}\ \emph {et~al.}(1992)\citenamefont {Poole},
  \citenamefont {Sciortino}, \citenamefont {Essmann},\ and\ \citenamefont
  {Stanley}}]{poole1992}%
  \BibitemOpen
  \bibfield  {author} {\bibinfo {author} {\bibfnamefont {P.~H.}\ \bibnamefont
  {Poole}}, \bibinfo {author} {\bibfnamefont {F.}~\bibnamefont {Sciortino}},
  \bibinfo {author} {\bibfnamefont {U.}~\bibnamefont {Essmann}}, \ and\
  \bibinfo {author} {\bibfnamefont {H.~E.}\ \bibnamefont {Stanley}},\
  }\href@noop {} {\bibfield  {journal} {\bibinfo  {journal} {Nature}\ }\textbf
  {\bibinfo {volume} {360}},\ \bibinfo {pages} {324} (\bibinfo {year}
  {1992})}\BibitemShut {NoStop}%
\bibitem [{\citenamefont {Mishima}\ and\ \citenamefont
  {Stanley}(1998{\natexlab{a}})}]{mishima1998review}%
  \BibitemOpen
  \bibfield  {author} {\bibinfo {author} {\bibfnamefont {O.}~\bibnamefont
  {Mishima}}\ and\ \bibinfo {author} {\bibfnamefont {H.~E.}\ \bibnamefont
  {Stanley}},\ }\href@noop {} {\bibfield  {journal} {\bibinfo  {journal}
  {Nature}\ }\textbf {\bibinfo {volume} {396}},\ \bibinfo {pages} {329}
  (\bibinfo {year} {1998}{\natexlab{a}})}\BibitemShut {NoStop}%
\bibitem [{\citenamefont {Sato}(1990)}]{sato1990}%
  \BibitemOpen
  \bibfield  {author} {\bibinfo {author} {\bibfnamefont {H.}~\bibnamefont
  {Sato}},\ }in\ \href@noop {} {\emph {\bibinfo {booktitle} {Properties of
  Water and Steam: Proceedings of the 11th International Conference}}},\
  \bibinfo {editor} {edited by\ \bibinfo {editor} {\bibfnamefont
  {M.}~\bibnamefont {Píchal}}\ and\ \bibinfo {editor} {\bibfnamefont
  {O.}~\bibnamefont {Šifner}}}\ (\bibinfo  {publisher} {Hemisphere},\ \bibinfo
  {address} {New York},\ \bibinfo {year} {1990})\ pp.\ \bibinfo {pages}
  {48--55}\BibitemShut {NoStop}%
\bibitem [{\citenamefont {Jeffery}\ and\ \citenamefont
  {Austin}(1997)}]{jeffery1997}%
  \BibitemOpen
  \bibfield  {author} {\bibinfo {author} {\bibfnamefont {C.~A.}\ \bibnamefont
  {Jeffery}}\ and\ \bibinfo {author} {\bibfnamefont {P.~H.}\ \bibnamefont
  {Austin}},\ }\href@noop {} {\bibfield  {journal} {\bibinfo  {journal} {J.
  Geophys. Res.: Atmos.}\ }\textbf {\bibinfo {volume} {102}},\ \bibinfo {pages}
  {25269} (\bibinfo {year} {1997})}\BibitemShut {NoStop}%
\bibitem [{\citenamefont {Jeffery}\ and\ \citenamefont
  {Austin}(1999)}]{jeffery1999}%
  \BibitemOpen
  \bibfield  {author} {\bibinfo {author} {\bibfnamefont {C.~A.}\ \bibnamefont
  {Jeffery}}\ and\ \bibinfo {author} {\bibfnamefont {P.~H.}\ \bibnamefont
  {Austin}},\ }\href@noop {} {\bibfield  {journal} {\bibinfo  {journal}
  {J.~Chem. Phys.}\ }\textbf {\bibinfo {volume} {110}},\ \bibinfo {pages} {484}
  (\bibinfo {year} {1999})}\BibitemShut {NoStop}%
\bibitem [{\citenamefont {Kiselev}\ and\ \citenamefont
  {Ely}(2002)}]{kiselev2002}%
  \BibitemOpen
  \bibfield  {author} {\bibinfo {author} {\bibfnamefont {S.~B.}\ \bibnamefont
  {Kiselev}}\ and\ \bibinfo {author} {\bibfnamefont {J.~F.}\ \bibnamefont
  {Ely}},\ }\href@noop {} {\bibfield  {journal} {\bibinfo  {journal} {J.~Chem.
  Phys.}\ }\textbf {\bibinfo {volume} {116}},\ \bibinfo {pages} {5657}
  (\bibinfo {year} {2002})}\BibitemShut {NoStop}%
\bibitem [{\citenamefont {Fuentevilla}\ and\ \citenamefont
  {Anisimov}(2006)}]{fuentevilla2006}%
  \BibitemOpen
  \bibfield  {author} {\bibinfo {author} {\bibfnamefont {D.~A.}\ \bibnamefont
  {Fuentevilla}}\ and\ \bibinfo {author} {\bibfnamefont {M.~A.}\ \bibnamefont
  {Anisimov}},\ }\href@noop {} {\bibfield  {journal} {\bibinfo  {journal}
  {Phys. Rev. Lett.}\ }\textbf {\bibinfo {volume} {97}},\ \bibinfo {pages}
  {195702} (\bibinfo {year} {2006})},\ \bibinfo {note} {erratum \textit{ibid.}
  \textbf{98}, 149904 (2007)}\BibitemShut {NoStop}%
\bibitem [{\citenamefont {Bertrand}\ and\ \citenamefont
  {Anisimov}(2011)}]{bertrand2011}%
  \BibitemOpen
  \bibfield  {author} {\bibinfo {author} {\bibfnamefont {C.~E.}\ \bibnamefont
  {Bertrand}}\ and\ \bibinfo {author} {\bibfnamefont {M.~A.}\ \bibnamefont
  {Anisimov}},\ }\href@noop {} {\bibfield  {journal} {\bibinfo  {journal}
  {J.~Phys. Chem.~B}\ }\textbf {\bibinfo {volume} {115}},\ \bibinfo {pages}
  {14099} (\bibinfo {year} {2011})}\BibitemShut {NoStop}%
\bibitem [{\citenamefont {Holten}\ \emph
  {et~al.}(2012{\natexlab{a}})\citenamefont {Holten}, \citenamefont {Bertrand},
  \citenamefont {Anisimov},\ and\ \citenamefont {Sengers}}]{holtenSCW}%
  \BibitemOpen
  \bibfield  {author} {\bibinfo {author} {\bibfnamefont {V.}~\bibnamefont
  {Holten}}, \bibinfo {author} {\bibfnamefont {C.~E.}\ \bibnamefont
  {Bertrand}}, \bibinfo {author} {\bibfnamefont {M.~A.}\ \bibnamefont
  {Anisimov}}, \ and\ \bibinfo {author} {\bibfnamefont {J.~V.}\ \bibnamefont
  {Sengers}},\ }\href@noop {} {\bibfield  {journal} {\bibinfo  {journal}
  {J.~Chem. Phys.}\ }\textbf {\bibinfo {volume} {136}},\ \bibinfo {pages}
  {094507} (\bibinfo {year} {2012}{\natexlab{a}})}\BibitemShut {NoStop}%
\bibitem [{\citenamefont {Holten}\ and\ \citenamefont
  {Anisimov}(2012{\natexlab{a}})}]{holtentwostate}%
  \BibitemOpen
  \bibfield  {author} {\bibinfo {author} {\bibfnamefont {V.}~\bibnamefont
  {Holten}}\ and\ \bibinfo {author} {\bibfnamefont {M.~A.}\ \bibnamefont
  {Anisimov}},\ }\href@noop {} {\bibfield  {journal} {\bibinfo  {journal} {Sci.
  Rep.}\ }\textbf {\bibinfo {volume} {2}},\ \bibinfo {pages} {713} (\bibinfo
  {year} {2012}{\natexlab{a}})}\BibitemShut {NoStop}%
\bibitem [{\citenamefont {Loerting}\ \emph {et~al.}(2011)\citenamefont
  {Loerting}, \citenamefont {Winkel}, \citenamefont {Seidl}, \citenamefont
  {Bauer}, \citenamefont {Mitterdorfer}, \citenamefont {Handle}, \citenamefont
  {Salzmann}, \citenamefont {Mayer}, \citenamefont {Finney},\ and\
  \citenamefont {Bowron}}]{loerting2011PCCP}%
  \BibitemOpen
  \bibfield  {author} {\bibinfo {author} {\bibfnamefont {T.}~\bibnamefont
  {Loerting}}, \bibinfo {author} {\bibfnamefont {K.}~\bibnamefont {Winkel}},
  \bibinfo {author} {\bibfnamefont {M.}~\bibnamefont {Seidl}}, \bibinfo
  {author} {\bibfnamefont {M.}~\bibnamefont {Bauer}}, \bibinfo {author}
  {\bibfnamefont {C.}~\bibnamefont {Mitterdorfer}}, \bibinfo {author}
  {\bibfnamefont {P.~H.}\ \bibnamefont {Handle}}, \bibinfo {author}
  {\bibfnamefont {C.~G.}\ \bibnamefont {Salzmann}}, \bibinfo {author}
  {\bibfnamefont {E.}~\bibnamefont {Mayer}}, \bibinfo {author} {\bibfnamefont
  {J.~L.}\ \bibnamefont {Finney}}, \ and\ \bibinfo {author} {\bibfnamefont
  {D.~T.}\ \bibnamefont {Bowron}},\ }\href@noop {} {\bibfield  {journal}
  {\bibinfo  {journal} {Phys. Chem. Chem. Phys.}\ }\textbf {\bibinfo {volume}
  {13}},\ \bibinfo {pages} {8783} (\bibinfo {year} {2011})}\BibitemShut
  {NoStop}%
\bibitem [{\citenamefont {Wagner}\ and\ \citenamefont
  {Pruß}(2002)}]{wag02nonote}%
  \BibitemOpen
  \bibfield  {author} {\bibinfo {author} {\bibfnamefont {W.}~\bibnamefont
  {Wagner}}\ and\ \bibinfo {author} {\bibfnamefont {A.}~\bibnamefont {Pruß}},\
  }\href@noop {} {\bibfield  {journal} {\bibinfo  {journal} {J.~Phys. Chem.
  Ref. Data}\ }\textbf {\bibinfo {volume} {31}},\ \bibinfo {pages} {387}
  (\bibinfo {year} {2002})}\BibitemShut {NoStop}%
\bibitem [{iap(2009)}]{iapws95}%
  \BibitemOpen
  \href@noop {} {\emph {\bibinfo {title} {Revised Release on the IAPWS
  Formulation 1995 for the Thermodynamic Properties of Ordinary Water Substance
  for General and Scientific Use}}},\ \bibinfo {organization} {International
  Association for the Properties of Water and Steam} (\bibinfo {year} {2009}),\
  \bibinfo {note} {available from www.iapws.org}\BibitemShut {NoStop}%
\bibitem [{\citenamefont {Angell}(1982)}]{angell1982book}%
  \BibitemOpen
  \bibfield  {author} {\bibinfo {author} {\bibfnamefont {C.~A.}\ \bibnamefont
  {Angell}},\ }in\ \href@noop {} {\emph {\bibinfo {booktitle} {Water and
  Aqueous Solutions at Subzero Temperatures}}},\ \bibinfo {series} {Water: A
  Comprehensive Treatise}, Vol.~\bibinfo {volume} {7},\ \bibinfo {editor}
  {edited by\ \bibinfo {editor} {\bibfnamefont {F.}~\bibnamefont {Franks}}}\
  (\bibinfo  {publisher} {Plenum},\ \bibinfo {address} {New York},\ \bibinfo
  {year} {1982})\ Chap.~\bibinfo {chapter} {1}, pp.\ \bibinfo {pages}
  {1--81}\BibitemShut {NoStop}%
\bibitem [{\citenamefont {Angell}(1983)}]{angell83}%
  \BibitemOpen
  \bibfield  {author} {\bibinfo {author} {\bibfnamefont {C.~A.}\ \bibnamefont
  {Angell}},\ }\href@noop {} {\bibfield  {journal} {\bibinfo  {journal} {Ann.
  Rev. Phys. Chem.}\ }\textbf {\bibinfo {volume} {34}},\ \bibinfo {pages} {593}
  (\bibinfo {year} {1983})}\BibitemShut {NoStop}%
\bibitem [{\citenamefont {Sato}\ \emph {et~al.}(1991)\citenamefont {Sato},
  \citenamefont {Watanabe}, \citenamefont {Levelt~Sengers}, \citenamefont
  {Gallagher}, \citenamefont {Hill}, \citenamefont {Straub},\ and\
  \citenamefont {Wagner}}]{sato1991}%
  \BibitemOpen
  \bibfield  {author} {\bibinfo {author} {\bibfnamefont {H.}~\bibnamefont
  {Sato}}, \bibinfo {author} {\bibfnamefont {K.}~\bibnamefont {Watanabe}},
  \bibinfo {author} {\bibfnamefont {J.~M.~H.}\ \bibnamefont {Levelt~Sengers}},
  \bibinfo {author} {\bibfnamefont {J.~S.}\ \bibnamefont {Gallagher}}, \bibinfo
  {author} {\bibfnamefont {P.~G.}\ \bibnamefont {Hill}}, \bibinfo {author}
  {\bibfnamefont {J.}~\bibnamefont {Straub}}, \ and\ \bibinfo {author}
  {\bibfnamefont {W.}~\bibnamefont {Wagner}},\ }\href@noop {} {\bibfield
  {journal} {\bibinfo  {journal} {J.~Phys. Chem. Ref. Data}\ }\textbf {\bibinfo
  {volume} {20}},\ \bibinfo {pages} {1023} (\bibinfo {year}
  {1991})}\BibitemShut {NoStop}%
\bibitem [{\citenamefont {Debenedetti}(2003)}]{deben03}%
  \BibitemOpen
  \bibfield  {author} {\bibinfo {author} {\bibfnamefont {P.~G.}\ \bibnamefont
  {Debenedetti}},\ }\href@noop {} {\bibfield  {journal} {\bibinfo  {journal}
  {J.~Phys.: Condens. Matter}\ }\textbf {\bibinfo {volume} {15}},\ \bibinfo
  {pages} {R1669} (\bibinfo {year} {2003})}\BibitemShut {NoStop}%
\bibitem [{\citenamefont {Holten}\ \emph
  {et~al.}(2012{\natexlab{b}})\citenamefont {Holten}, \citenamefont {Kalová},
  \citenamefont {Anisimov},\ and\ \citenamefont {Sengers}}]{holtenMF2012}%
  \BibitemOpen
  \bibfield  {author} {\bibinfo {author} {\bibfnamefont {V.}~\bibnamefont
  {Holten}}, \bibinfo {author} {\bibfnamefont {J.}~\bibnamefont {Kalová}},
  \bibinfo {author} {\bibfnamefont {M.~A.}\ \bibnamefont {Anisimov}}, \ and\
  \bibinfo {author} {\bibfnamefont {J.~V.}\ \bibnamefont {Sengers}},\
  }\href@noop {} {\bibfield  {journal} {\bibinfo  {journal} {Int. J.
  Thermophys.}\ }\textbf {\bibinfo {volume} {33}},\ \bibinfo {pages} {758}
  (\bibinfo {year} {2012}{\natexlab{b}})}\BibitemShut {NoStop}%
\bibitem [{\citenamefont {Adams}(1931)}]{adams1931}%
  \BibitemOpen
  \bibfield  {author} {\bibinfo {author} {\bibfnamefont {L.~H.}\ \bibnamefont
  {Adams}},\ }\href@noop {} {\bibfield  {journal} {\bibinfo  {journal} {J. Am.
  Chem. Soc.}\ }\textbf {\bibinfo {volume} {53}},\ \bibinfo {pages} {3769}
  (\bibinfo {year} {1931})}\BibitemShut {NoStop}%
\bibitem [{\citenamefont {Grindley}\ and\ \citenamefont
  {Lind}(1971)}]{grindley1971}%
  \BibitemOpen
  \bibfield  {author} {\bibinfo {author} {\bibfnamefont {T.}~\bibnamefont
  {Grindley}}\ and\ \bibinfo {author} {\bibfnamefont {J.~E.}\ \bibnamefont
  {Lind}, \bibfnamefont {Jr.}},\ }\href@noop {} {\bibfield  {journal} {\bibinfo
   {journal} {J.~Chem. Phys.}\ }\textbf {\bibinfo {volume} {54}},\ \bibinfo
  {pages} {3983} (\bibinfo {year} {1971})}\BibitemShut {NoStop}%
\bibitem [{\citenamefont {Borzunov}, \citenamefont {Razumikhin},\ and\
  \citenamefont {Stekol'nikov}(1974)}]{borzunov1974}%
  \BibitemOpen
  \bibfield  {author} {\bibinfo {author} {\bibfnamefont {V.~A.}\ \bibnamefont
  {Borzunov}}, \bibinfo {author} {\bibfnamefont {V.~N.}\ \bibnamefont
  {Razumikhin}}, \ and\ \bibinfo {author} {\bibfnamefont {V.~A.}\ \bibnamefont
  {Stekol'nikov}},\ }in\ \href@noop {} {\emph {\bibinfo {booktitle}
  {Thermophysical properties of matter and substances}}},\ Vol.~\bibinfo
  {volume} {2},\ \bibinfo {editor} {edited by\ \bibinfo {editor} {\bibfnamefont
  {V.~A.}\ \bibnamefont {Rabinovich}}}\ (\bibinfo  {publisher} {Amerind
  Publishing Co.},\ \bibinfo {address} {New Delhi},\ \bibinfo {year} {1974})\
  pp.\ \bibinfo {pages} {187--195}\BibitemShut {NoStop}%
\bibitem [{\citenamefont {Kell}\ and\ \citenamefont
  {Whalley}(1975)}]{kellwhalley1975}%
  \BibitemOpen
  \bibfield  {author} {\bibinfo {author} {\bibfnamefont {G.~S.}\ \bibnamefont
  {Kell}}\ and\ \bibinfo {author} {\bibfnamefont {E.}~\bibnamefont {Whalley}},\
  }\href@noop {} {\bibfield  {journal} {\bibinfo  {journal} {J.~Chem. Phys.}\
  }\textbf {\bibinfo {volume} {62}},\ \bibinfo {pages} {3496} (\bibinfo {year}
  {1975})}\BibitemShut {NoStop}%
\bibitem [{\citenamefont {Bradshaw}\ and\ \citenamefont
  {Schleicher}(1976)}]{bradshaw1976}%
  \BibitemOpen
  \bibfield  {author} {\bibinfo {author} {\bibfnamefont {A.}~\bibnamefont
  {Bradshaw}}\ and\ \bibinfo {author} {\bibfnamefont {K.}~\bibnamefont
  {Schleicher}},\ }\href@noop {} {\bibfield  {journal} {\bibinfo  {journal}
  {Deep-Sea Res.}\ }\textbf {\bibinfo {volume} {23}},\ \bibinfo {pages} {583}
  (\bibinfo {year} {1976})}\BibitemShut {NoStop}%
\bibitem [{\citenamefont {Aleksandrov}, \citenamefont {Khasanshin},\ and\
  \citenamefont {Larkin}(1976{\natexlab{a}})}]{aleksandrov1976rhorussian}%
  \BibitemOpen
  \bibfield  {author} {\bibinfo {author} {\bibfnamefont {A.~A.}\ \bibnamefont
  {Aleksandrov}}, \bibinfo {author} {\bibfnamefont {T.~S.}\ \bibnamefont
  {Khasanshin}}, \ and\ \bibinfo {author} {\bibfnamefont {D.~K.}\ \bibnamefont
  {Larkin}},\ }\href@noop {} {\bibfield  {journal} {\bibinfo  {journal} {Zh.
  Fiz. Khim.}\ }\textbf {\bibinfo {volume} {50}},\ \bibinfo {pages} {394}
  (\bibinfo {year} {1976}{\natexlab{a}})}\BibitemShut {NoStop}%
\bibitem [{\citenamefont {Aleksandrov}, \citenamefont {Khasanshin},\ and\
  \citenamefont {Larkin}(1976{\natexlab{b}})}]{aleksandrov1976rho}%
  \BibitemOpen
  \bibfield  {author} {\bibinfo {author} {\bibfnamefont {A.~A.}\ \bibnamefont
  {Aleksandrov}}, \bibinfo {author} {\bibfnamefont {T.~S.}\ \bibnamefont
  {Khasanshin}}, \ and\ \bibinfo {author} {\bibfnamefont {D.~K.}\ \bibnamefont
  {Larkin}},\ }\href@noop {} {\bibfield  {journal} {\bibinfo  {journal} {Russ.
  J. Phys. Chem.}\ }\textbf {\bibinfo {volume} {50}},\ \bibinfo {pages} {231}
  (\bibinfo {year} {1976}{\natexlab{b}})}\BibitemShut {NoStop}%
\bibitem [{\citenamefont {Hilbert}, \citenamefont {Tödheide},\ and\
  \citenamefont {Franck}(1981)}]{hilbert1981}%
  \BibitemOpen
  \bibfield  {author} {\bibinfo {author} {\bibfnamefont {R.}~\bibnamefont
  {Hilbert}}, \bibinfo {author} {\bibfnamefont {K.}~\bibnamefont {Tödheide}}, \
  and\ \bibinfo {author} {\bibfnamefont {E.~U.}\ \bibnamefont {Franck}},\
  }\href@noop {} {\bibfield  {journal} {\bibinfo  {journal} {Ber. Bunsenges.
  Phys. Chem.}\ }\textbf {\bibinfo {volume} {85}},\ \bibinfo {pages} {636}
  (\bibinfo {year} {1981})}\BibitemShut {NoStop}%
\bibitem [{\citenamefont {Hare}\ and\ \citenamefont {Sorensen}(1987)}]{hare87}%
  \BibitemOpen
  \bibfield  {author} {\bibinfo {author} {\bibfnamefont {D.~E.}\ \bibnamefont
  {Hare}}\ and\ \bibinfo {author} {\bibfnamefont {C.~M.}\ \bibnamefont
  {Sorensen}},\ }\href@noop {} {\bibfield  {journal} {\bibinfo  {journal}
  {J.~Chem. Phys.}\ }\textbf {\bibinfo {volume} {87}},\ \bibinfo {pages} {4840}
  (\bibinfo {year} {1987})}\BibitemShut {NoStop}%
\bibitem [{\citenamefont {Sotani}\ \emph {et~al.}(2000)\citenamefont {Sotani},
  \citenamefont {Arabas}, \citenamefont {Kubota},\ and\ \citenamefont
  {Kijima}}]{sotani2000}%
  \BibitemOpen
  \bibfield  {author} {\bibinfo {author} {\bibfnamefont {T.}~\bibnamefont
  {Sotani}}, \bibinfo {author} {\bibfnamefont {J.}~\bibnamefont {Arabas}},
  \bibinfo {author} {\bibfnamefont {H.}~\bibnamefont {Kubota}}, \ and\ \bibinfo
  {author} {\bibfnamefont {M.}~\bibnamefont {Kijima}},\ }\href@noop {}
  {\bibfield  {journal} {\bibinfo  {journal} {High Temp.-High Press.}\ }\textbf
  {\bibinfo {volume} {32}},\ \bibinfo {pages} {433} (\bibinfo {year}
  {2000})}\BibitemShut {NoStop}%
\bibitem [{\citenamefont {Tanaka}\ \emph {et~al.}(2001)\citenamefont {Tanaka},
  \citenamefont {Girard}, \citenamefont {Davis}, \citenamefont {Peuto},\ and\
  \citenamefont {Bignell}}]{tanaka2001}%
  \BibitemOpen
  \bibfield  {author} {\bibinfo {author} {\bibfnamefont {M.}~\bibnamefont
  {Tanaka}}, \bibinfo {author} {\bibfnamefont {G.}~\bibnamefont {Girard}},
  \bibinfo {author} {\bibfnamefont {R.}~\bibnamefont {Davis}}, \bibinfo
  {author} {\bibfnamefont {A.}~\bibnamefont {Peuto}}, \ and\ \bibinfo {author}
  {\bibfnamefont {N.}~\bibnamefont {Bignell}},\ }\href@noop {} {\bibfield
  {journal} {\bibinfo  {journal} {Metrologia}\ }\textbf {\bibinfo {volume}
  {38}},\ \bibinfo {pages} {301} (\bibinfo {year} {2001})}\BibitemShut
  {NoStop}%
\bibitem [{\citenamefont {Asada}\ \emph {et~al.}(2002)\citenamefont {Asada},
  \citenamefont {Sotani}, \citenamefont {Arabas}, \citenamefont {Kubota},
  \citenamefont {Matsuo},\ and\ \citenamefont {Tanaka}}]{asada2002}%
  \BibitemOpen
  \bibfield  {author} {\bibinfo {author} {\bibfnamefont {S.}~\bibnamefont
  {Asada}}, \bibinfo {author} {\bibfnamefont {T.}~\bibnamefont {Sotani}},
  \bibinfo {author} {\bibfnamefont {J.}~\bibnamefont {Arabas}}, \bibinfo
  {author} {\bibfnamefont {H.}~\bibnamefont {Kubota}}, \bibinfo {author}
  {\bibfnamefont {S.}~\bibnamefont {Matsuo}}, \ and\ \bibinfo {author}
  {\bibfnamefont {Y.}~\bibnamefont {Tanaka}},\ }\href@noop {} {\bibfield
  {journal} {\bibinfo  {journal} {J. Phys.: Condens. Matter}\ }\textbf
  {\bibinfo {volume} {14}},\ \bibinfo {pages} {11447} (\bibinfo {year}
  {2002})}\BibitemShut {NoStop}%
\bibitem [{\citenamefont {Mishima}(2010)}]{mishima2010}%
  \BibitemOpen
  \bibfield  {author} {\bibinfo {author} {\bibfnamefont {O.}~\bibnamefont
  {Mishima}},\ }\href@noop {} {\bibfield  {journal} {\bibinfo  {journal}
  {J.~Chem. Phys.}\ }\textbf {\bibinfo {volume} {133}},\ \bibinfo {pages}
  {144503} (\bibinfo {year} {2010})}\BibitemShut {NoStop}%
\bibitem [{\citenamefont {Guignon}, \citenamefont {Aparicio},\ and\
  \citenamefont {Sanz}(2010)}]{guignon2010}%
  \BibitemOpen
  \bibfield  {author} {\bibinfo {author} {\bibfnamefont {B.}~\bibnamefont
  {Guignon}}, \bibinfo {author} {\bibfnamefont {C.}~\bibnamefont {Aparicio}}, \
  and\ \bibinfo {author} {\bibfnamefont {P.~D.}\ \bibnamefont {Sanz}},\
  }\href@noop {} {\bibfield  {journal} {\bibinfo  {journal} {J.~Chem. Eng.
  Data}\ }\textbf {\bibinfo {volume} {55}},\ \bibinfo {pages} {3338} (\bibinfo
  {year} {2010})}\BibitemShut {NoStop}%
\bibitem [{iap(2011)}]{iapwsmeltsub2011}%
  \BibitemOpen
  \href@noop {} {\emph {\bibinfo {title} {Revised Release on the Pressure along
  the Melting and Sublimation Curves of Ordinary Water Substance}}},\ \bibinfo
  {organization} {International Association for the Properties of Water and
  Steam} (\bibinfo {year} {2011}),\ \bibinfo {note} {available from
  www.iapws.org}\BibitemShut {NoStop}%
\bibitem [{\citenamefont {Wagner}\ \emph {et~al.}(2011)\citenamefont {Wagner},
  \citenamefont {Riethmann}, \citenamefont {Feistel},\ and\ \citenamefont
  {Harvey}}]{wagner2011}%
  \BibitemOpen
  \bibfield  {author} {\bibinfo {author} {\bibfnamefont {W.}~\bibnamefont
  {Wagner}}, \bibinfo {author} {\bibfnamefont {T.}~\bibnamefont {Riethmann}},
  \bibinfo {author} {\bibfnamefont {R.}~\bibnamefont {Feistel}}, \ and\
  \bibinfo {author} {\bibfnamefont {A.~H.}\ \bibnamefont {Harvey}},\
  }\href@noop {} {\bibfield  {journal} {\bibinfo  {journal} {J.~Phys. Chem.
  Ref. Data}\ }\textbf {\bibinfo {volume} {40}},\ \bibinfo {pages} {043103}
  (\bibinfo {year} {2011})}\BibitemShut {NoStop}%
\bibitem [{\citenamefont {Bridgman}(1912)}]{bridgman1912}%
  \BibitemOpen
  \bibfield  {author} {\bibinfo {author} {\bibfnamefont {P.~W.}\ \bibnamefont
  {Bridgman}},\ }\href@noop {} {\bibfield  {journal} {\bibinfo  {journal}
  {Proc. Am. Acad. Arts Sci.}\ }\textbf {\bibinfo {volume} {47}},\ \bibinfo
  {pages} {441} (\bibinfo {year} {1912})}\BibitemShut {NoStop}%
\bibitem [{\citenamefont {Kell}\ and\ \citenamefont
  {Whalley}(1968)}]{kell1968}%
  \BibitemOpen
  \bibfield  {author} {\bibinfo {author} {\bibfnamefont {G.~S.}\ \bibnamefont
  {Kell}}\ and\ \bibinfo {author} {\bibfnamefont {E.}~\bibnamefont {Whalley}},\
  }\href@noop {} {\bibfield  {journal} {\bibinfo  {journal} {J.~Chem. Phys.}\
  }\textbf {\bibinfo {volume} {48}},\ \bibinfo {pages} {2359} (\bibinfo {year}
  {1968})}\BibitemShut {NoStop}%
\bibitem [{\citenamefont {Ter~Minassian}, \citenamefont {Pruzan},\ and\
  \citenamefont {Soulard}(1981)}]{terminassian1981}%
  \BibitemOpen
  \bibfield  {author} {\bibinfo {author} {\bibfnamefont {L.}~\bibnamefont
  {Ter~Minassian}}, \bibinfo {author} {\bibfnamefont {P.}~\bibnamefont
  {Pruzan}}, \ and\ \bibinfo {author} {\bibfnamefont {A.}~\bibnamefont
  {Soulard}},\ }\href@noop {} {\bibfield  {journal} {\bibinfo  {journal}
  {J.~Chem. Phys.}\ }\textbf {\bibinfo {volume} {75}},\ \bibinfo {pages} {3064}
  (\bibinfo {year} {1981})}\BibitemShut {NoStop}%
\bibitem [{\citenamefont {Caldwell}(1978)}]{caldwell1978}%
  \BibitemOpen
  \bibfield  {author} {\bibinfo {author} {\bibfnamefont {D.~R.}\ \bibnamefont
  {Caldwell}},\ }\href@noop {} {\bibfield  {journal} {\bibinfo  {journal}
  {Deep-Sea Res.}\ }\textbf {\bibinfo {volume} {25}},\ \bibinfo {pages} {175}
  (\bibinfo {year} {1978})}\BibitemShut {NoStop}%
\bibitem [{\citenamefont {Speedy}\ and\ \citenamefont
  {Angell}(1976)}]{speedy1976}%
  \BibitemOpen
  \bibfield  {author} {\bibinfo {author} {\bibfnamefont {R.~J.}\ \bibnamefont
  {Speedy}}\ and\ \bibinfo {author} {\bibfnamefont {C.~A.}\ \bibnamefont
  {Angell}},\ }\href@noop {} {\bibfield  {journal} {\bibinfo  {journal}
  {J.~Chem. Phys.}\ }\textbf {\bibinfo {volume} {65}},\ \bibinfo {pages} {851}
  (\bibinfo {year} {1976})}\BibitemShut {NoStop}%
\bibitem [{\citenamefont {Kanno}\ and\ \citenamefont
  {Angell}(1979)}]{kanno1979}%
  \BibitemOpen
  \bibfield  {author} {\bibinfo {author} {\bibfnamefont {H.}~\bibnamefont
  {Kanno}}\ and\ \bibinfo {author} {\bibfnamefont {C.~A.}\ \bibnamefont
  {Angell}},\ }\href@noop {} {\bibfield  {journal} {\bibinfo  {journal}
  {J.~Chem. Phys.}\ }\textbf {\bibinfo {volume} {70}},\ \bibinfo {pages} {4008}
  (\bibinfo {year} {1979})}\BibitemShut {NoStop}%
\bibitem [{\citenamefont {Sotani}, \citenamefont {Kubota},\ and\ \citenamefont
  {Sakata}(1998)}]{sotani1998}%
  \BibitemOpen
  \bibfield  {author} {\bibinfo {author} {\bibfnamefont {T.}~\bibnamefont
  {Sotani}}, \bibinfo {author} {\bibfnamefont {H.}~\bibnamefont {Kubota}}, \
  and\ \bibinfo {author} {\bibfnamefont {A.}~\bibnamefont {Sakata}},\
  }\href@noop {} {\bibfield  {journal} {\bibinfo  {journal} {High Temp.-High
  Press.}\ }\textbf {\bibinfo {volume} {30}},\ \bibinfo {pages} {509} (\bibinfo
  {year} {1998})}\BibitemShut {NoStop}%
\bibitem [{\citenamefont {Wagner}\ and\ \citenamefont
  {Thol}(2013)}]{wagnerthol2013}%
  \BibitemOpen
  \bibfield  {author} {\bibinfo {author} {\bibfnamefont {W.}~\bibnamefont
  {Wagner}}\ and\ \bibinfo {author} {\bibfnamefont {M.}~\bibnamefont {Thol}},\
  }\href@noop {} {\enquote {\bibinfo {title} {The behavior of {IAPWS}-95 at
  temperatures from 250 {K} to 300 {K} and pressures up to 400 {MPa}},}\
  }\bibinfo {type} {Report prepared for the Task Group Subcooled Water and the
  Working Group Thermophysical Properties of Water and Steam, International
  Association for the Properties of Water and Steam}\ (\bibinfo  {institution}
  {Chair of Thermodynamics, Ruhr-University Bochum, Germany},\ \bibinfo {year}
  {2013})\BibitemShut {NoStop}%
\bibitem [{\citenamefont {Teká\v{c}}, \citenamefont {Cibulka},\ and\
  \citenamefont {Holub}(1985)}]{tekac1985}%
  \BibitemOpen
  \bibfield  {author} {\bibinfo {author} {\bibfnamefont {V.}~\bibnamefont
  {Teká\v{c}}}, \bibinfo {author} {\bibfnamefont {I.}~\bibnamefont {Cibulka}},
  \ and\ \bibinfo {author} {\bibfnamefont {R.}~\bibnamefont {Holub}},\
  }\href@noop {} {\bibfield  {journal} {\bibinfo  {journal} {Fluid Phase
  Equilib.}\ }\textbf {\bibinfo {volume} {19}},\ \bibinfo {pages} {33}
  (\bibinfo {year} {1985})}\BibitemShut {NoStop}%
\bibitem [{\citenamefont {Hiro}, \citenamefont {Wada},\ and\ \citenamefont
  {Kumagai}(2014)}]{hiro2013}%
  \BibitemOpen
  \bibfield  {author} {\bibinfo {author} {\bibfnamefont {K.}~\bibnamefont
  {Hiro}}, \bibinfo {author} {\bibfnamefont {T.}~\bibnamefont {Wada}}, \ and\
  \bibinfo {author} {\bibfnamefont {S.}~\bibnamefont {Kumagai}},\ }\href
  {\doibase10.1080/00319104.2013.793598} {\bibfield  {journal} {\bibinfo
  {journal} {Phys. Chem. Liq.}\ }\textbf {\bibinfo {volume} {52}},\ \bibinfo
  {pages} {37} (\bibinfo {year} {2014})}\BibitemShut {NoStop}%
\bibitem [{\citenamefont {Lin}\ and\ \citenamefont
  {Trusler}(2012)}]{lintrusler2012}%
  \BibitemOpen
  \bibfield  {author} {\bibinfo {author} {\bibfnamefont {C.-W.}\ \bibnamefont
  {Lin}}\ and\ \bibinfo {author} {\bibfnamefont {J.~P.~M.}\ \bibnamefont
  {Trusler}},\ }\href@noop {} {\bibfield  {journal} {\bibinfo  {journal}
  {J.~Chem. Phys.}\ }\textbf {\bibinfo {volume} {136}},\ \bibinfo {pages}
  {094511} (\bibinfo {year} {2012})}\BibitemShut {NoStop}%
\bibitem [{\citenamefont {Smith}\ and\ \citenamefont
  {Lawson}(1954)}]{smithlawson1954}%
  \BibitemOpen
  \bibfield  {author} {\bibinfo {author} {\bibfnamefont {A.~H.}\ \bibnamefont
  {Smith}}\ and\ \bibinfo {author} {\bibfnamefont {A.~W.}\ \bibnamefont
  {Lawson}},\ }\href {\doibase10.1063/1.1740074} {\bibfield  {journal}
  {\bibinfo  {journal} {J.~Chem. Phys.}\ }\textbf {\bibinfo {volume} {22}},\
  \bibinfo {pages} {351} (\bibinfo {year} {1954})}\BibitemShut {NoStop}%
\bibitem [{\citenamefont {Holton}\ \emph {et~al.}(1968)\citenamefont {Holton},
  \citenamefont {Hagelberg}, \citenamefont {Kao},\ and\ \citenamefont
  {Johnson}}]{holton1968}%
  \BibitemOpen
  \bibfield  {author} {\bibinfo {author} {\bibfnamefont {G.}~\bibnamefont
  {Holton}}, \bibinfo {author} {\bibfnamefont {M.~P.}\ \bibnamefont
  {Hagelberg}}, \bibinfo {author} {\bibfnamefont {S.}~\bibnamefont {Kao}}, \
  and\ \bibinfo {author} {\bibfnamefont {W.~H.}\ \bibnamefont {Johnson},
  \bibfnamefont {Jr.}},\ }\href {\doibase10.1121/1.1910739} {\bibfield
  {journal} {\bibinfo  {journal} {J. Acoust. Soc. Am.}\ }\textbf {\bibinfo
  {volume} {43}},\ \bibinfo {pages} {102} (\bibinfo {year} {1968})}\BibitemShut
  {NoStop}%
\bibitem [{\citenamefont {Belogol'skii}\ \emph {et~al.}(1999)\citenamefont
  {Belogol'skii}, \citenamefont {Sekoyan}, \citenamefont {Samorukova},
  \citenamefont {Stefanov},\ and\ \citenamefont {Levtsov}}]{belogolskii1999}%
  \BibitemOpen
  \bibfield  {author} {\bibinfo {author} {\bibfnamefont {V.~A.}\ \bibnamefont
  {Belogol'skii}}, \bibinfo {author} {\bibfnamefont {S.~S.}\ \bibnamefont
  {Sekoyan}}, \bibinfo {author} {\bibfnamefont {L.~M.}\ \bibnamefont
  {Samorukova}}, \bibinfo {author} {\bibfnamefont {S.~R.}\ \bibnamefont
  {Stefanov}}, \ and\ \bibinfo {author} {\bibfnamefont {V.~I.}\ \bibnamefont
  {Levtsov}},\ }\href@noop {} {\bibfield  {journal} {\bibinfo  {journal} {Meas.
  Tech.}\ }\textbf {\bibinfo {volume} {42}},\ \bibinfo {pages} {406} (\bibinfo
  {year} {1999})}\BibitemShut {NoStop}%
\bibitem [{\citenamefont {Leroy}, \citenamefont {Robinson},\ and\ \citenamefont
  {Goldsmith}(2008)}]{leroy2008}%
  \BibitemOpen
  \bibfield  {author} {\bibinfo {author} {\bibfnamefont {C.~C.}\ \bibnamefont
  {Leroy}}, \bibinfo {author} {\bibfnamefont {S.~P.}\ \bibnamefont {Robinson}},
  \ and\ \bibinfo {author} {\bibfnamefont {M.~J.}\ \bibnamefont {Goldsmith}},\
  }\href@noop {} {\bibfield  {journal} {\bibinfo  {journal} {J. Acoust. Soc.
  Am.}\ }\textbf {\bibinfo {volume} {124}},\ \bibinfo {pages} {2774} (\bibinfo
  {year} {2008})}\BibitemShut {NoStop}%
\bibitem [{\citenamefont {Aleksandrov}\ and\ \citenamefont
  {Larkin}(1976{\natexlab{a}})}]{aleksandrov1976}%
  \BibitemOpen
  \bibfield  {author} {\bibinfo {author} {\bibfnamefont {A.~A.}\ \bibnamefont
  {Aleksandrov}}\ and\ \bibinfo {author} {\bibfnamefont {D.~K.}\ \bibnamefont
  {Larkin}},\ }\href@noop {} {\bibfield  {journal} {\bibinfo  {journal}
  {Thermal Eng.}\ }\textbf {\bibinfo {volume} {23}},\ \bibinfo {pages} {72}
  (\bibinfo {year} {1976}{\natexlab{a}})}\BibitemShut {NoStop}%
\bibitem [{\citenamefont {Aleksandrov}\ and\ \citenamefont
  {Larkin}(1976{\natexlab{b}})}]{aleksandrov1976russian}%
  \BibitemOpen
  \bibfield  {author} {\bibinfo {author} {\bibfnamefont {A.~A.}\ \bibnamefont
  {Aleksandrov}}\ and\ \bibinfo {author} {\bibfnamefont {D.~K.}\ \bibnamefont
  {Larkin}},\ }\href@noop {} {\bibfield  {journal} {\bibinfo  {journal}
  {Teploenergetika}\ }\textbf {\bibinfo {volume} {23}},\ \bibinfo {pages} {75}
  (\bibinfo {year} {1976}{\natexlab{b}})}\BibitemShut {NoStop}%
\bibitem [{\citenamefont {Mamedov}(1979{\natexlab{a}})}]{mamedov1979}%
  \BibitemOpen
  \bibfield  {author} {\bibinfo {author} {\bibfnamefont {A.~M.}\ \bibnamefont
  {Mamedov}},\ }\href {\doibase10.1007/BF00861303} {\bibfield  {journal}
  {\bibinfo  {journal} {J. Eng. Phys.}\ }\textbf {\bibinfo {volume} {36}},\
  \bibinfo {pages} {113} (\bibinfo {year} {1979}{\natexlab{a}})}\BibitemShut
  {NoStop}%
\bibitem [{\citenamefont {Mamedov}(1979{\natexlab{b}})}]{mamedov1979russian}%
  \BibitemOpen
  \bibfield  {author} {\bibinfo {author} {\bibfnamefont {A.~M.}\ \bibnamefont
  {Mamedov}},\ }\href@noop {} {\bibfield  {journal} {\bibinfo  {journal} {Inzh.
  Fiz. Zh.}\ }\textbf {\bibinfo {volume} {36}},\ \bibinfo {pages} {156}
  (\bibinfo {year} {1979}{\natexlab{b}})}\BibitemShut {NoStop}%
\bibitem [{\citenamefont {Aleksandrov}\ and\ \citenamefont
  {Kochetkov}(1979{\natexlab{a}})}]{aleksandrov1979}%
  \BibitemOpen
  \bibfield  {author} {\bibinfo {author} {\bibfnamefont {A.~A.}\ \bibnamefont
  {Aleksandrov}}\ and\ \bibinfo {author} {\bibfnamefont {A.~I.}\ \bibnamefont
  {Kochetkov}},\ }\href@noop {} {\bibfield  {journal} {\bibinfo  {journal}
  {Thermal Eng.}\ }\textbf {\bibinfo {volume} {26}},\ \bibinfo {pages} {558}
  (\bibinfo {year} {1979}{\natexlab{a}})}\BibitemShut {NoStop}%
\bibitem [{\citenamefont {Aleksandrov}\ and\ \citenamefont
  {Kochetkov}(1979{\natexlab{b}})}]{aleksandrov1979russian}%
  \BibitemOpen
  \bibfield  {author} {\bibinfo {author} {\bibfnamefont {A.~A.}\ \bibnamefont
  {Aleksandrov}}\ and\ \bibinfo {author} {\bibfnamefont {A.~I.}\ \bibnamefont
  {Kochetkov}},\ }\href@noop {} {\bibfield  {journal} {\bibinfo  {journal}
  {Teploenergetika}\ }\textbf {\bibinfo {volume} {26}},\ \bibinfo {pages} {65}
  (\bibinfo {year} {1979}{\natexlab{b}})}\BibitemShut {NoStop}%
\bibitem [{\citenamefont {Vance}\ and\ \citenamefont
  {Brown}(2010)}]{vance2010}%
  \BibitemOpen
  \bibfield  {author} {\bibinfo {author} {\bibfnamefont {S.}~\bibnamefont
  {Vance}}\ and\ \bibinfo {author} {\bibfnamefont {J.~M.}\ \bibnamefont
  {Brown}},\ }\href@noop {} {\bibfield  {journal} {\bibinfo  {journal} {J.
  Acoust. Soc. Am.}\ }\textbf {\bibinfo {volume} {127}},\ \bibinfo {pages}
  {174} (\bibinfo {year} {2010})}\BibitemShut {NoStop}%
\bibitem [{\citenamefont {Hidalgo~Baltasar}\ \emph {et~al.}(2011)\citenamefont
  {Hidalgo~Baltasar}, \citenamefont {Taravillo}, \citenamefont {Baonza},
  \citenamefont {Sanz},\ and\ \citenamefont {Guignon}}]{hidalgobaltasar2011}%
  \BibitemOpen
  \bibfield  {author} {\bibinfo {author} {\bibfnamefont {E.}~\bibnamefont
  {Hidalgo~Baltasar}}, \bibinfo {author} {\bibfnamefont {M.}~\bibnamefont
  {Taravillo}}, \bibinfo {author} {\bibfnamefont {V.~G.}\ \bibnamefont
  {Baonza}}, \bibinfo {author} {\bibfnamefont {P.~D.}\ \bibnamefont {Sanz}}, \
  and\ \bibinfo {author} {\bibfnamefont {B.}~\bibnamefont {Guignon}},\
  }\href@noop {} {\bibfield  {journal} {\bibinfo  {journal} {J.~Chem. Eng.
  Data}\ }\textbf {\bibinfo {volume} {56}},\ \bibinfo {pages} {4800} (\bibinfo
  {year} {2011})}\BibitemShut {NoStop}%
\bibitem [{\citenamefont {Taschin}\ \emph {et~al.}(2011)\citenamefont
  {Taschin}, \citenamefont {Cucini}, \citenamefont {Bartolini},\ and\
  \citenamefont {Torre}}]{taschin2011}%
  \BibitemOpen
  \bibfield  {author} {\bibinfo {author} {\bibfnamefont {A.}~\bibnamefont
  {Taschin}}, \bibinfo {author} {\bibfnamefont {R.}~\bibnamefont {Cucini}},
  \bibinfo {author} {\bibfnamefont {P.}~\bibnamefont {Bartolini}}, \ and\
  \bibinfo {author} {\bibfnamefont {R.}~\bibnamefont {Torre}},\ }\href@noop {}
  {\bibfield  {journal} {\bibinfo  {journal} {Phil. Mag.}\ }\textbf {\bibinfo
  {volume} {91}},\ \bibinfo {pages} {1796} (\bibinfo {year}
  {2011})}\BibitemShut {NoStop}%
\bibitem [{\citenamefont {Wilson}(1959)}]{wilson1959}%
  \BibitemOpen
  \bibfield  {author} {\bibinfo {author} {\bibfnamefont {W.~D.}\ \bibnamefont
  {Wilson}},\ }\href@noop {} {\bibfield  {journal} {\bibinfo  {journal} {J.
  Acoust. Soc. Am.}\ }\textbf {\bibinfo {volume} {31}},\ \bibinfo {pages}
  {1067} (\bibinfo {year} {1959})}\BibitemShut {NoStop}%
\bibitem [{\citenamefont {Del~Grosso}\ and\ \citenamefont
  {Mader}(1972)}]{delgrosso1972}%
  \BibitemOpen
  \bibfield  {author} {\bibinfo {author} {\bibfnamefont {V.~A.}\ \bibnamefont
  {Del~Grosso}}\ and\ \bibinfo {author} {\bibfnamefont {C.~W.}\ \bibnamefont
  {Mader}},\ }\href@noop {} {\bibfield  {journal} {\bibinfo  {journal} {J.
  Acoust. Soc. Am.}\ }\textbf {\bibinfo {volume} {52}},\ \bibinfo {pages}
  {1442} (\bibinfo {year} {1972})}\BibitemShut {NoStop}%
\bibitem [{\citenamefont {Trinh}\ and\ \citenamefont
  {Apfel}(1978{\natexlab{a}})}]{trinhapfel1978JASA}%
  \BibitemOpen
  \bibfield  {author} {\bibinfo {author} {\bibfnamefont {E.}~\bibnamefont
  {Trinh}}\ and\ \bibinfo {author} {\bibfnamefont {R.~E.}\ \bibnamefont
  {Apfel}},\ }\href@noop {} {\bibfield  {journal} {\bibinfo  {journal} {J.
  Acoust. Soc. Am.}\ }\textbf {\bibinfo {volume} {63}},\ \bibinfo {pages} {777}
  (\bibinfo {year} {1978}{\natexlab{a}})}\BibitemShut {NoStop}%
\bibitem [{\citenamefont {Trinh}\ and\ \citenamefont
  {Apfel}(1978{\natexlab{b}})}]{trinhapfel1978JCP}%
  \BibitemOpen
  \bibfield  {author} {\bibinfo {author} {\bibfnamefont {E.}~\bibnamefont
  {Trinh}}\ and\ \bibinfo {author} {\bibfnamefont {R.~E.}\ \bibnamefont
  {Apfel}},\ }\href@noop {} {\bibfield  {journal} {\bibinfo  {journal}
  {J.~Chem. Phys.}\ }\textbf {\bibinfo {volume} {69}},\ \bibinfo {pages} {4245}
  (\bibinfo {year} {1978}{\natexlab{b}})}\BibitemShut {NoStop}%
\bibitem [{\citenamefont {Bacri}\ and\ \citenamefont
  {Rajaonarison}(1979)}]{bacri1979}%
  \BibitemOpen
  \bibfield  {author} {\bibinfo {author} {\bibfnamefont {J.-C.}\ \bibnamefont
  {Bacri}}\ and\ \bibinfo {author} {\bibfnamefont {R.}~\bibnamefont
  {Rajaonarison}},\ }\href@noop {} {\bibfield  {journal} {\bibinfo  {journal}
  {J. Physique Lett.}\ }\textbf {\bibinfo {volume} {40}},\ \bibinfo {pages}
  {L403} (\bibinfo {year} {1979})}\BibitemShut {NoStop}%
\bibitem [{\citenamefont {Trinh}\ and\ \citenamefont
  {Apfel}(1980)}]{trinhapfel1980}%
  \BibitemOpen
  \bibfield  {author} {\bibinfo {author} {\bibfnamefont {E.}~\bibnamefont
  {Trinh}}\ and\ \bibinfo {author} {\bibfnamefont {R.~E.}\ \bibnamefont
  {Apfel}},\ }\href@noop {} {\bibfield  {journal} {\bibinfo  {journal}
  {J.~Chem. Phys.}\ }\textbf {\bibinfo {volume} {72}},\ \bibinfo {pages} {6731}
  (\bibinfo {year} {1980})}\BibitemShut {NoStop}%
\bibitem [{\citenamefont {Petitet}, \citenamefont {Tufeu},\ and\ \citenamefont
  {Le~Neindre}(1983)}]{petitet1983}%
  \BibitemOpen
  \bibfield  {author} {\bibinfo {author} {\bibfnamefont {J.~P.}\ \bibnamefont
  {Petitet}}, \bibinfo {author} {\bibfnamefont {R.}~\bibnamefont {Tufeu}}, \
  and\ \bibinfo {author} {\bibfnamefont {B.}~\bibnamefont {Le~Neindre}},\
  }\href@noop {} {\bibfield  {journal} {\bibinfo  {journal} {Int. J.
  Thermophys.}\ }\textbf {\bibinfo {volume} {4}},\ \bibinfo {pages} {35}
  (\bibinfo {year} {1983})}\BibitemShut {NoStop}%
\bibitem [{\citenamefont {Fujii}\ and\ \citenamefont
  {Masui}(1993)}]{fujii1993}%
  \BibitemOpen
  \bibfield  {author} {\bibinfo {author} {\bibfnamefont {K.}~\bibnamefont
  {Fujii}}\ and\ \bibinfo {author} {\bibfnamefont {R.}~\bibnamefont {Masui}},\
  }\href {\doibase10.1121/1.405661} {\bibfield  {journal} {\bibinfo  {journal}
  {J. Acoust. Soc. Am.}\ }\textbf {\bibinfo {volume} {93}},\ \bibinfo {pages}
  {276} (\bibinfo {year} {1993})}\BibitemShut {NoStop}%
\bibitem [{\citenamefont {Benedetto}\ \emph {et~al.}(2005)\citenamefont
  {Benedetto}, \citenamefont {Gavioso}, \citenamefont {Giuliano~Albo},
  \citenamefont {Lago}, \citenamefont {Madonna~Ripa},\ and\ \citenamefont
  {Spagnolo}}]{benedetto2005}%
  \BibitemOpen
  \bibfield  {author} {\bibinfo {author} {\bibfnamefont {G.}~\bibnamefont
  {Benedetto}}, \bibinfo {author} {\bibfnamefont {R.~M.}\ \bibnamefont
  {Gavioso}}, \bibinfo {author} {\bibfnamefont {P.~A.}\ \bibnamefont
  {Giuliano~Albo}}, \bibinfo {author} {\bibfnamefont {S.}~\bibnamefont {Lago}},
  \bibinfo {author} {\bibfnamefont {D.}~\bibnamefont {Madonna~Ripa}}, \ and\
  \bibinfo {author} {\bibfnamefont {R.}~\bibnamefont {Spagnolo}},\ }\href@noop
  {} {\bibfield  {journal} {\bibinfo  {journal} {Int. J. Thermophys.}\ }\textbf
  {\bibinfo {volume} {26}},\ \bibinfo {pages} {1667} (\bibinfo {year}
  {2005})}\BibitemShut {NoStop}%
\bibitem [{\citenamefont {Taschin}\ \emph {et~al.}(2006)\citenamefont
  {Taschin}, \citenamefont {Bartolini}, \citenamefont {Eramo},\ and\
  \citenamefont {Torre}}]{taschin2006}%
  \BibitemOpen
  \bibfield  {author} {\bibinfo {author} {\bibfnamefont {A.}~\bibnamefont
  {Taschin}}, \bibinfo {author} {\bibfnamefont {P.}~\bibnamefont {Bartolini}},
  \bibinfo {author} {\bibfnamefont {R.}~\bibnamefont {Eramo}}, \ and\ \bibinfo
  {author} {\bibfnamefont {R.}~\bibnamefont {Torre}},\ }\href@noop {}
  {\bibfield  {journal} {\bibinfo  {journal} {Phys. Rev. E}\ }\textbf {\bibinfo
  {volume} {74}},\ \bibinfo {pages} {031502} (\bibinfo {year}
  {2006})}\BibitemShut {NoStop}%
\bibitem [{\citenamefont {Angell}, \citenamefont {Oguni},\ and\ \citenamefont
  {Sichina}(1982)}]{angell1982}%
  \BibitemOpen
  \bibfield  {author} {\bibinfo {author} {\bibfnamefont {C.~A.}\ \bibnamefont
  {Angell}}, \bibinfo {author} {\bibfnamefont {M.}~\bibnamefont {Oguni}}, \
  and\ \bibinfo {author} {\bibfnamefont {W.~J.}\ \bibnamefont {Sichina}},\
  }\href@noop {} {\bibfield  {journal} {\bibinfo  {journal} {J. Phys. Chem.}\
  }\textbf {\bibinfo {volume} {86}},\ \bibinfo {pages} {998} (\bibinfo {year}
  {1982})}\BibitemShut {NoStop}%
\bibitem [{\citenamefont {Archer}(1993)}]{archer1993}%
  \BibitemOpen
  \bibfield  {author} {\bibinfo {author} {\bibfnamefont {D.~G.}\ \bibnamefont
  {Archer}},\ }\href@noop {} {\bibfield  {journal} {\bibinfo  {journal}
  {J.~Phys. Chem. Ref. Data}\ }\textbf {\bibinfo {volume} {22}},\ \bibinfo
  {pages} {1441} (\bibinfo {year} {1993})}\BibitemShut {NoStop}%
\bibitem [{\citenamefont {Sirota}, \citenamefont {Grishkov},\ and\
  \citenamefont {Tomishko}(1970)}]{sirota1970}%
  \BibitemOpen
  \bibfield  {author} {\bibinfo {author} {\bibfnamefont {A.~M.}\ \bibnamefont
  {Sirota}}, \bibinfo {author} {\bibfnamefont {A.~{\relax Ya}.}\ \bibnamefont
  {Grishkov}}, \ and\ \bibinfo {author} {\bibfnamefont {A.~G.}\ \bibnamefont
  {Tomishko}},\ }\href@noop {} {\bibfield  {journal} {\bibinfo  {journal}
  {Thermal Eng.}\ }\textbf {\bibinfo {volume} {17}},\ \bibinfo {pages} {90}
  (\bibinfo {year} {1970})}\BibitemShut {NoStop}%
\bibitem [{\citenamefont {Czarnota}(1984)}]{czarnota1984}%
  \BibitemOpen
  \bibfield  {author} {\bibinfo {author} {\bibfnamefont {I.}~\bibnamefont
  {Czarnota}},\ }\href@noop {} {\bibfield  {journal} {\bibinfo  {journal} {High
  Temp.-High Press.}\ }\textbf {\bibinfo {volume} {16}},\ \bibinfo {pages}
  {295} (\bibinfo {year} {1984})}\BibitemShut {NoStop}%
\bibitem [{\citenamefont {Manyà}\ \emph {et~al.}(2011)\citenamefont {Manyà},
  \citenamefont {Antal}, \citenamefont {Kinoshita},\ and\ \citenamefont
  {Masutani}}]{manya2011}%
  \BibitemOpen
  \bibfield  {author} {\bibinfo {author} {\bibfnamefont {J.~J.}\ \bibnamefont
  {Manyà}}, \bibinfo {author} {\bibfnamefont {M.~J.}\ \bibnamefont {Antal},
  \bibfnamefont {Jr.}}, \bibinfo {author} {\bibfnamefont {C.~K.}\ \bibnamefont
  {Kinoshita}}, \ and\ \bibinfo {author} {\bibfnamefont {S.~M.}\ \bibnamefont
  {Masutani}},\ }\href@noop {} {\bibfield  {journal} {\bibinfo  {journal} {Ind.
  Eng. Chem. Res.}\ }\textbf {\bibinfo {volume} {50}},\ \bibinfo {pages} {6470}
  (\bibinfo {year} {2011})}\BibitemShut {NoStop}%
\bibitem [{\citenamefont {Osborne}, \citenamefont {Stimson},\ and\
  \citenamefont {Ginnings}(1939)}]{osborne1939}%
  \BibitemOpen
  \bibfield  {author} {\bibinfo {author} {\bibfnamefont {N.~S.}\ \bibnamefont
  {Osborne}}, \bibinfo {author} {\bibfnamefont {H.~F.}\ \bibnamefont
  {Stimson}}, \ and\ \bibinfo {author} {\bibfnamefont {D.~C.}\ \bibnamefont
  {Ginnings}},\ }\href@noop {} {\bibfield  {journal} {\bibinfo  {journal} {J.
  Res. Natl. Bur. Stand.}\ }\textbf {\bibinfo {volume} {23}},\ \bibinfo {pages}
  {197} (\bibinfo {year} {1939})}\BibitemShut {NoStop}%
\bibitem [{\citenamefont {Anisimov}\ \emph {et~al.}(1972)\citenamefont
  {Anisimov}, \citenamefont {Voronel'}, \citenamefont {Zaugol'nikova},\ and\
  \citenamefont {Ovodov}}]{anisimov1972}%
  \BibitemOpen
  \bibfield  {author} {\bibinfo {author} {\bibfnamefont {M.~A.}\ \bibnamefont
  {Anisimov}}, \bibinfo {author} {\bibfnamefont {A.~V.}\ \bibnamefont
  {Voronel'}}, \bibinfo {author} {\bibfnamefont {N.~S.}\ \bibnamefont
  {Zaugol'nikova}}, \ and\ \bibinfo {author} {\bibfnamefont {G.~I.}\
  \bibnamefont {Ovodov}},\ }\href@noop {} {\bibfield  {journal} {\bibinfo
  {journal} {JETP Lett.}\ }\textbf {\bibinfo {volume} {15}},\ \bibinfo {pages}
  {317} (\bibinfo {year} {1972})}\BibitemShut {NoStop}%
\bibitem [{\citenamefont {Angell}, \citenamefont {Shuppert},\ and\
  \citenamefont {Tucker}(1973)}]{angell1973}%
  \BibitemOpen
  \bibfield  {author} {\bibinfo {author} {\bibfnamefont {C.~A.}\ \bibnamefont
  {Angell}}, \bibinfo {author} {\bibfnamefont {J.}~\bibnamefont {Shuppert}}, \
  and\ \bibinfo {author} {\bibfnamefont {J.~C.}\ \bibnamefont {Tucker}},\
  }\href@noop {} {\bibfield  {journal} {\bibinfo  {journal} {J. Phys. Chem.}\
  }\textbf {\bibinfo {volume} {77}},\ \bibinfo {pages} {3092} (\bibinfo {year}
  {1973})}\BibitemShut {NoStop}%
\bibitem [{\citenamefont {Bertolini}, \citenamefont {Cassettari},\ and\
  \citenamefont {Salvetti}(1985)}]{bertolini1985}%
  \BibitemOpen
  \bibfield  {author} {\bibinfo {author} {\bibfnamefont {D.}~\bibnamefont
  {Bertolini}}, \bibinfo {author} {\bibfnamefont {M.}~\bibnamefont
  {Cassettari}}, \ and\ \bibinfo {author} {\bibfnamefont {G.}~\bibnamefont
  {Salvetti}},\ }\href@noop {} {\bibfield  {journal} {\bibinfo  {journal}
  {Chem. Phys. Lett.}\ }\textbf {\bibinfo {volume} {199}},\ \bibinfo {pages}
  {553} (\bibinfo {year} {1985})}\BibitemShut {NoStop}%
\bibitem [{\citenamefont {Tombari}, \citenamefont {Ferrari},\ and\
  \citenamefont {Salvetti}(1999)}]{tombari1999}%
  \BibitemOpen
  \bibfield  {author} {\bibinfo {author} {\bibfnamefont {E.}~\bibnamefont
  {Tombari}}, \bibinfo {author} {\bibfnamefont {C.}~\bibnamefont {Ferrari}}, \
  and\ \bibinfo {author} {\bibfnamefont {G.}~\bibnamefont {Salvetti}},\
  }\href@noop {} {\bibfield  {journal} {\bibinfo  {journal} {Chem. Phys.
  Lett.}\ }\textbf {\bibinfo {volume} {300}},\ \bibinfo {pages} {749} (\bibinfo
  {year} {1999})}\BibitemShut {NoStop}%
\bibitem [{\citenamefont {Archer}\ and\ \citenamefont {Carter}(2000)}]{arc00}%
  \BibitemOpen
  \bibfield  {author} {\bibinfo {author} {\bibfnamefont {D.~G.}\ \bibnamefont
  {Archer}}\ and\ \bibinfo {author} {\bibfnamefont {R.~W.}\ \bibnamefont
  {Carter}},\ }\href@noop {} {\bibfield  {journal} {\bibinfo  {journal}
  {J.~Phys. Chem.~B}\ }\textbf {\bibinfo {volume} {104}},\ \bibinfo {pages}
  {8563} (\bibinfo {year} {2000})}\BibitemShut {NoStop}%
\bibitem [{\citenamefont {Preston-Thomas}(1990)}]{prestonthomas1990}%
  \BibitemOpen
  \bibfield  {author} {\bibinfo {author} {\bibfnamefont {H.}~\bibnamefont
  {Preston-Thomas}},\ }\href@noop {} {\bibfield  {journal} {\bibinfo  {journal}
  {Metrologia}\ }\textbf {\bibinfo {volume} {27}},\ \bibinfo {pages} {3}
  (\bibinfo {year} {1990})}\BibitemShut {NoStop}%
\bibitem [{\citenamefont {Rusby}(1991)}]{rusby91}%
  \BibitemOpen
  \bibfield  {author} {\bibinfo {author} {\bibfnamefont {R.~L.}\ \bibnamefont
  {Rusby}},\ }\href@noop {} {\bibfield  {journal} {\bibinfo  {journal} {J.
  Chem. Thermodyn.}\ }\textbf {\bibinfo {volume} {23}},\ \bibinfo {pages}
  {1153} (\bibinfo {year} {1991})}\BibitemShut {NoStop}%
\bibitem [{\citenamefont {Bedford}\ and\ \citenamefont
  {Kirby}(1969)}]{bedford1969}%
  \BibitemOpen
  \bibfield  {author} {\bibinfo {author} {\bibfnamefont {R.~E.}\ \bibnamefont
  {Bedford}}\ and\ \bibinfo {author} {\bibfnamefont {C.~G.~M.}\ \bibnamefont
  {Kirby}},\ }\href@noop {} {\bibfield  {journal} {\bibinfo  {journal}
  {Metrologia}\ }\textbf {\bibinfo {volume} {5}},\ \bibinfo {pages} {83}
  (\bibinfo {year} {1969})}\BibitemShut {NoStop}%
\bibitem [{\citenamefont {Douglas}(1969)}]{douglas1969}%
  \BibitemOpen
  \bibfield  {author} {\bibinfo {author} {\bibfnamefont {T.~B.}\ \bibnamefont
  {Douglas}},\ }\href@noop {} {\bibfield  {journal} {\bibinfo  {journal} {J.
  Res. Nat. Bur. Stand.}\ }\textbf {\bibinfo {volume} {73A}},\ \bibinfo {pages}
  {451} (\bibinfo {year} {1969})}\BibitemShut {NoStop}%
\bibitem [{\citenamefont {Goldberg}\ and\ \citenamefont
  {Weir}(1992)}]{goldberg1992}%
  \BibitemOpen
  \bibfield  {author} {\bibinfo {author} {\bibfnamefont {R.~N.}\ \bibnamefont
  {Goldberg}}\ and\ \bibinfo {author} {\bibfnamefont {R.~D.}\ \bibnamefont
  {Weir}},\ }\href@noop {} {\bibfield  {journal} {\bibinfo  {journal} {Pure
  Appl. Chem.}\ }\textbf {\bibinfo {volume} {64}},\ \bibinfo {pages} {1545}
  (\bibinfo {year} {1992})}\BibitemShut {NoStop}%
\bibitem [{\citenamefont {Weir}\ and\ \citenamefont
  {Goldberg}(1996)}]{weir1996}%
  \BibitemOpen
  \bibfield  {author} {\bibinfo {author} {\bibfnamefont {R.~D.}\ \bibnamefont
  {Weir}}\ and\ \bibinfo {author} {\bibfnamefont {R.~N.}\ \bibnamefont
  {Goldberg}},\ }\href@noop {} {\bibfield  {journal} {\bibinfo  {journal} {J.
  Chem. Thermodyn.}\ }\textbf {\bibinfo {volume} {28}},\ \bibinfo {pages} {261}
  (\bibinfo {year} {1996})}\BibitemShut {NoStop}%
\bibitem [{\citenamefont {Bottomley}(1978)}]{bottomley1978}%
  \BibitemOpen
  \bibfield  {author} {\bibinfo {author} {\bibfnamefont {G.~A.}\ \bibnamefont
  {Bottomley}},\ }\href {\doibase10.1071/CH9781177} {\bibfield  {journal}
  {\bibinfo  {journal} {Aust. J. Chem.}\ }\textbf {\bibinfo {volume} {31}},\
  \bibinfo {pages} {1177} (\bibinfo {year} {1978})}\BibitemShut {NoStop}%
\bibitem [{SCW()}]{SCWsupplement2013}%
  \BibitemOpen
  \href@noop {} {}\bibinfo {note} {See supplemental material at [URL will be
  inserted by AIP] for tables with experimental data and computer code for the
  equation of state.}\BibitemShut {Stop}%
\bibitem [{\citenamefont {Nilsson}, \citenamefont {Huang},\ and\ \citenamefont
  {Pettersson}(2012)}]{nilsson2012}%
  \BibitemOpen
  \bibfield  {author} {\bibinfo {author} {\bibfnamefont {A.}~\bibnamefont
  {Nilsson}}, \bibinfo {author} {\bibfnamefont {C.}~\bibnamefont {Huang}}, \
  and\ \bibinfo {author} {\bibfnamefont {L.~G.}\ \bibnamefont {Pettersson}},\
  }\href@noop {} {\bibfield  {journal} {\bibinfo  {journal} {J. Mol. Liq.}\
  }\textbf {\bibinfo {volume} {176}},\ \bibinfo {pages} {2} (\bibinfo {year}
  {2012})}\BibitemShut {NoStop}%
\bibitem [{\citenamefont {Taschin}\ \emph {et~al.}(2013)\citenamefont
  {Taschin}, \citenamefont {Bartolini}, \citenamefont {Eramo}, \citenamefont
  {Righini},\ and\ \citenamefont {Torre}}]{taschin2013}%
  \BibitemOpen
  \bibfield  {author} {\bibinfo {author} {\bibfnamefont {A.}~\bibnamefont
  {Taschin}}, \bibinfo {author} {\bibfnamefont {P.}~\bibnamefont {Bartolini}},
  \bibinfo {author} {\bibfnamefont {R.}~\bibnamefont {Eramo}}, \bibinfo
  {author} {\bibfnamefont {R.}~\bibnamefont {Righini}}, \ and\ \bibinfo
  {author} {\bibfnamefont {R.}~\bibnamefont {Torre}},\ }\href@noop {}
  {\bibfield  {journal} {\bibinfo  {journal} {Nat. Commun.}\ }\textbf {\bibinfo
  {volume} {4}},\ \bibinfo {pages} {2401} (\bibinfo {year} {2013})}\BibitemShut
  {NoStop}%
\bibitem [{\citenamefont {Mishima}(2000)}]{mishima2000}%
  \BibitemOpen
  \bibfield  {author} {\bibinfo {author} {\bibfnamefont {O.}~\bibnamefont
  {Mishima}},\ }\href@noop {} {\bibfield  {journal} {\bibinfo  {journal} {Phys.
  Rev. Lett.}\ }\textbf {\bibinfo {volume} {85}},\ \bibinfo {pages} {334}
  (\bibinfo {year} {2000})}\BibitemShut {NoStop}%
\bibitem [{\citenamefont {Holten}\ and\ \citenamefont
  {Anisimov}(2012{\natexlab{b}})}]{holtentwostatesupplement}%
  \BibitemOpen
  \bibfield  {author} {\bibinfo {author} {\bibfnamefont {V.}~\bibnamefont
  {Holten}}\ and\ \bibinfo {author} {\bibfnamefont {M.~A.}\ \bibnamefont
  {Anisimov}},\ }\href@noop {} {\bibfield  {journal} {\bibinfo  {journal} {Sci.
  Rep.}\ }\textbf {\bibinfo {volume} {2}},\ \bibinfo {pages} {713} (\bibinfo
  {year} {2012}{\natexlab{b}})},\ \bibinfo {note} {supplementary
  information}\BibitemShut {NoStop}%
\bibitem [{\citenamefont {Lemmon}, \citenamefont {McLinden},\ and\
  \citenamefont {Wagner}(2009)}]{lemmon2009}%
  \BibitemOpen
  \bibfield  {author} {\bibinfo {author} {\bibfnamefont {E.~W.}\ \bibnamefont
  {Lemmon}}, \bibinfo {author} {\bibfnamefont {M.~O.}\ \bibnamefont
  {McLinden}}, \ and\ \bibinfo {author} {\bibfnamefont {W.}~\bibnamefont
  {Wagner}},\ }\href@noop {} {\bibfield  {journal} {\bibinfo  {journal}
  {J.~Chem. Eng. Data}\ }\textbf {\bibinfo {volume} {54}},\ \bibinfo {pages}
  {3141} (\bibinfo {year} {2009})}\BibitemShut {NoStop}%
\bibitem [{\citenamefont {Henderson}\ and\ \citenamefont
  {Speedy}(1987{\natexlab{a}})}]{henderson1987a}%
  \BibitemOpen
  \bibfield  {author} {\bibinfo {author} {\bibfnamefont {S.~J.}\ \bibnamefont
  {Henderson}}\ and\ \bibinfo {author} {\bibfnamefont {R.~J.}\ \bibnamefont
  {Speedy}},\ }\href@noop {} {\bibfield  {journal} {\bibinfo  {journal} {J.
  Phys. Chem.}\ }\textbf {\bibinfo {volume} {91}},\ \bibinfo {pages} {3062}
  (\bibinfo {year} {1987}{\natexlab{a}})}\BibitemShut {NoStop}%
\bibitem [{\citenamefont {Millero}, \citenamefont {Curry},\ and\ \citenamefont
  {Drost-Hansen}(1969)}]{millero1969}%
  \BibitemOpen
  \bibfield  {author} {\bibinfo {author} {\bibfnamefont {F.~J.}\ \bibnamefont
  {Millero}}, \bibinfo {author} {\bibfnamefont {R.~W.}\ \bibnamefont {Curry}},
  \ and\ \bibinfo {author} {\bibfnamefont {W.}~\bibnamefont {Drost-Hansen}},\
  }\href@noop {} {\bibfield  {journal} {\bibinfo  {journal} {J.~Chem. Eng.
  Data}\ }\textbf {\bibinfo {volume} {14}},\ \bibinfo {pages} {422} (\bibinfo
  {year} {1969})}\BibitemShut {NoStop}%
\bibitem [{\citenamefont {Wang}\ \emph {et~al.}(2013)\citenamefont {Wang},
  \citenamefont {Liu}, \citenamefont {Zhou}, \citenamefont {Song},
  \citenamefont {Bi}, \citenamefont {Liu},\ and\ \citenamefont
  {Xie}}]{wang2013}%
  \BibitemOpen
  \bibfield  {author} {\bibinfo {author} {\bibfnamefont {Z.-G.}\ \bibnamefont
  {Wang}}, \bibinfo {author} {\bibfnamefont {Y.-G.}\ \bibnamefont {Liu}},
  \bibinfo {author} {\bibfnamefont {W.-G.}\ \bibnamefont {Zhou}}, \bibinfo
  {author} {\bibfnamefont {W.}~\bibnamefont {Song}}, \bibinfo {author}
  {\bibfnamefont {Y.}~\bibnamefont {Bi}}, \bibinfo {author} {\bibfnamefont
  {L.}~\bibnamefont {Liu}}, \ and\ \bibinfo {author} {\bibfnamefont {H.-S.}\
  \bibnamefont {Xie}},\ }\href@noop {} {\bibfield  {journal} {\bibinfo
  {journal} {Chin. Phys. Lett.}\ }\textbf {\bibinfo {volume} {30}},\ \bibinfo
  {pages} {054302} (\bibinfo {year} {2013})}\BibitemShut {NoStop}%
\bibitem [{\citenamefont {Abramson}\ and\ \citenamefont
  {Brown}(2004)}]{abramson2004}%
  \BibitemOpen
  \bibfield  {author} {\bibinfo {author} {\bibfnamefont {E.~H.}\ \bibnamefont
  {Abramson}}\ and\ \bibinfo {author} {\bibfnamefont {J.~M.}\ \bibnamefont
  {Brown}},\ }\href@noop {} {\bibfield  {journal} {\bibinfo  {journal}
  {Geochim. Cosmochim. Acta}\ }\textbf {\bibinfo {volume} {68}},\ \bibinfo
  {pages} {1827} (\bibinfo {year} {2004})}\BibitemShut {NoStop}%
\bibitem [{\citenamefont {Feistel}\ and\ \citenamefont
  {Wagner}(2006)}]{feistel2006}%
  \BibitemOpen
  \bibfield  {author} {\bibinfo {author} {\bibfnamefont {R.}~\bibnamefont
  {Feistel}}\ and\ \bibinfo {author} {\bibfnamefont {W.}~\bibnamefont
  {Wagner}},\ }\href@noop {} {\bibfield  {journal} {\bibinfo  {journal}
  {J.~Phys. Chem. Ref. Data}\ }\textbf {\bibinfo {volume} {35}},\ \bibinfo
  {pages} {1021} (\bibinfo {year} {2006})}\BibitemShut {NoStop}%
\bibitem [{\citenamefont {Feistel}\ \emph {et~al.}(2008)\citenamefont
  {Feistel}, \citenamefont {Wright}, \citenamefont {Miyagawa}, \citenamefont
  {Harvey}, \citenamefont {Hruby}, \citenamefont {Jackett}, \citenamefont
  {McDougall},\ and\ \citenamefont {Wagner}}]{feistel2008}%
  \BibitemOpen
  \bibfield  {author} {\bibinfo {author} {\bibfnamefont {R.}~\bibnamefont
  {Feistel}}, \bibinfo {author} {\bibfnamefont {D.~G.}\ \bibnamefont {Wright}},
  \bibinfo {author} {\bibfnamefont {K.}~\bibnamefont {Miyagawa}}, \bibinfo
  {author} {\bibfnamefont {A.~H.}\ \bibnamefont {Harvey}}, \bibinfo {author}
  {\bibfnamefont {J.}~\bibnamefont {Hruby}}, \bibinfo {author} {\bibfnamefont
  {D.~R.}\ \bibnamefont {Jackett}}, \bibinfo {author} {\bibfnamefont {T.~J.}\
  \bibnamefont {McDougall}}, \ and\ \bibinfo {author} {\bibfnamefont
  {W.}~\bibnamefont {Wagner}},\ }\href@noop {} {\bibfield  {journal} {\bibinfo
  {journal} {Ocean Sci.}\ }\textbf {\bibinfo {volume} {4}},\ \bibinfo {pages}
  {275} (\bibinfo {year} {2008})}\BibitemShut {NoStop}%
\bibitem [{\citenamefont {Guildner}, \citenamefont {Johnson},\ and\
  \citenamefont {Jones}(1976)}]{guildner1976}%
  \BibitemOpen
  \bibfield  {author} {\bibinfo {author} {\bibfnamefont {L.~A.}\ \bibnamefont
  {Guildner}}, \bibinfo {author} {\bibfnamefont {D.~P.}\ \bibnamefont
  {Johnson}}, \ and\ \bibinfo {author} {\bibfnamefont {F.~E.}\ \bibnamefont
  {Jones}},\ }\href@noop {} {\bibfield  {journal} {\bibinfo  {journal} {J. Res.
  Natl. Bur. Stand.}\ }\textbf {\bibinfo {volume} {80A}},\ \bibinfo {pages}
  {505} (\bibinfo {year} {1976})}\BibitemShut {NoStop}%
\bibitem [{\citenamefont {Bignell}\ and\ \citenamefont
  {Bean}(1988)}]{bignell1988}%
  \BibitemOpen
  \bibfield  {author} {\bibinfo {author} {\bibfnamefont {N.}~\bibnamefont
  {Bignell}}\ and\ \bibinfo {author} {\bibfnamefont {V.~E.}\ \bibnamefont
  {Bean}},\ }\href@noop {} {\bibfield  {journal} {\bibinfo  {journal}
  {Metrologia}\ }\textbf {\bibinfo {volume} {25}},\ \bibinfo {pages} {205}
  (\bibinfo {year} {1988})}\BibitemShut {NoStop}%
\bibitem [{\citenamefont {Babb}(1963)}]{babb1963}%
  \BibitemOpen
  \bibfield  {author} {\bibinfo {author} {\bibfnamefont {S.~E.}\ \bibnamefont
  {Babb}, \bibfnamefont {Jr.}},\ }in\ \href@noop {} {\emph {\bibinfo
  {booktitle} {High-Pressure Measurement}}},\ \bibinfo {editor} {edited by\
  \bibinfo {editor} {\bibfnamefont {A.~A.}\ \bibnamefont {Giardini}}\ and\
  \bibinfo {editor} {\bibfnamefont {E.~C.}\ \bibnamefont {Lloyd}}}\ (\bibinfo
  {publisher} {Butterworths},\ \bibinfo {address} {Washington},\ \bibinfo
  {year} {1963})\ pp.\ \bibinfo {pages} {115--124}\BibitemShut {NoStop}%
\bibitem [{\citenamefont {La~Mori}(1965)}]{lamori1965}%
  \BibitemOpen
  \bibfield  {author} {\bibinfo {author} {\bibfnamefont {P.~N.}\ \bibnamefont
  {La~Mori}},\ }in\ \href@noop {} {\emph {\bibinfo {booktitle} {1964 Symposium
  on High-Pressure Technology}}},\ \bibinfo {editor} {edited by\ \bibinfo
  {editor} {\bibfnamefont {E.~C.}\ \bibnamefont {Lloyd}}\ and\ \bibinfo
  {editor} {\bibfnamefont {A.~A.}\ \bibnamefont {Giardini}}}\ (\bibinfo
  {publisher} {American Society of Mechanical Engineers},\ \bibinfo {address}
  {New York},\ \bibinfo {year} {1965})\ \bibinfo {note} {paper
  64-WA/PT-25}\BibitemShut {NoStop}%
\bibitem [{\citenamefont {Molinar}\ \emph {et~al.}(1980)\citenamefont
  {Molinar}, \citenamefont {Bean}, \citenamefont {Houck},\ and\ \citenamefont
  {Welch}}]{molinar1980}%
  \BibitemOpen
  \bibfield  {author} {\bibinfo {author} {\bibfnamefont {G.~F.}\ \bibnamefont
  {Molinar}}, \bibinfo {author} {\bibfnamefont {V.}~\bibnamefont {Bean}},
  \bibinfo {author} {\bibfnamefont {J.}~\bibnamefont {Houck}}, \ and\ \bibinfo
  {author} {\bibfnamefont {B.}~\bibnamefont {Welch}},\ }\href@noop {}
  {\bibfield  {journal} {\bibinfo  {journal} {Metrologia}\ }\textbf {\bibinfo
  {volume} {16}},\ \bibinfo {pages} {21} (\bibinfo {year} {1980})}\BibitemShut
  {NoStop}%
\bibitem [{\citenamefont {Evans}(1967)}]{evans1967}%
  \BibitemOpen
  \bibfield  {author} {\bibinfo {author} {\bibfnamefont {L.~F.}\ \bibnamefont
  {Evans}},\ }\href@noop {} {\bibfield  {journal} {\bibinfo  {journal} {J.
  Appl. Phys.}\ }\textbf {\bibinfo {volume} {38}},\ \bibinfo {pages} {4930}
  (\bibinfo {year} {1967})}\BibitemShut {NoStop}%
\bibitem [{\citenamefont {Henderson}\ and\ \citenamefont
  {Speedy}(1987{\natexlab{b}})}]{henderson1987b}%
  \BibitemOpen
  \bibfield  {author} {\bibinfo {author} {\bibfnamefont {S.~J.}\ \bibnamefont
  {Henderson}}\ and\ \bibinfo {author} {\bibfnamefont {R.~J.}\ \bibnamefont
  {Speedy}},\ }\href@noop {} {\bibfield  {journal} {\bibinfo  {journal} {J.
  Phys. Chem.}\ }\textbf {\bibinfo {volume} {91}},\ \bibinfo {pages} {3069}
  (\bibinfo {year} {1987}{\natexlab{b}})}\BibitemShut {NoStop}%
\bibitem [{\citenamefont {Nordell}(1990)}]{nordell1990}%
  \BibitemOpen
  \bibfield  {author} {\bibinfo {author} {\bibfnamefont {B.}~\bibnamefont
  {Nordell}},\ }\href@noop {} {\bibfield  {journal} {\bibinfo  {journal} {Cold
  Reg. Sci. Technol.}\ }\textbf {\bibinfo {volume} {19}},\ \bibinfo {pages}
  {83} (\bibinfo {year} {1990})}\BibitemShut {NoStop}%
\bibitem [{\citenamefont {Maruyama}(2005)}]{maruyama2005}%
  \BibitemOpen
  \bibfield  {author} {\bibinfo {author} {\bibfnamefont {M.}~\bibnamefont
  {Maruyama}},\ }\href@noop {} {\bibfield  {journal} {\bibinfo  {journal} {J.
  Crystal Growth}\ }\textbf {\bibinfo {volume} {275}},\ \bibinfo {pages} {598}
  (\bibinfo {year} {2005})}\BibitemShut {NoStop}%
\bibitem [{\citenamefont {Mishima}(1996)}]{mishima1996}%
  \BibitemOpen
  \bibfield  {author} {\bibinfo {author} {\bibfnamefont {O.}~\bibnamefont
  {Mishima}},\ }\href@noop {} {\bibfield  {journal} {\bibinfo  {journal}
  {Nature}\ }\textbf {\bibinfo {volume} {384}},\ \bibinfo {pages} {546}
  (\bibinfo {year} {1996})}\BibitemShut {NoStop}%
\bibitem [{\citenamefont {Mishima}(2011)}]{mishima2011}%
  \BibitemOpen
  \bibfield  {author} {\bibinfo {author} {\bibfnamefont {O.}~\bibnamefont
  {Mishima}},\ }\href@noop {} {\bibfield  {journal} {\bibinfo  {journal}
  {J.~Phys. Chem.~B}\ }\textbf {\bibinfo {volume} {115}},\ \bibinfo {pages}
  {14064} (\bibinfo {year} {2011})}\BibitemShut {NoStop}%
\bibitem [{\citenamefont {Kishimoto}\ and\ \citenamefont
  {Maruyama}(1998)}]{kishimoto1998}%
  \BibitemOpen
  \bibfield  {author} {\bibinfo {author} {\bibfnamefont {Y.}~\bibnamefont
  {Kishimoto}}\ and\ \bibinfo {author} {\bibfnamefont {M.}~\bibnamefont
  {Maruyama}},\ }\href@noop {} {\bibfield  {journal} {\bibinfo  {journal} {Rev.
  High Pressure Sci. Technol.}\ }\textbf {\bibinfo {volume} {7}},\ \bibinfo
  {pages} {1144} (\bibinfo {year} {1998})}\BibitemShut {NoStop}%
\bibitem [{\citenamefont {Murphy}\ and\ \citenamefont {Koop}(2005)}]{mur05}%
  \BibitemOpen
  \bibfield  {author} {\bibinfo {author} {\bibfnamefont {D.~M.}\ \bibnamefont
  {Murphy}}\ and\ \bibinfo {author} {\bibfnamefont {T.}~\bibnamefont {Koop}},\
  }\href@noop {} {\bibfield  {journal} {\bibinfo  {journal} {Q.~J.~R. Meteorol.
  Soc.}\ }\textbf {\bibinfo {volume} {131}},\ \bibinfo {pages} {1539} (\bibinfo
  {year} {2005})}\BibitemShut {NoStop}%
\bibitem [{\citenamefont {Scheel}\ and\ \citenamefont
  {Heuse}(1909)}]{scheel1909}%
  \BibitemOpen
  \bibfield  {author} {\bibinfo {author} {\bibfnamefont {K.}~\bibnamefont
  {Scheel}}\ and\ \bibinfo {author} {\bibfnamefont {W.}~\bibnamefont {Heuse}},\
  }\href {\doibase10.1002/andp.19093340906} {\bibfield  {journal} {\bibinfo
  {journal} {Ann. Phys.}\ }\textbf {\bibinfo {volume} {334}},\ \bibinfo {pages}
  {723} (\bibinfo {year} {1909})}\BibitemShut {NoStop}%
\bibitem [{\citenamefont {Kraus}\ and\ \citenamefont
  {Greer}(1984)}]{kraus1984}%
  \BibitemOpen
  \bibfield  {author} {\bibinfo {author} {\bibfnamefont {G.~F.}\ \bibnamefont
  {Kraus}}\ and\ \bibinfo {author} {\bibfnamefont {S.~C.}\ \bibnamefont
  {Greer}},\ }\href {\doibase10.1021/j150664a067} {\bibfield  {journal}
  {\bibinfo  {journal} {J.~Phys. Chem.}\ }\textbf {\bibinfo {volume} {88}},\
  \bibinfo {pages} {4781} (\bibinfo {year} {1984})}\BibitemShut {NoStop}%
\bibitem [{\citenamefont {Fukuta}\ and\ \citenamefont
  {Gramada}(2003)}]{fukuta2003}%
  \BibitemOpen
  \bibfield  {author} {\bibinfo {author} {\bibfnamefont {N.}~\bibnamefont
  {Fukuta}}\ and\ \bibinfo {author} {\bibfnamefont {C.~M.}\ \bibnamefont
  {Gramada}},\ }\href {\doibase10.1175/1520-0469(2003)060<1871:VPMOSW>2.0.CO;2}
  {\bibfield  {journal} {\bibinfo  {journal} {J. Atmos. Sci.}\ }\textbf
  {\bibinfo {volume} {60}},\ \bibinfo {pages} {1871} (\bibinfo {year}
  {2003})}\BibitemShut {NoStop}%
\bibitem [{\citenamefont {Cantrell}\ \emph {et~al.}(2008)\citenamefont
  {Cantrell}, \citenamefont {Ochshorn}, \citenamefont {Kostinski},\ and\
  \citenamefont {Bozin}}]{cantrell2008}%
  \BibitemOpen
  \bibfield  {author} {\bibinfo {author} {\bibfnamefont {W.}~\bibnamefont
  {Cantrell}}, \bibinfo {author} {\bibfnamefont {E.}~\bibnamefont {Ochshorn}},
  \bibinfo {author} {\bibfnamefont {A.}~\bibnamefont {Kostinski}}, \ and\
  \bibinfo {author} {\bibfnamefont {K.}~\bibnamefont {Bozin}},\ }\href
  {\doibase10.1175/2008/JTECHA1028.1} {\bibfield  {journal} {\bibinfo
  {journal} {J. Atmos. Oceanic Technol.}\ }\textbf {\bibinfo {volume} {25}},\
  \bibinfo {pages} {1724} (\bibinfo {year} {2008})}\BibitemShut {NoStop}%
\bibitem [{\citenamefont {Cantrell}\ \emph {et~al.}(2009)\citenamefont
  {Cantrell}, \citenamefont {Ochshorn}, \citenamefont {Kostinski},\ and\
  \citenamefont {Bozin}}]{cantrell2009}%
  \BibitemOpen
  \bibfield  {author} {\bibinfo {author} {\bibfnamefont {W.}~\bibnamefont
  {Cantrell}}, \bibinfo {author} {\bibfnamefont {E.}~\bibnamefont {Ochshorn}},
  \bibinfo {author} {\bibfnamefont {A.}~\bibnamefont {Kostinski}}, \ and\
  \bibinfo {author} {\bibfnamefont {K.}~\bibnamefont {Bozin}},\ }\href
  {\doibase10.1175/2008JTECHA1269.1} {\bibfield  {journal} {\bibinfo  {journal}
  {J. Atmos. Oceanic Technol.}\ }\textbf {\bibinfo {volume} {26}},\ \bibinfo
  {pages} {853} (\bibinfo {year} {2009})}\BibitemShut {NoStop}%
\bibitem [{\citenamefont {Duan}, \citenamefont {Thompson},\ and\ \citenamefont
  {Ward}(2008)}]{duan2008}%
  \BibitemOpen
  \bibfield  {author} {\bibinfo {author} {\bibfnamefont {F.}~\bibnamefont
  {Duan}}, \bibinfo {author} {\bibfnamefont {I.}~\bibnamefont {Thompson}}, \
  and\ \bibinfo {author} {\bibfnamefont {C.~A.}\ \bibnamefont {Ward}},\
  }\href@noop {} {\bibfield  {journal} {\bibinfo  {journal} {J.~Phys. Chem.~B}\
  }\textbf {\bibinfo {volume} {112}},\ \bibinfo {pages} {8605} (\bibinfo {year}
  {2008})}\BibitemShut {NoStop}%
\bibitem [{\citenamefont {Xans}\ and\ \citenamefont
  {Barnaud}(1975)}]{xans1975}%
  \BibitemOpen
  \bibfield  {author} {\bibinfo {author} {\bibfnamefont {P.}~\bibnamefont
  {Xans}}\ and\ \bibinfo {author} {\bibfnamefont {G.}~\bibnamefont {Barnaud}},\
  }\href@noop {} {\bibfield  {journal} {\bibinfo  {journal} {C. R. Acad. Sci.
  Paris}\ }\textbf {\bibinfo {volume} {280}},\ \bibinfo {pages} {25} (\bibinfo
  {year} {1975})}\BibitemShut {NoStop}%
\bibitem [{\citenamefont {Kanno}, \citenamefont {Speedy},\ and\ \citenamefont
  {Angell}(1975)}]{kanno1975}%
  \BibitemOpen
  \bibfield  {author} {\bibinfo {author} {\bibfnamefont {H.}~\bibnamefont
  {Kanno}}, \bibinfo {author} {\bibfnamefont {R.~J.}\ \bibnamefont {Speedy}}, \
  and\ \bibinfo {author} {\bibfnamefont {C.~A.}\ \bibnamefont {Angell}},\
  }\href@noop {} {\bibfield  {journal} {\bibinfo  {journal} {Science}\ }\textbf
  {\bibinfo {volume} {189}},\ \bibinfo {pages} {880} (\bibinfo {year}
  {1975})}\BibitemShut {NoStop}%
\bibitem [{\citenamefont {Kanno}\ and\ \citenamefont
  {Miyata}(2006)}]{kanno2006}%
  \BibitemOpen
  \bibfield  {author} {\bibinfo {author} {\bibfnamefont {H.}~\bibnamefont
  {Kanno}}\ and\ \bibinfo {author} {\bibfnamefont {K.}~\bibnamefont {Miyata}},\
  }\href@noop {} {\bibfield  {journal} {\bibinfo  {journal} {Chem. Phys.
  Lett.}\ }\textbf {\bibinfo {volume} {422}},\ \bibinfo {pages} {507} (\bibinfo
  {year} {2006})}\BibitemShut {NoStop}%
\bibitem [{\citenamefont {Mishima}\ and\ \citenamefont
  {Stanley}(1998{\natexlab{b}})}]{mishima1998}%
  \BibitemOpen
  \bibfield  {author} {\bibinfo {author} {\bibfnamefont {O.}~\bibnamefont
  {Mishima}}\ and\ \bibinfo {author} {\bibfnamefont {H.~E.}\ \bibnamefont
  {Stanley}},\ }\href@noop {} {\bibfield  {journal} {\bibinfo  {journal}
  {Nature}\ }\textbf {\bibinfo {volume} {392}},\ \bibinfo {pages} {164}
  (\bibinfo {year} {1998}{\natexlab{b}})}\BibitemShut {NoStop}%
\bibitem [{\citenamefont {Kanno}\ and\ \citenamefont
  {Angell}(1977)}]{kanno1977}%
  \BibitemOpen
  \bibfield  {author} {\bibinfo {author} {\bibfnamefont {H.}~\bibnamefont
  {Kanno}}\ and\ \bibinfo {author} {\bibfnamefont {C.~A.}\ \bibnamefont
  {Angell}},\ }\href@noop {} {\bibfield  {journal} {\bibinfo  {journal} {J.
  Phys. Chem.}\ }\textbf {\bibinfo {volume} {81}},\ \bibinfo {pages} {2639}
  (\bibinfo {year} {1977})}\BibitemShut {NoStop}%
\bibitem [{\citenamefont {Wagner}, \citenamefont {Saul},\ and\ \citenamefont
  {Pruß}(1994)}]{wagner1994}%
  \BibitemOpen
  \bibfield  {author} {\bibinfo {author} {\bibfnamefont {W.}~\bibnamefont
  {Wagner}}, \bibinfo {author} {\bibfnamefont {A.}~\bibnamefont {Saul}}, \ and\
  \bibinfo {author} {\bibfnamefont {A.}~\bibnamefont {Pruß}},\ }\href@noop {}
  {\bibfield  {journal} {\bibinfo  {journal} {J.~Phys. Chem. Ref. Data}\
  }\textbf {\bibinfo {volume} {23}},\ \bibinfo {pages} {515} (\bibinfo {year}
  {1994})}\BibitemShut {NoStop}%
\bibitem [{\citenamefont {Mohr}, \citenamefont {Taylor},\ and\ \citenamefont
  {Newell}(2012)}]{mohr2012}%
  \BibitemOpen
  \bibfield  {author} {\bibinfo {author} {\bibfnamefont {P.~J.}\ \bibnamefont
  {Mohr}}, \bibinfo {author} {\bibfnamefont {B.~N.}\ \bibnamefont {Taylor}}, \
  and\ \bibinfo {author} {\bibfnamefont {D.~B.}\ \bibnamefont {Newell}},\
  }\href@noop {} {\bibfield  {journal} {\bibinfo  {journal} {Rev. Mod. Phys.}\
  }\textbf {\bibinfo {volume} {84}},\ \bibinfo {pages} {1527} (\bibinfo {year}
  {2012})}\BibitemShut {NoStop}%
\end{thebibliography}%

\end{document}